\documentclass[trackchanges,twocolumn]{aastex62}
\usepackage{natbib}
\usepackage{amssymb,amsmath}
\usepackage{graphicx}
\usepackage{color}
\usepackage{fancyhdr}
\usepackage[T1]{fontenc}
\usepackage{lipsum, babel}
\usepackage{hyperref}

\newcommand{\V}[1]{\mathbf{#1}} 
\newcommand{\B}{\mathbf{B}}
\newcommand{\J}{\mathbf{J}}
\newcommand{\gkeyll}{{\tt Gkeyll}}
\newcommand{\Alfven}{Alfv\'{e}n}
\newcommand{\figref}[1]{Fig.~\ref{#1}}   
\newcommand{\uofaMaths}{\affiliation{Department of Applied Mathematics, University of Arizona, Tucson, AZ 85721, USA}}
\newcommand{\uofa}{\affiliation{Lunar and Planetary Laboratory, University of Arizona, Tucson, AZ 85721, USA}}
\newcommand{\princeton}{\affiliation{Department of Astrophysical Sciences, Peyton Hall, Princeton University, Princeton, NJ 08544, USA}}
\newcommand{\oaw}{\affiliation{Space Research Institute, Austrian Academy of Sciences, Schmiedlstrasse 6, 8042 Graz, AT}}
\newcommand{\ucl}{\affiliation{Mullard Space Science Laboratory, University College London, Holmbury St. Mary Dorking, Surrey RH5 6NT, UK}}
\newcommand{\unh}{\affiliation{Space Science Center, University of New Hampshire, Durham, NH 03824, USA}}

\begin{document}

\title{Magnetic Field Reconstruction for a Realistic Multi-Point, Multi-Scale Spacecraft Observatory}

\author[0000-0002-2649-020X]{T. Broeren}\uofaMaths
\author[0000-0001-6038-1923]{K.~G. Klein}\uofa
\author[0000-0003-0143-951X]{J.~M. TenBarge} \princeton
\author{Ivan Dors}
\author[0000-0002-3913-1353]{O.~W. Roberts} \oaw
\author{D. Verscharen} \ucl,\unh

\begin{abstract}
Future in situ space plasma investigations will likely involve spatially distributed observatories comprised of multiple spacecraft, beyond the four and five spacecraft configurations currently in operation. Inferring the magnetic field structure across the observatory, and not simply at the observation points, is a necessary step towards characterizing fundamental plasma processes using these unique multi-point, multi-scale data sets. We propose improvements upon the classic first-order reconstruction method, as well as a second-order method, utilizing magnetometer measurements from a realistic nine-spacecraft observatory. The improved first-order method, which averages over select ensembles of four spacecraft, reconstructs the magnetic field associated with simple current sheets and numerical simulations of turbulence accurately over larger volumes compared to second-order methods or first-order methods using a single regular tetrahedron. Using this averaging method on data sets with fewer than nine measurement points, the volume of accurate reconstruction compared to a known magnetic vector field improves approximately linearly with the number of measurement points. 

\end{abstract}

\section{Introduction}
\label{sec:intro}

Plasmas, which are ubiquitous throughout the universe, are readily available for study in the natural laboratory of space. Electromagnetic fields play a fundamental role in the transport, heating, and acceleration of charged particles that compose plasmas. 
In order to characterize fundamental processes governing heliospheric plasmas, the space plasma community has utilized in-situ spacecraft measurements of electromagnetic fields and charged particles. These in-situ measurements include the characterization of the vector magnetic field $\B$ at a spacecraft via magnetometers; see \S 2.4 of \cite{Verscharen:2019}. 

Knowledge of $\B$ from a single magnetometer is limited; single-point measurements can not construct the full three-dimensional structure characteristic of processes such as magnetic reconnection and plasma turbulence. To avoid this shortcoming, ESA's CLUSTER \citep{Escoubet:2001}, NASA's THEMIS \citep{Angelopoulos:2008} and MMS \citep{Burch:2016} missions have employed four- and five-spacecraft configurations, where each spacecraft is equipped with an instrument suite that includes a magnetometer. These missions study the boundaries of the Earth's magnetosphere, including how magnetic reconnection transfers magnetic energy into kinetic energy of plasma particles.

Analysis techniques have been created for multi-spacecraft missions, such as CLUSTER, which search for specific types of plasma waves (\cite{Constantinescu:2006}) and which analyze current sheet structure (\cite{Narita:2013}) for a four spacecraft configuration. Knowledge of the direction of wave propagation allows us to use multi-spacecraft filtering (\cite{Motschmann:1998}) to determine the general polarisation properties of any multi-point measurement of a wave field in space plasmas. Measurements from exactly four spacecraft (e.g. a tetrahedron of spacecraft) can be used to estimate current density via the Curlometer technique (\cite{Robert:1998}). 

The Cluster and MMS missions have also utilized the Curlometer technique to interpolate the value of the magnetic field over a region near the tetrahedron's barycenter, regardless of the field's geometry. However, these interpolations are limited to measuring fluctuations on a scale on the same order as that of their inter-spacecraft distances \cite[e.g.][]{Robert:1998,Forsyth:2011}. To study multi-scale processes, such as plasma turbulence, with structures on characteristic length scales that cover many orders of magnitude, we must employ measurements from more than four spacecraft. Therefore, we develop a method which extends the magnetic field reconstruction technique Curlometer to configurations of more than four spacecraft.

Many such multi-spacecraft missions have been proposed, e.g. Cross-scale \citep{Schwartz:2009}, AME \citep{Dai:2020} and HelioSwarm \citep{Klein:2019:WP}, but in order to optimize such missions, it is urgent to robustly quantify the impact of particular spacecraft configurations on multi-point analysis methods, capturing the effects of the physical scales spanned by the spacecraft in the observatory and the geometry of the polyhedra that can be drawn from the constituent spacecraft. Such quantification will help demonstrate that a proposed mission will be able to usefully analyze a large number of magnetometer measurements made in the pristine solar wind, magnetosphere, and magnetosheath. It will also assist in the optimization of spacecraft configurations and quantification of errors derived from multi-point, multi-scale measurements. In this paper, we focus on the fidelity of the reproduction of the magnetic field using a sparsely sampled set of measurements whose spatial configuration is based upon realistic configurations of the proposed nine-spacecraft HelioSwarm observatory, described for instance by \cite{Plice:2020}. 

The reconstruction method is described in \S \ref{sec:method}, the results are applied to two magnetic field models, including a numerical simulation of turbulence, in \S \ref{sec:results}, with a concluding discussion in \S \ref{sec:conclusion}.

\section{Methodology}
\label{sec:method}

\subsection{Geometrical Definitions}
\label{ssec:geometry}
Given $N$ spacecraft, we identify $C(N,k)$ polyhedra with $k$ vertices. As spatial divergence analysis methods, \citep[e.g.][]{Paschmann:1998,Paschmann:2008,Dunlop:1988} require at least four vertices to resolve three-dimensional structure, we only consider polyhedra with at least four vertices, known as tetrahedra. For $N=9$, there are $126$ (i.e. 9 choose 4) tetrahedra, $126$ polyhedra with 5 vertices, $84$ with 6 vertices, $36$ with 7 vertices, $9$ with 8 vertices, and $1$ with 9 vertices, for a total of $382$ polyhedra with at least 4 vertices.

Each polyhedron is characterized in terms of its size and shape. Because measurements from all $d$ spacecraft are weighted equally, we define the \textit{barycenter} of the $q^{\textrm{th}}$ polyhedron with set $\V{D}$ of $d$ vertices drawn from the $N\ge d$ spacecraft positions $\V{x}_i$ as
\begin{equation}
    \V{x}^{q,d}_{0} =\frac{1}{d} \sum_{i\in \V{D}} \V{x}_i.
\end{equation}

Given the barycenter, we then define the volumetric tensor of the $q^{\textrm{th}}$ polyhedra with set $\V{D}$ of $d$ vertices as
\begin{equation}
    R^{q,d}_{jk} = \frac{1}{d} \sum_{i \in \V{D}}
    \left(x_{ij}- x^{q,d}_{0j}\right)
    \left(x_{ik}- x^{q,d}_{0k} \right).
    \label{eqn:R_tensor}
\end{equation}
Here $x_{ij}$ represents the $j^{th} \in \{x,y,z\}$ component of the position vector for the $i^{th}$ spacecraft. The eigenvectors of the tensor $\V{R}^{q,d}$ represent the three semi-axes of the polyhedra and are associated with the eigenvalues $a^{q,d} = \sqrt{{R}^{q,d}_1}$ (major axis), $b^{q,d} = \sqrt{{R}^{q,d}_2}$ (middle axis), and $c^{q,d} = \sqrt{{R}^{q,d}_3}$ (minor axis), where $a\ge b \ge c$ \cite[a more detailed analysis of the eigenvalues can be found in Chapt 12 of][]{Paschmann:1998}.

To provide a useful geometric interpretation of these shapes, we define a \textit{characteristic size} $L$, as well as an \textit{elongation} $E$ and a \textit{planarity} $P$ \cite[see chapter 16.3 of][]{Paschmann:1998}\footnote{Note that there is some discrepancy in the community about if the elongation should be defined as $E=1-b/a$ or $E = \sqrt{1-(b/a)^2}$, with a similarly subjective choice for planarity. Both definitions span the same range, and we have opted for the former definition.}:
\begin{align}
    L &=2a \nonumber \\
    E &=1 - b/a \label{eqn:LEP} \\
    P &= 1 - c/b. \nonumber
\end{align}

\subsection{Reconstruction Techniques}
\label{ssec:reconstruction}

\subsubsection{First-Order Method}
\label{sssec:Order_1}
In a first-order Taylor series expansion, we use the values of the magnetic field, $\B$, measured at four spacecraft positions $\textbf{x}_{i}$ to estimate the value of $\B$ (and its corresponding directional derivatives) at any other point in space, $\xi$ \citep{Fu:2015,Fu:2020}. The Taylor expansion is:
\begin{align}
\hat{B}_m^{i} &= B_m + \sum_{k \in \{x,y,z\}} \partial_k B_{m} r_{k}^{i} \label{eqn:curl_O1} \\
& \forall i \in \{1,2,3,4\},m \in \{x,y,z\}. \nonumber
\end{align}
In this equation $\hat{B}_m^{i}$ is the measured $m^{th}$ component of $\B$ at the $i^{th}$ spacecraft, $B_m$ is the computed $m^{th}$ component of $\B$ at $\xi$, $\partial_k B_{m}$ is the computed derivative of the $m^{th}$ component of $\B$ with respect to the $k^{th}$ direction at $\xi$, and $r_k^{i}$ is the relative position of spacecraft $i$ with respect to $\xi$. In other words, if $x_{ik}$ is the $k^{th}$ component of spacecraft $i$'s location, then $r_k^{i} := x_{ik} - \xi_k$.

This is a system of 12 equations with 12 unknowns, where the 12 equations represent the $x$, $y$, and $z$ components of $\B$ for each of the four spacecraft. The 12 unknowns are the $x$, $y$, and $z$ components of $\B$ at $\xi$ and the nine terms in the Jacobian of $\B$ at $\xi$.

This system can be reformatted into linear ($A\V{x}=\V{b}$) form and solved with a common linear system solver. This 12-dimensional linear system (shown in full detail in Appendix \ref{sec:appendix.method_O1}, Equation \ref{eqn:order_1_system}) comprises the first-order reconstruction method. 

This magnetic field reconstruction method is related to the Curlometer method \citep{Dunlop:1988,Robert:1998}, which utilizes Ampère's law to calculate the current density $\J$ as the curl of $\B$. The Curlometer solves the same set of equations, but uses the partial derivatives to estimate the current density at the center of each tetrahedron.
The Curlometer method has been widely applied to four-spacecraft magnetic field measurements made for instance by Cluster and MMS, \cite[c.f. Chapter 16.2 of][]{Paschmann:1998}. Future missions, such as the proposed HelioSwarm Observatory \citep{Klein:2019:WP}, will have more than four spacecraft. Therefore, for every reconstructed point, $\xi$, we can apply this reconstruction method for each of the $C(N,4)$ tetrahedra and average the reconstructed values, yielding a statistically larger base of estimates and improving the accuracy of the reconstruction.

\subsubsection{Second-Order Method}
\label{sssec:Order_2}
Because the proposed HelioSwarm Observatory has nine spacecraft, we can use measurements of $\B$ from all nine spatial points simultaneously to apply a second-order reconstruction method. This method, also based on a Taylor series expansion, is more accurate for values located near the center of the expansion (i.e. near the barycenter of the nine-spacecraft constellation) than a single implementation of the first-order method. Following the work of \cite{Torbert:2020}, we write:

{\footnotesize
\begin{align}
\hat{B}_m^{i} &= B_m + \sum_{k \in \{x,y,z\}} \partial_k B_{m} r_{k}^{i} + \frac{1}{2}\sum_{j,k \in \{x,y,z\}} \partial_j \partial_k B_{m} r_{k}^{i} r_{j}^{i} \label{eqn:curl_O2} \\
& \forall i \in \{1,...,9\},m \in \{x,y,z\}. \nonumber
\end{align}
}
These terms are the same as in the first-order method, with the addition of $\partial_j \partial_k B_{m}$, the second derivative of the $m^{th}$ component of $\B$ with respect to the $k^{th}$ and $j^{th}$ directions, at $\xi$.

This is a system of 31 equations with 30 unknowns. 27 of these equations are associated with the $x$, $y$, and $z$ components of $\B$ from the nine spacecraft. There are four additional constraints, imposed by the magnetic field having zero divergence, as well as the divergence of the magnetic field having zero gradient. The 30 unknowns are the $x$, $y$, and $z$ components $\B$ at $\xi$, the nine terms in the Jacobian of $\B$ at $\xi$, and the 18 terms in the Hessian of $\B$ at $\xi$ (excluding the 9 redundant terms).

This system can be reformatted into linear ($A\V{x}=\V{b}$) form where $A$ is a $31 \times 30$ matrix (shown in full detail in Appendix \ref{sec:appendix.method_O2}, equation \ref{eqn:taylor2_system}). This system is over-determined, therefore in general, an exact solution does not exist. However, we can find an approximate solution via the method of ordinary least squares. This method finds the solution to the problem $A \V{x} = \V{b}$ which minimizes the two-norm of the error, i.e. 
\begin{equation}
\V{x} = \underset{\V{x}}{\operatorname{argmin}} \| A\V{x} - \V{b}\|_2 .
\label{eqn:lstqs}
\end{equation}
This second-order reconstruction method is referred to as $M_{2}$ throughout this paper.

\subsubsection{Quantifying Error}
\label{sssec:error_quant}
We define the error at any point in space, $\xi$, as:
\begin{equation}
    \theta = 100\frac{\|\B_{calc}(\xi)-\B_{true}(\xi)\|_2}{\|\B_{true}(\xi)\|_2} 
    \label{eqn:B_error}
\end{equation}
where $\B_{true}(\xi)$ is the magnetic field vector at point $\xi$ and $\B_{calc}(\xi)$ is the reconstructed magnetic field vector at point $\xi$. 

Given that we can determine the value of this error at all points in a simulation or for a given analytic field, we also define $\epsilon(\theta)$ as the proportion of the volume that is reconstructed with less than $\theta$\% error. For a sufficiently dense grid of $N_x \times N_y \times N_z$ uniformly-spaced points, $\epsilon(\theta)$ can be estimated as
\begin{equation}
    \epsilon(\theta) = \frac{\text{\# points with $\leq \theta \%$ error}}{N_x N_y N_z}.
    \label{eqn:epsilon_3}
\end{equation}
To define the physical volume in which a given magnetic field reconstruction is accurate, $\epsilon(\theta)$ is translated into a dimensional quantity by multiplying it by the total volume covered by the $N_x \times N_y \times N_z$ grid.

\subsubsection{Error Minimization Techniques}
\label{sssec:tetra_inclusion}
As the first-order method (\S \ref{sssec:Order_1}) only requires a single tetrahedron of spacecraft to estimate $\B$, in this paper we will test four selection methods for using a subset of the 126 tetrahedra to improve the reconstruction. These methods combine the statistically large set of tetrahedra with our knowledge of the spacecraft positions relative to $\xi$ and the geometry of all 126 tetrahedra.

For the first method, $M_{1.1}$, at each point in space we reconstruct the magnetic field using all 126 tetrahedra to produce 126 estimates for $\B(\xi)$. We then average over these $\B(\xi)$ values component-wise to estimate $\B(\xi)$.

For the second method, $M_{1.2}$, we perform the same averaging as method one, but only include tetrahedra whose barycenter are within a characteristic distance of $\xi$. i.e. for each reconstructed point $\xi$, only include tetrahedra $j$ in the average which satisfy
\begin{equation}
    \|(r_0)_j - \xi \|_2 < L_j,
    \label{eqn:near_tetra}
\end{equation}
where $L_j$ is the characteristic size and $(r_0)_j$ is the barycenter of the $j^{\text{th}}$ tetrahedron.

For the third method, $M_{1.3}$, we perform the same selection as method two, but with the added restriction that the shape of tetrahedron $j$ must be quasi-regular. In terms of the geometric quantities of the spacecraft configuration (defined in equation \ref{eqn:LEP}), this translates to elongation $E$ and planarity $P$ being sufficiently small. Because E and P are symmetric with respect to orientation, we will define a composite geometric parameter $\chi_j$
\begin{equation}
    \chi_j = \sqrt{E_j^2 + P_j^2}.
    \label{eqn:chi}
\end{equation}
Small $\chi_j$ implies that both the elongation and planarity of tetrahedron $j$ are small. For method $M_{1.3}$, we restrict our averaging to only include tetrahedra where
\begin{align}
    \chi_j &\leq 1 \nonumber \\
     \|(r_0)_j - \xi \|_2 &< L_j .\label{eqn:method_3}
\end{align}

For the fourth method, $M_{1.4}$, we perform the same selection of tetrahedra as method three, but require the tetrahedra included in the averaging to be more regular. For method $M_{1.4}$, our shape and position requirement is 
\begin{align}
    \chi_j &\leq 0.6 \nonumber \\
     \|(r_0)_j - \xi \|_2 &< L_j .\label{eqn:method_4}
\end{align}
The value of 0.6 was selected because page 408 of \cite{Paschmann:1998} shows it be a threshold value for elongation and planarity which separates the well performing ‘pseudo-sphere type’ and ‘potato type’ spacecraft configurations from the poorer performing ‘knife blade type’, ‘cigar type’, and ‘pancake type’ configurations.

The first-order methods $M_{1.1},M_{1.2},M_{1.3},M_{1.4}$ will be compared to the second-order method $M_2$ as well as the first-order method applied to a single regular (i.e. $\chi=0$) tetrahedron of spacecraft. This single regular tetrahedron will have the same characteristic scale as the nine-spacecraft configuration it is compared to.

\subsection{Models}
\label{ssec:models}
To validate and quantify the errors of our reconstruction, we implement our reconstruction methods on two magnetic field models, a simple current sheet and a numerical simulation of turbulence.

\subsubsection{Simple Current Sheet}
\label{sssec:simple}
 For our first model, we define a magnetic field where $\B$ is analytically defined at all spatial points. This field, which represents a simple current sheet, can be described in cylindrical coordinates as
\begin{equation}
\B(r) = \mu_0 J_0 \sigma \left[ \sigma/r - e^{-r\sigma}(1 + \sigma/r) \right] \hat{\theta} .
\label{eqn:B_simple}
\end{equation}
The variable $\sigma$ represents the current sheet characteristic width and $J_0$ represents the magnitude of the current at its center.

\subsubsection{Turbulence Simulation}
\label{sssec:turb_sim}
Physically realistic fields, such as those generated by turbulence in the solar wind, are significantly more complex than the simple current sheet model of equation \ref{eqn:B_simple}. We therefore test our reconstruction techniques on magnetic fields drawn from numerical simulations of turbulence. In particular, we utilize the magnetic fields from a fully developed turbulence simulation performed with the five  moment, multi-fluid solver within the \gkeyll\ simulation framework \citep{Hakim:2006,Wang:2015,Wang:2020}. This turbulence simulation is designed to represent plasma behavior in the pristine solar wind at 1AU.

We use the five moment ($n_s, \V{u}_s, p_s$), two fluid ($s=p,e$) plasma model to evolve a proton-electron plasma. We note that the five moment, two fluid model formally reduces to Hall MHD in the limit $m_e \rightarrow 0$ and $\epsilon_0 \rightarrow 0$ \citep{Srinivasan:2011}, where $\epsilon_0$ is the vacuum permittivity. We use a reduced (proton to electron) mass ratio of $m_p / m_e = 100$, a temperature ratio of $T_p / T_e = 1$, Alfvén velocity of $v_{A} /c =  B/\sqrt{\mu_0 n_p m_p c^2} =0.02$, plasma beta (ratio of plasma thermal pressure to magnetic pressure) of $\beta_p = 2\mu_0 n_p T_p/B^2 = 1$, and adiabatic index $\gamma = 5/3$. We employ an elongated domain $L_x = L_y = 0.2 L_z = 100 \pi \rho_p$ with resolution $n_x = n_y = n_z = 448$. Lengths are normalized to the proton gyroradius $\rho_p = v_{tp}/\Omega_p$, the ratio of the proton thermal speed $v_{tp}=\sqrt{2 T_p/m_p}$ and the proton cyclotron frequency $\Omega_p=q_p B/m_p$.
We choose a uniform background density and magnetic field, $\V{B}_0 = B_0 \hat{\V{z}}$, and initialize the simulation with the three dimensional extension of the Orszag-Tang vortex \citep{Orszag:1979} described in \cite{Li:2015}
\begin{equation}
    \begin{split}
        \frac{\V{z}_1^+}{v_A} &= -\frac{2z_0}{v_A} \sin{(k_\perp y - k_z z)} \hat{\V{x}}, \hfill  \frac{z_1^-}{v_A} = 0 \\
        \frac{\V{z}_2^\pm}{v_A} &= \frac{z_0}{v_A} \sin{(k_\perp x \mp k_z z)} \hat{\V{y}}\\
        \frac{\V{z}_3^\pm}{v_A} &= \pm \frac{z_0}{v_A} \sin{(2k_\perp x \mp k_z z)} \hat{\V{y}},
    \end{split}
\end{equation}
where $\V{z}^\pm = \delta \V{u} \pm \delta \V{B}/\sqrt{\mu_0 \rho_0}$ are the Elsasser variables \citep{Elsasser:1950}, $k_{x,y} = 2\pi / L_{x,y}$, and $k_z = 2\pi / L_z$. The initial amplitude, $z_0 = 0.2$, is chosen to satisfy the critical balance condition, $k_x z_0 / k_z v_A = 1$ \citep{Goldreich:1995}.  

\begin{figure}
\hspace*{-1cm}
        \includegraphics[width=1.1\columnwidth]{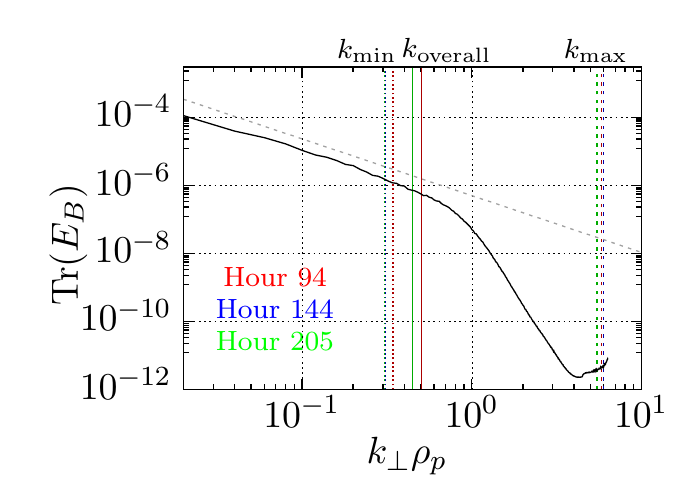}
    \caption{Trace magnetic field spectrum (solid black) from the \gkeyll\ simulation computed at $t = 1500 / \Omega_{p}$, with a $k_\perp^{-5/3}$ dashed line plotted for reference. The characteristic scales associated with the three spacecraft configurations, Hours 94, 144, and 205, drawn from the HelioSwarm DRM are shown as vertical colored lines.}
    \label{fig:spectrum}
\end{figure}

The simulation is run for one \Alfven\ crossing time, $t_A = 1500 / \Omega_{p}$, at which point the turbulence has fully developed and reached a steady state. In \figref{fig:spectrum}, we plot the trace magnetic energy spectrum as a function of $k_\perp = \sqrt{k_x^2 + k_y^2}$, with a $k_\perp^{-5/3}$ dashed line plotted for reference. The steep roll-over in the spectrum at $k_\perp \rho_p \simeq 1$ is due to numerical diffusion from the finite volume scheme employed by \gkeyll.

To compare the simulation to the selected spacecraft configurations with separations in physical units, we note that the proton gyroradius can be written as
\begin{equation}
    \rho_p = \sqrt{\frac{m_p}{m_e}}\sqrt{\beta_p} \frac{c}{\omega_{pe}}.
\end{equation}
With the constants in the turbulence simulation of $m_p/m_e=100$, $\beta_p = 1$, $\omega_{pe} = \frac{c}{d_e} = 5.64\times 10^{4}\sqrt{n}$, we set $n_e=0.2829$ cm$^{-3}$, so that $\rho_p = 100$ km.

We extract from this simulation a 3-dimensional grid of values representing the plasma's physical parameters at different points in space. From this grid, we use trilinear interpolation to estimate the value of $\B$ at any point in the simulation volume.

\subsubsection{Spacecraft Configurations}
\label{sssec:sc_configs}
\begin{figure}
\hspace*{-1cm}
    \includegraphics[width=1.1\columnwidth]{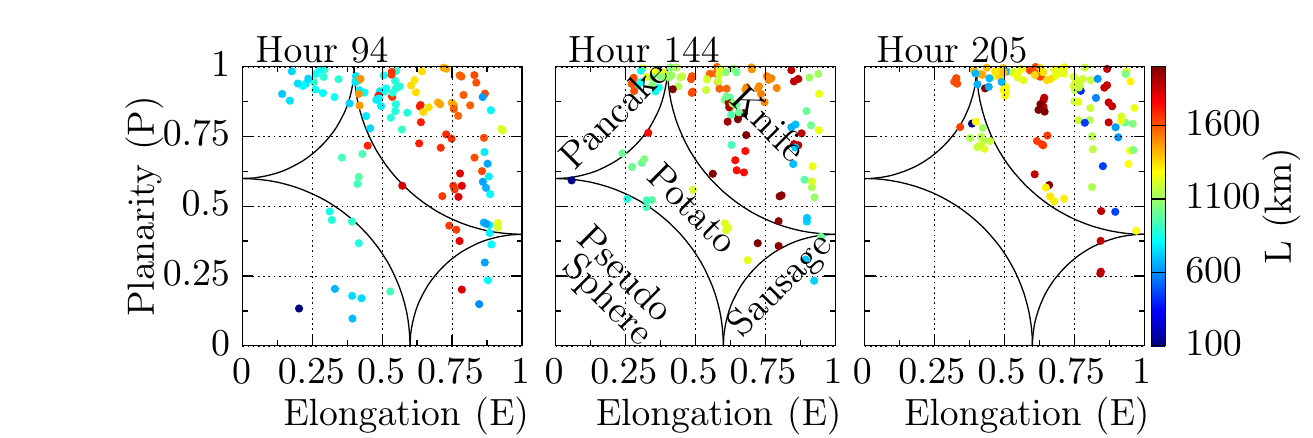}
    \caption{Elongation and planarity of the 126 tetrahedron associated with the three nine-spacecraft configurations under consideration, with characteristic lengths shown in color.}
    \label{fig:config}
\end{figure}

To illustrate our reconstruction methods for realistic spacecraft configurations, we study these methods using three different nine-spacecraft configurations. The spacecraft configurations are selected from the phase A design reference mission (DRM) of the proposed HelioSwarm Observatory concept, corresponding to hours 94, 144, and 205 of the science phase. These hours are selected because they represent a selection of spacecraft tetrahedra that have significantly different distributions of their elongation, planarity, and length. In Table \ref{tab:configs} we note the geometric characteristics of the overall nine-vertex polyhedra for each of the three configurations. We also calculate the size, elongation, and planarity of all 126 tetrahedron in each configuration and display them in Fig.~\ref{fig:config}, noting the minimum and maximum values of these three parameters for each configuration in Table \ref{tab:configs}. The wavelengths associated with the overall, minimum, and maximum scales, $k \rho_p = 2 \pi \rho_p/L$, are overlaid on Fig.~\ref{fig:spectrum}, using a fiducial value of $\rho_p=100 $ km. 

\begin{table}
\centering
    \begin{tabular}{ll|lll}
    Hour &         & L(km)    & E    & P    \\ \hline
    94   & Overall & 1245 & 0.48 & 0.60 \\
         & Min     & 108  & 0.14 & 0.10 \\
         & Max     & 1834 & 0.93 & 0.99 \\ \hline
    144  & Overall & 1395 & 0.42 & 0.70 \\
         & Min     & 108  & 0.06 & 0.23 \\
         & Max     & 2030 & 0.95 & 0.99 \\ \hline
    205  & Overall & 1401 & 0.45 & 0.75 \\
         & Min     & 115  & 0.32 & 0.26 \\
         & Max     & 2045 & 0.97 & 0.99
    \end{tabular}
    \caption{Characteristic geometric parameters for the three nine-spacecraft configurations under consideration and the minimum/maximum characteristic geometric parameters created from choosing any four of the nine spacecraft of each configuration.}
    \label{tab:configs}
\end{table}

\section{Application of Reconstruction}
\label{sec:results}
To find the expected error at all points in space near a particular spacecraft configuration, we take a Monte Carlo approach and place the barycenter of each nine-spacecraft configuration into a known magnetic field at random locations. We then reconstruct the magnetic field on a grid of points centered at the barycenter of the nine-spacecraft configuration using the first- and second-order reconstruction methods. The location of each point in the reconstructed grid is constant with respect to the spacecraft configuration. Therefore, we find the average of the errors, $\theta$, at all reconstructed grid points for all elements of the Monte Carlo ensemble, allowing the calculation of the expected value of error at each point on the grid.

 Additionally, we compare the divergence found on a grid of points sampled from the baseline current sheet and turbulence simulation magnetic fields with that of the same points sampled from the fields reconstructed using our first-order reconstruction methods. This comparison yields divergence values of similar magnitude in the baseline and reconstructed fields, which indicates that our reconstruction methods do not introduce nonphysical values of divergence.

\subsection{Current Sheet}
\label{ssec:results.cs}
\begin{figure}[ht]
    \includegraphics[width=\columnwidth]{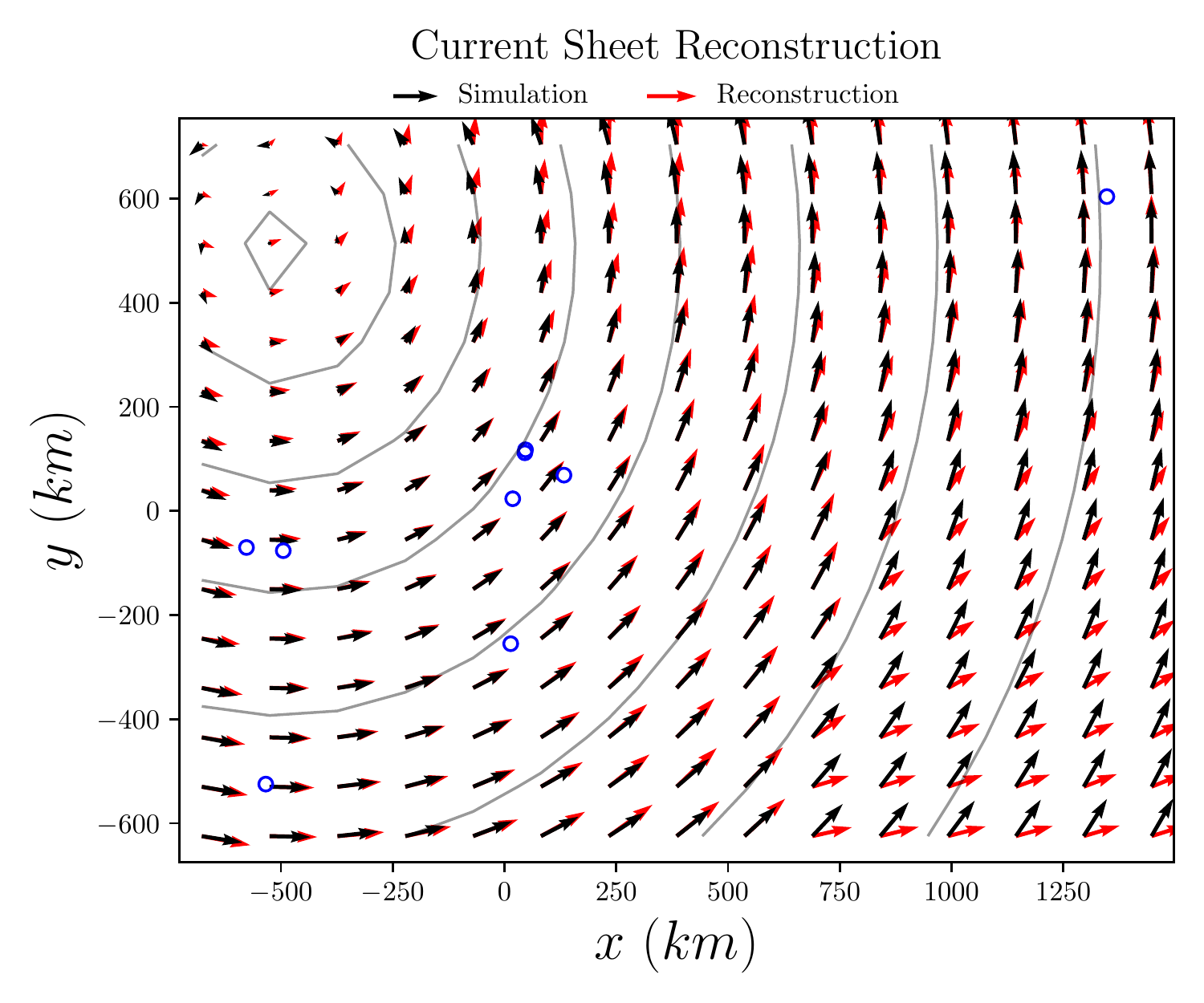}
    \caption{An example of the spacecraft configuration at hour 94, pictured as the blue circles, reconstructing the magnetic field associated with a simple current sheet using first-order method $M_{1.3}$. The true magnetic field is shown as black arrows, and the reconstructed magnetic field is shown as red arrows. This current sheet, centered at $(-500,500)$ km, has characteristic width $\sigma = 2000$ km. Contour lines of the $\hat{z}$ component of current density $\J$ are shown in gray.}
    \label{fig:B_reconst_CS}
\end{figure}

We present an example magnetic field reconstruction of the simple current sheet model (\S \ref{sssec:simple}) in Figure \ref{fig:B_reconst_CS}. Here, we use the first-order method $M_{1.3}$ to reconstruct the magnetic field in the $z=0$ plane for the simple current sheet, Eqn\ref{eqn:B_simple} with $\sigma=2000$ km using the hour 94 spacecraft configuration. There is little difference between the reconstructed and original fields near the center of the spacecraft configuration, and the difference in vectors increases with distance from the spacecraft configuration's center.

We perform 200 Monte Carlo iterations of reconstruction using each method, observing that 200 was sufficient to point-wise converge in error. The characteristic width of the current sheet is chosen as a uniform random variable $\sigma \sim U[500,5000]$ km, while the barycenter of the nine-spacecraft configuration is selected as a 3D uniform random variable $r_0\sim U[-1000,1000]^3$ km. We reconstruct a $30\times 30\times 30$ grid of points $\xi$ that extends $100$ km past the furthest spacecraft in all directions.

\begin{figure*}
\centering
\textbf{Errors in Simple Current Sheet Model: Hour 94} 
\begin{tabular}{ccc}
\includegraphics[width=0.3\textwidth]{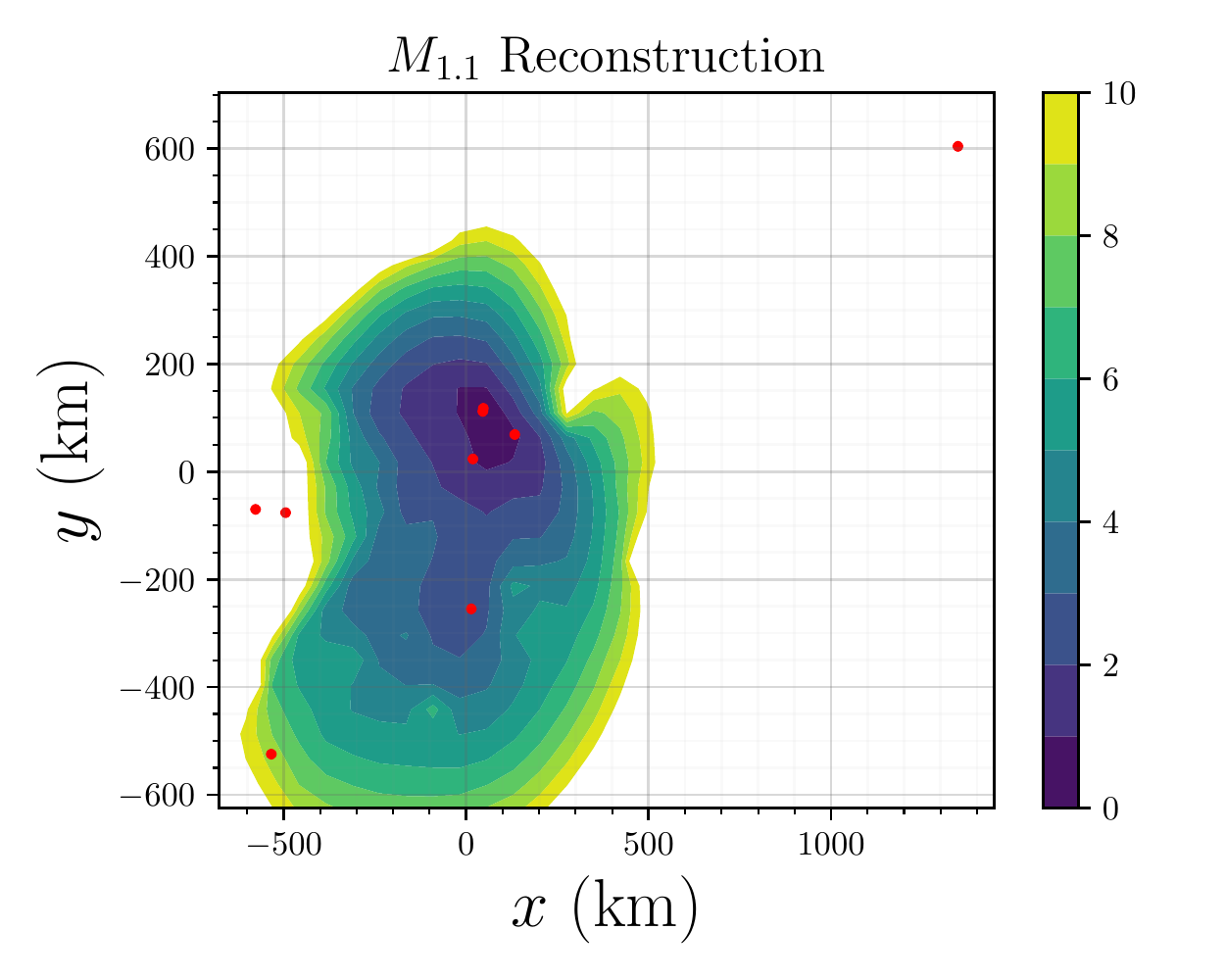}&
\includegraphics[width=0.3\textwidth]{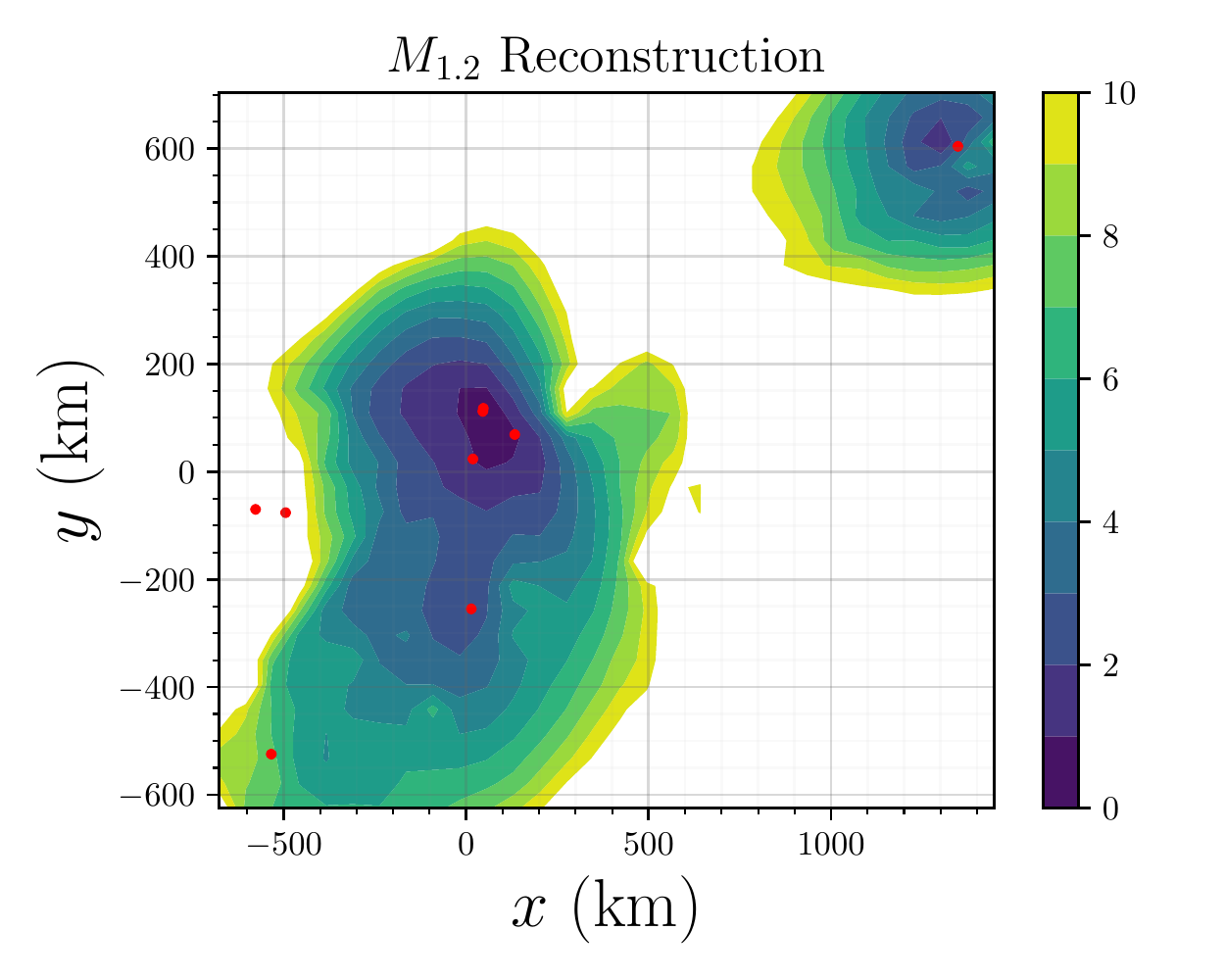}&
\includegraphics[width=0.3\textwidth]{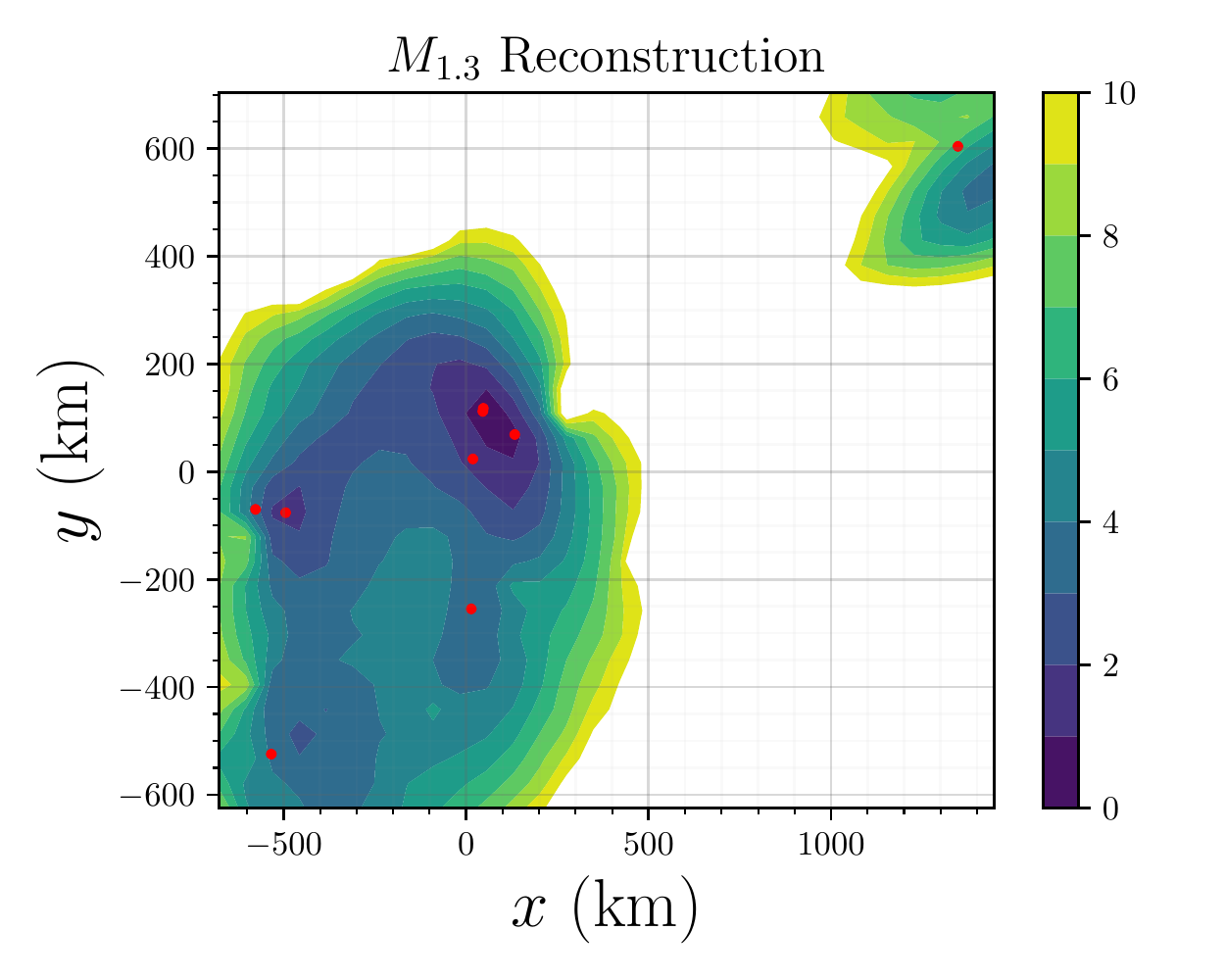}\\
\includegraphics[width=0.3\textwidth]{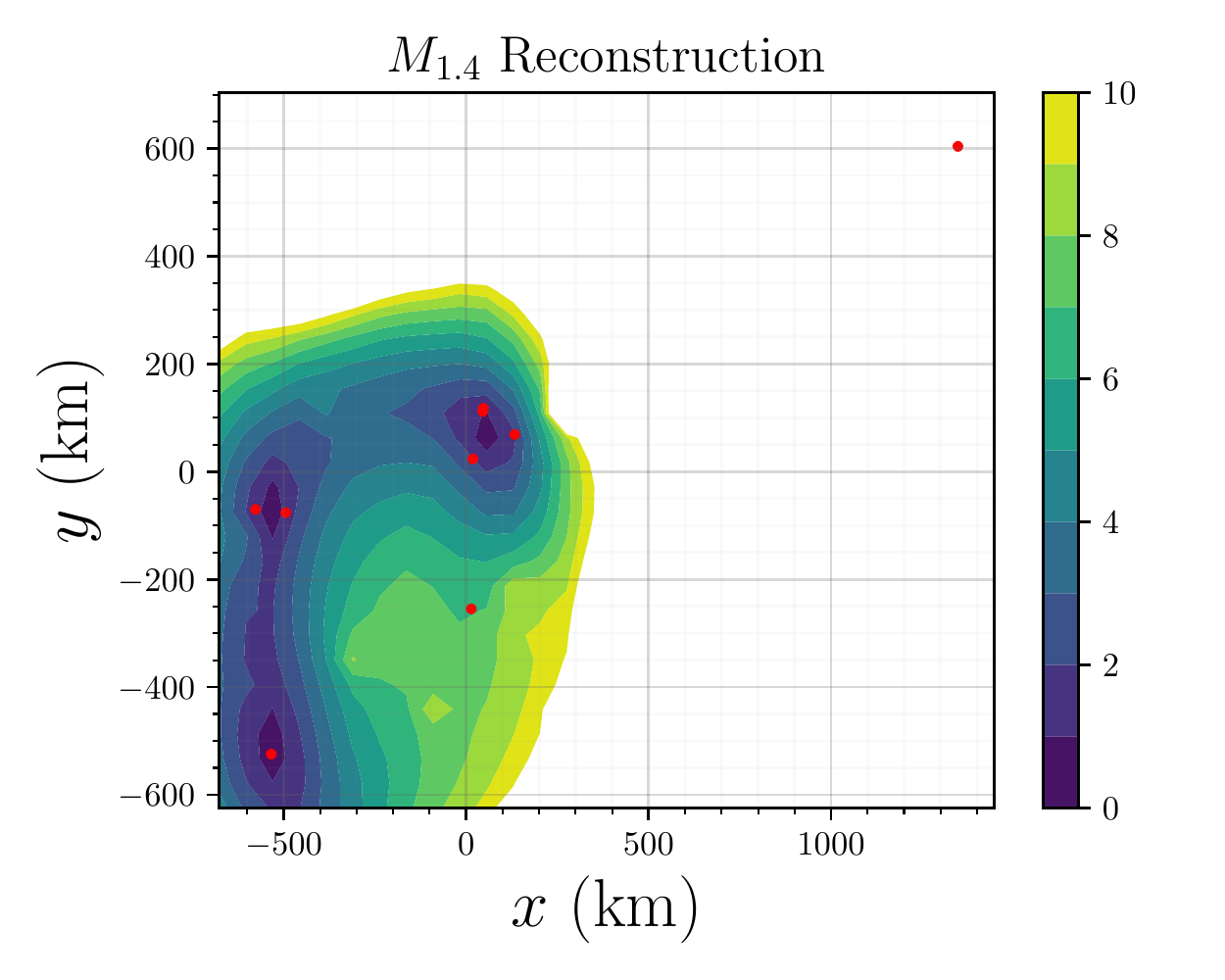}&
\includegraphics[width=0.3\textwidth]{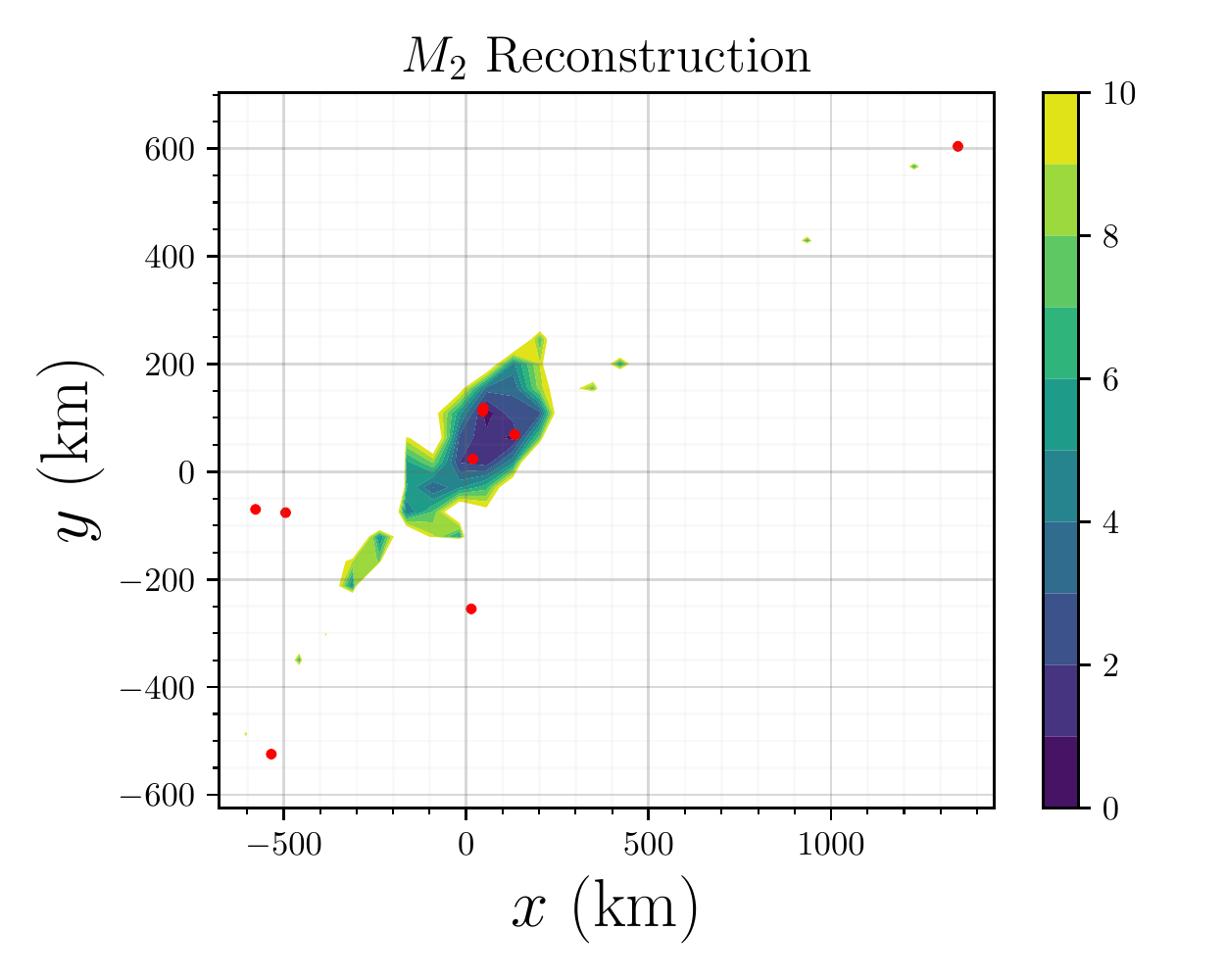}&
\includegraphics[width=0.3\textwidth]{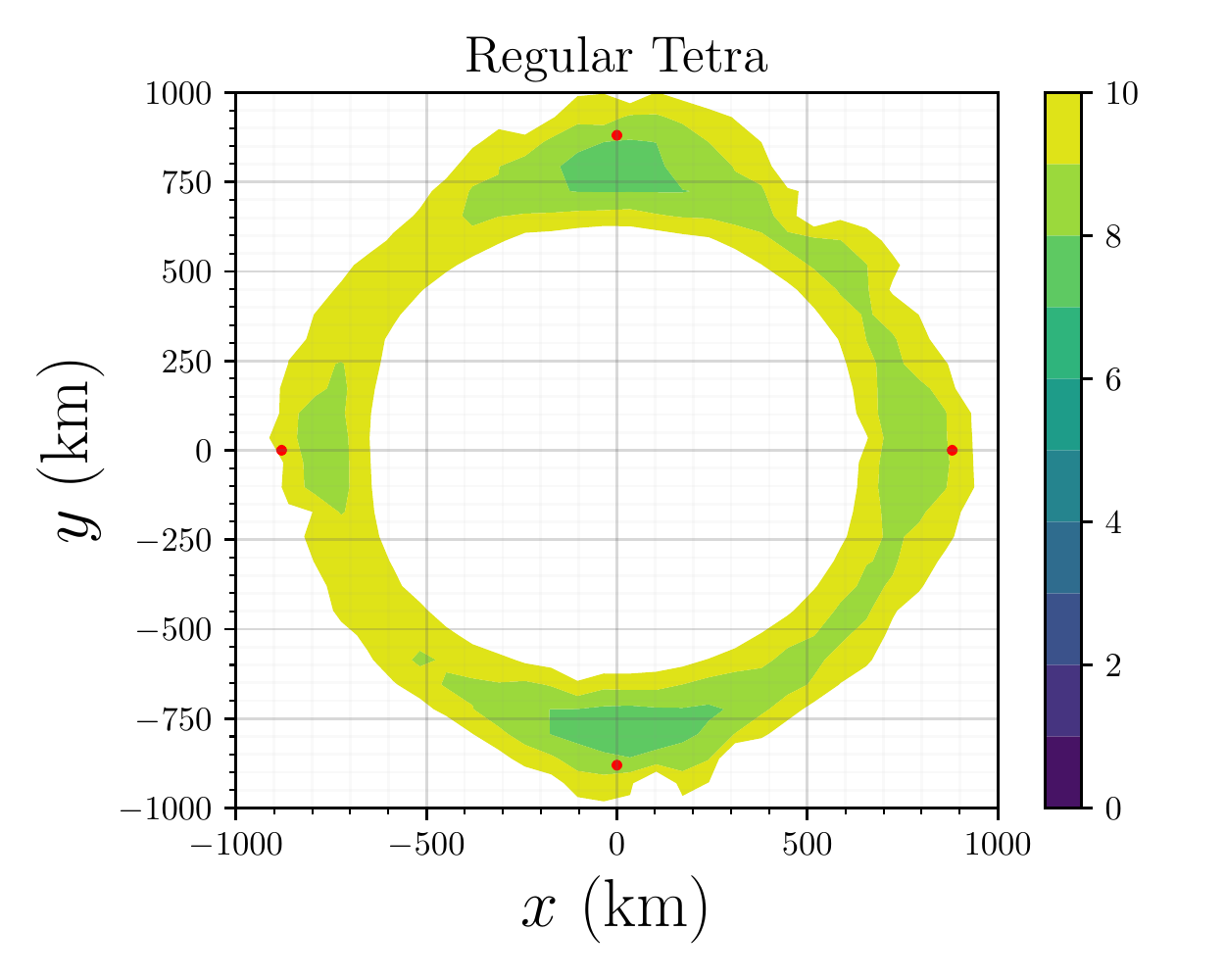}
\end{tabular}
\caption{Computation error (defined in equation \ref{eqn:B_error}) at all points on the $z=0$ plane of the simple current sheet model, using the swarm configuration at hour 94 of the HelioSwarm DRM using first-order methods $M_{1,1}$, $M_{1,2}$, $M_{1,3}$ and $M_{1,4}$, the second-order method, $M_{2}$ and a single regular tetrahedron. The red points represent the spacecraft locations. Areas in white either have a reconstruction error above $10\%$, or have no tetrahedron satisfying reconstruction method condition, resulting in no reconstructed field values.}
\label{fig:simp_94} 
\end{figure*}

\begin{figure*}
\centering
\textbf{Errors in Simple Current Sheet Model: Hour 144} 
\begin{tabular}{ccc}
\includegraphics[width=0.3\textwidth]{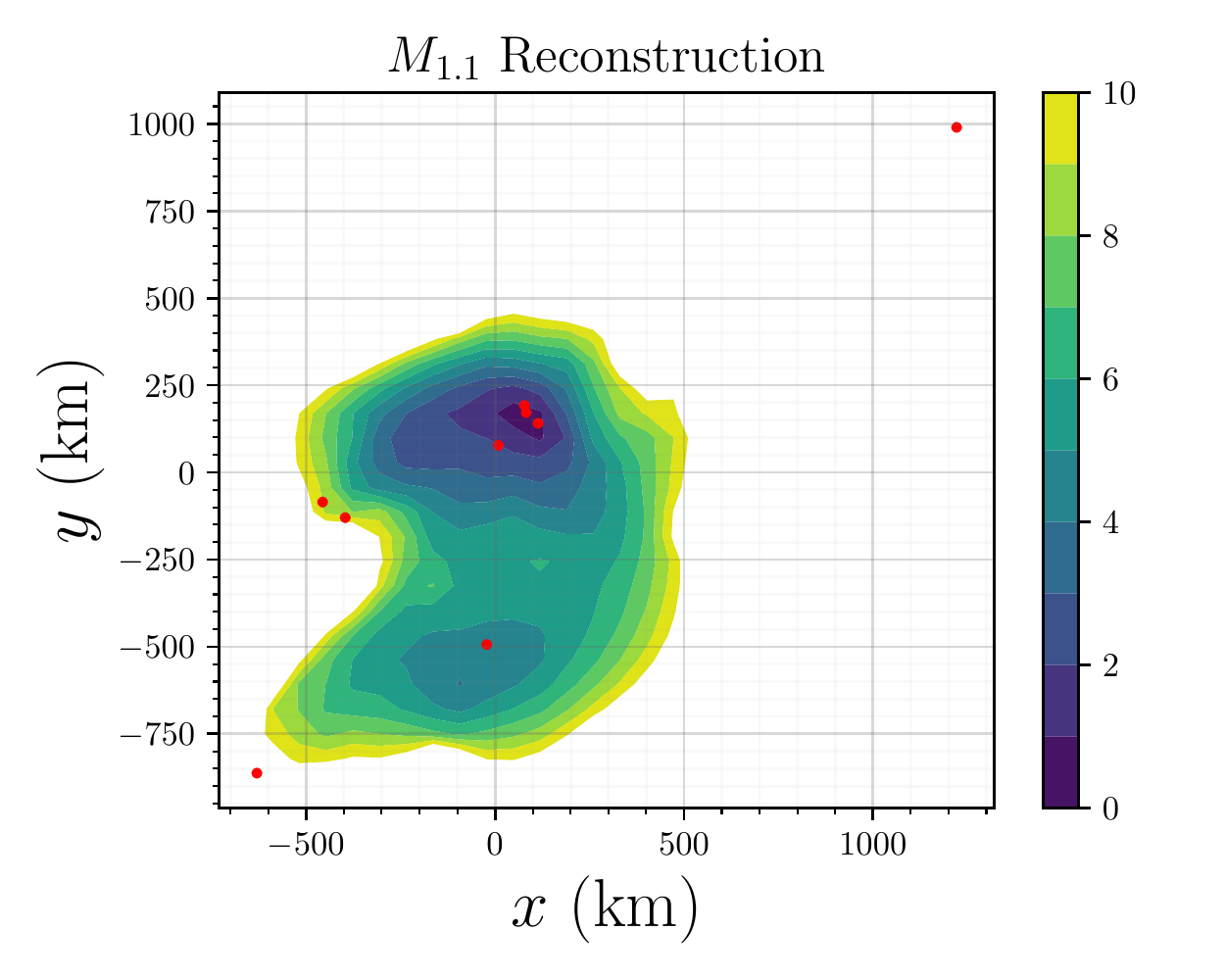}&
\includegraphics[width=0.3\textwidth]{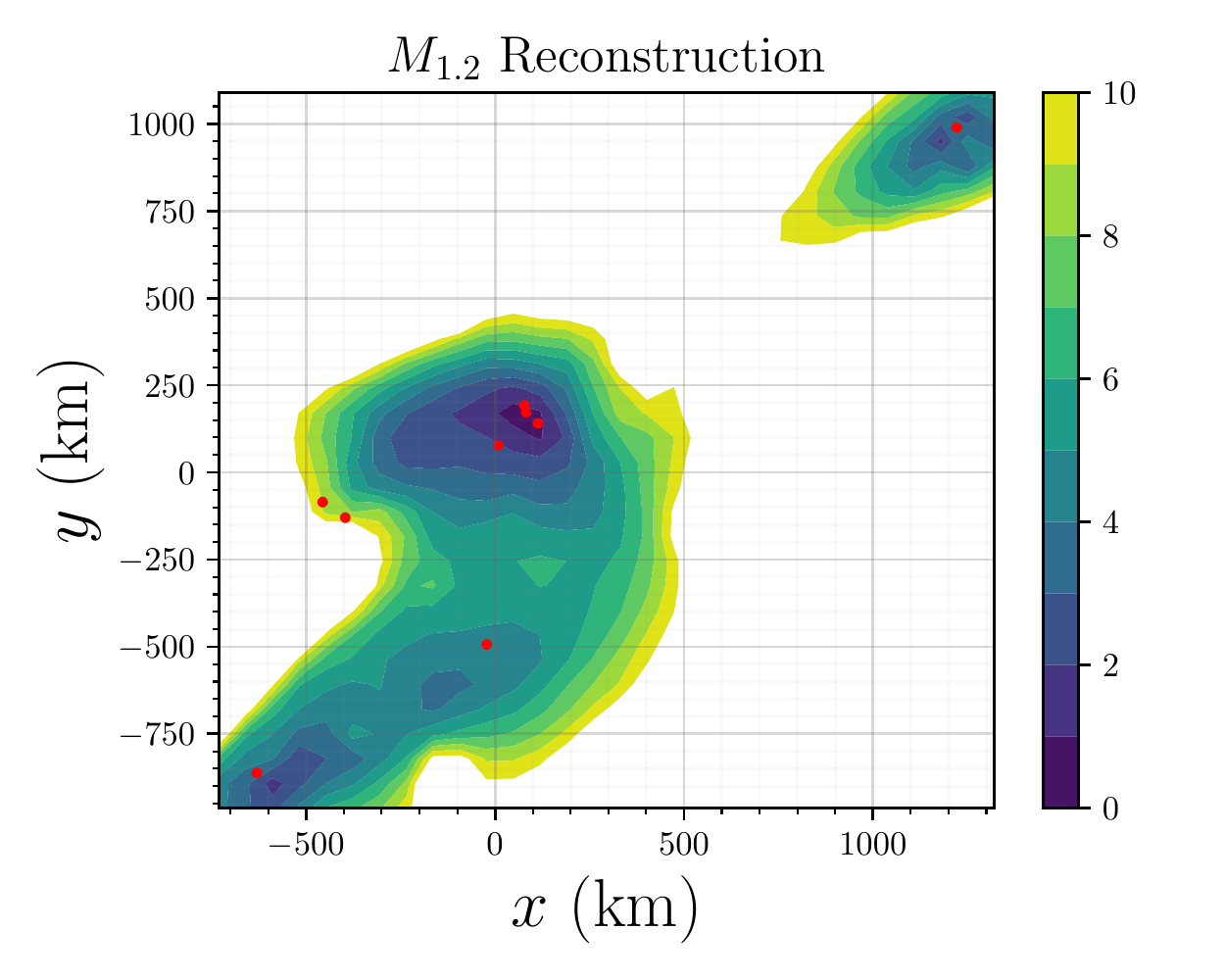}&
\includegraphics[width=0.3\textwidth]{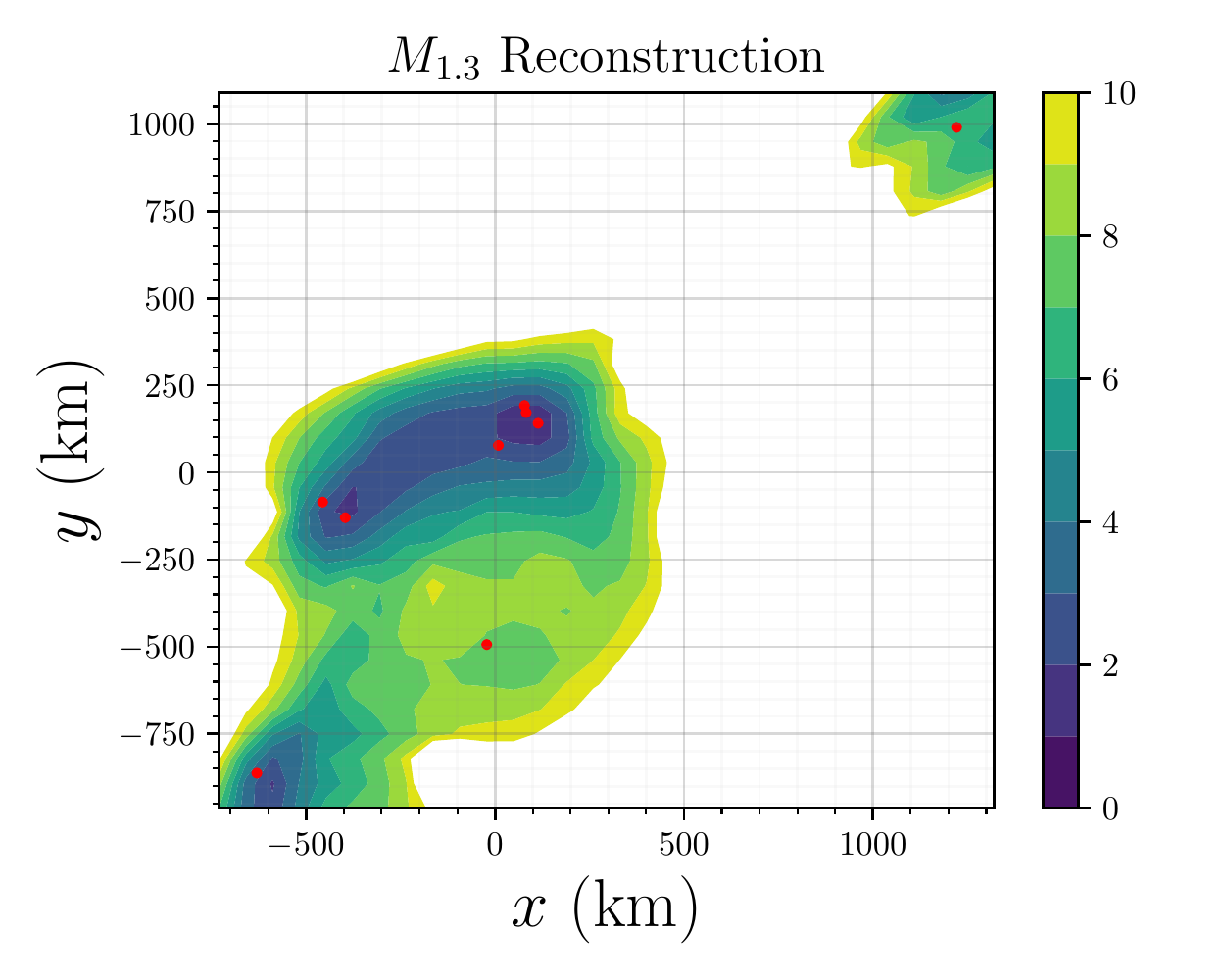}\\
\includegraphics[width=0.3\textwidth]{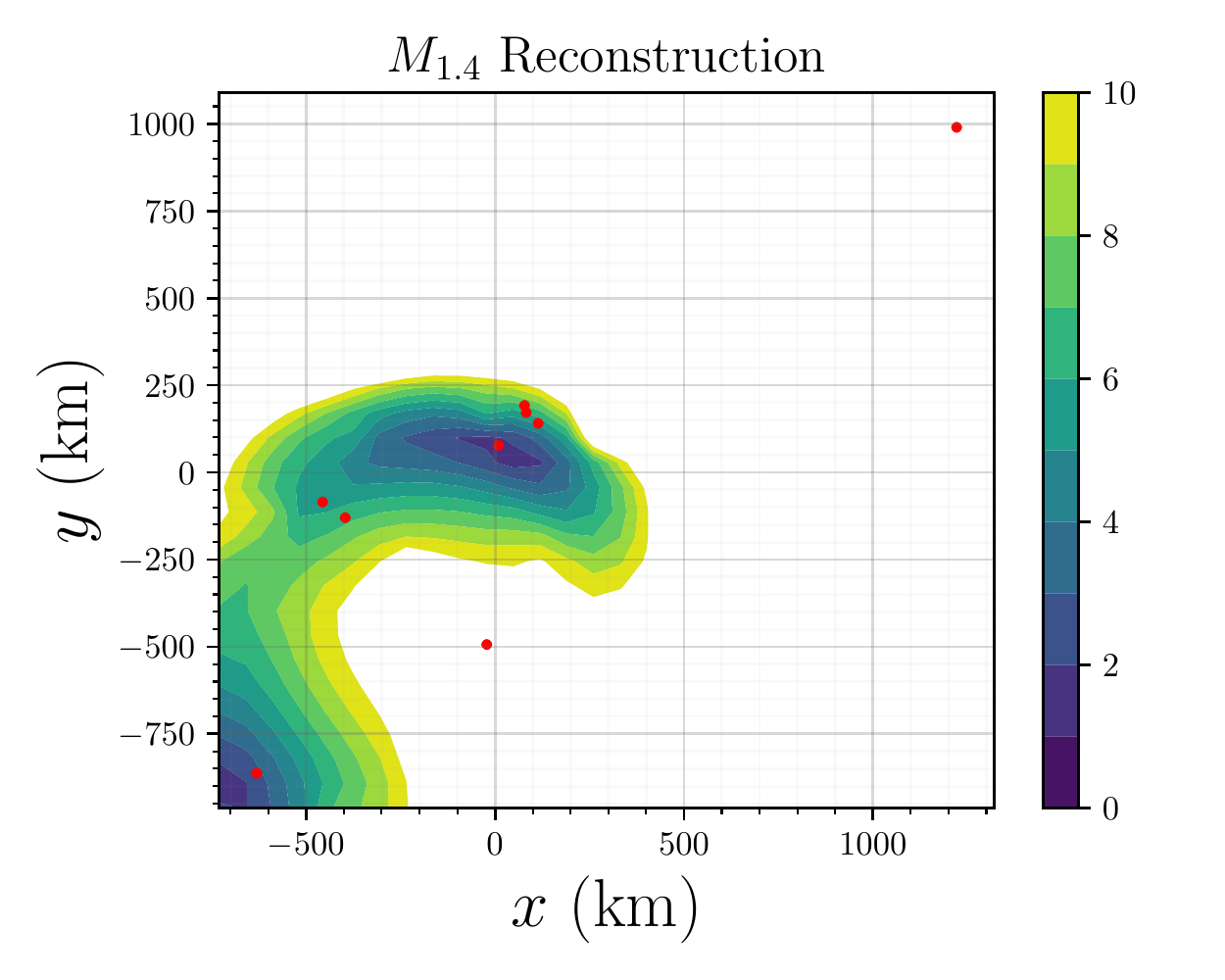}&
\includegraphics[width=0.3\textwidth]{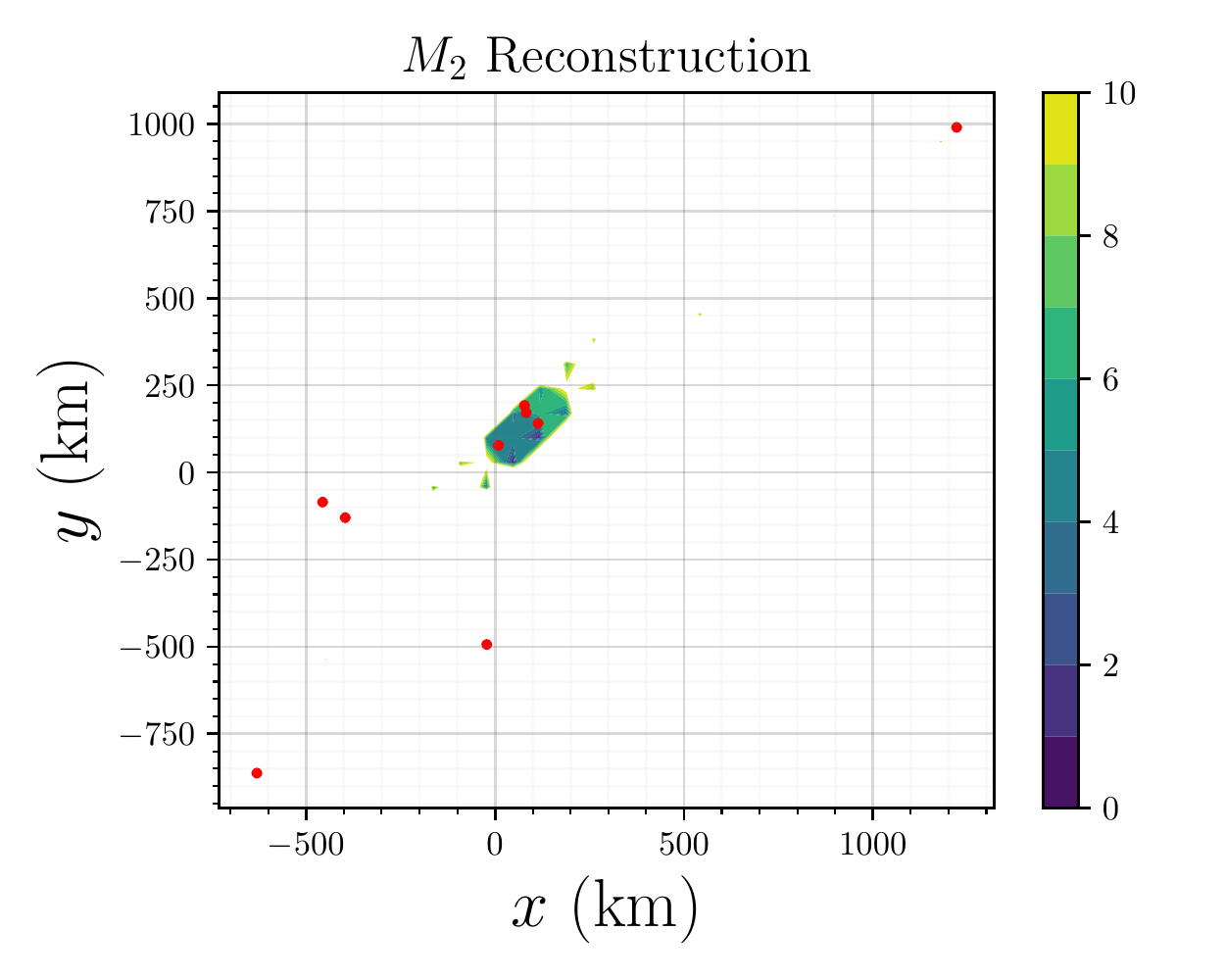}&
\includegraphics[width=0.3\textwidth]{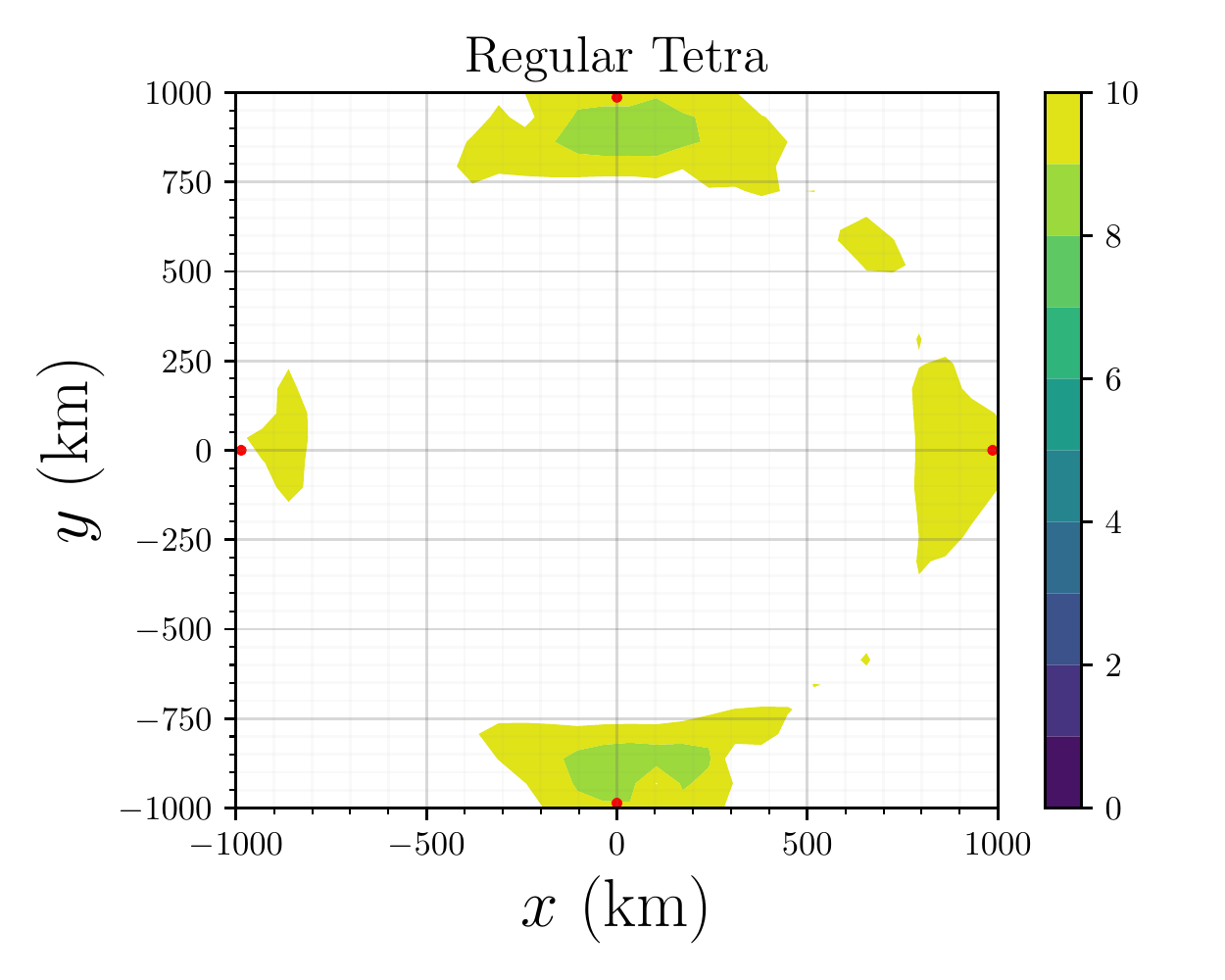}
\end{tabular}
\caption{Computation error (defined in equation \ref{eqn:B_error}) at all points on the $z=0$ plane of the simple current sheet model, using the swarm configuration at hour 144 of the HelioSwarm DRM. 
The layout is identical to Fig.~\ref{fig:simp_94}.}
\label{fig:simp_144} 
\end{figure*}

\begin{figure*}
\centering
\textbf{Errors in Simple Current Sheet Model: Hour 205} 
\begin{tabular}{ccc}
\includegraphics[width=0.3\textwidth]{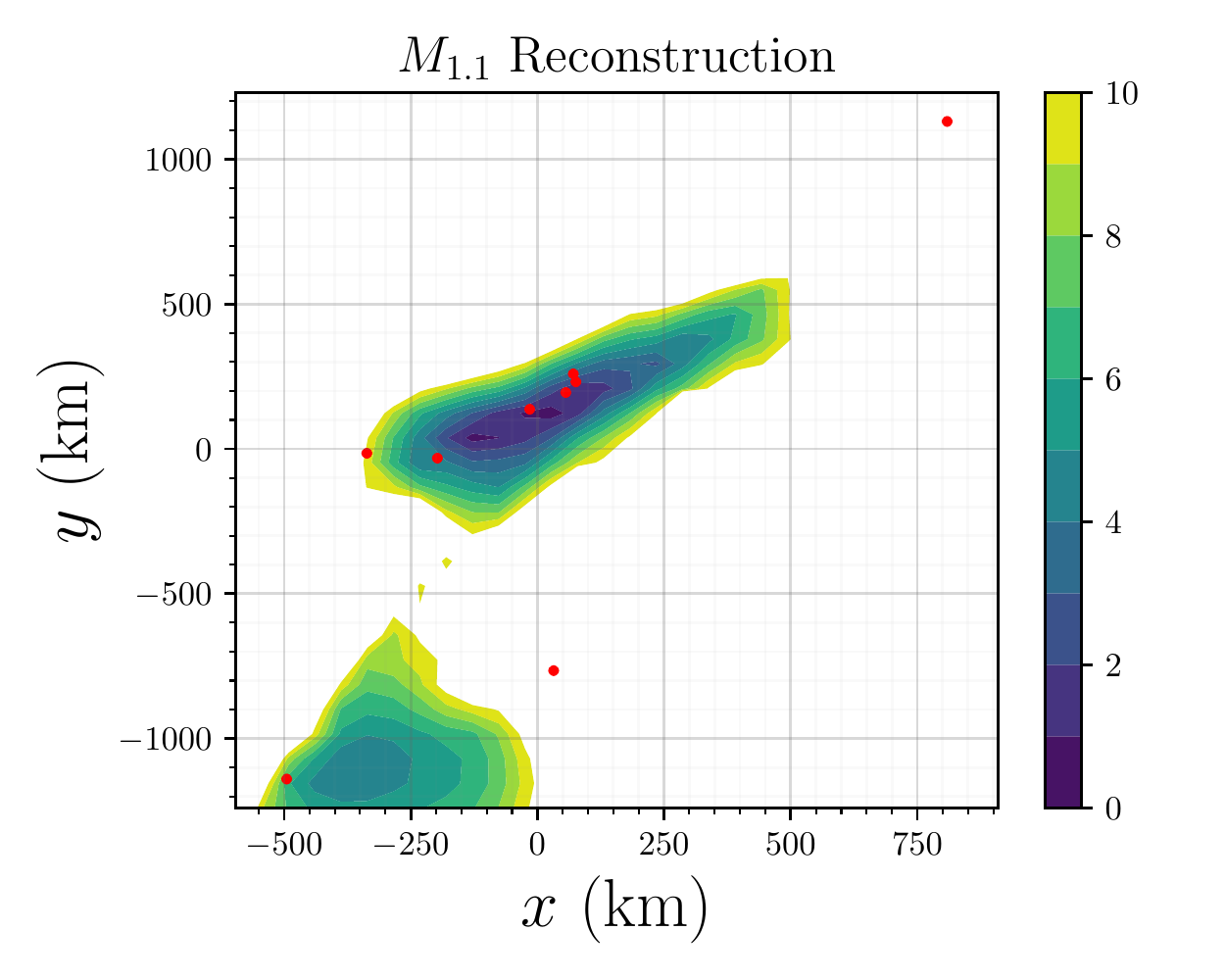}&
\includegraphics[width=0.3\textwidth]{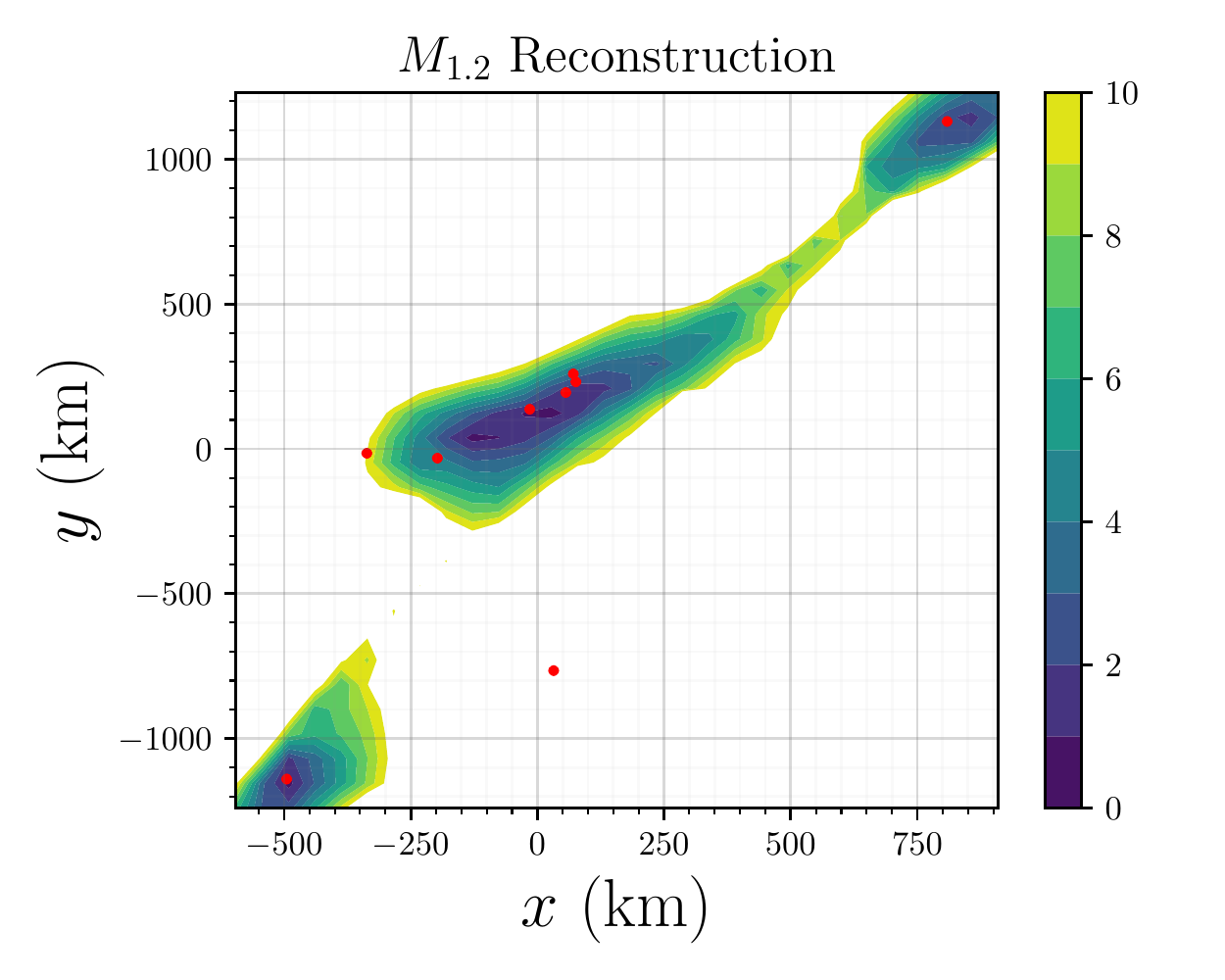}&
\includegraphics[width=0.3\textwidth]{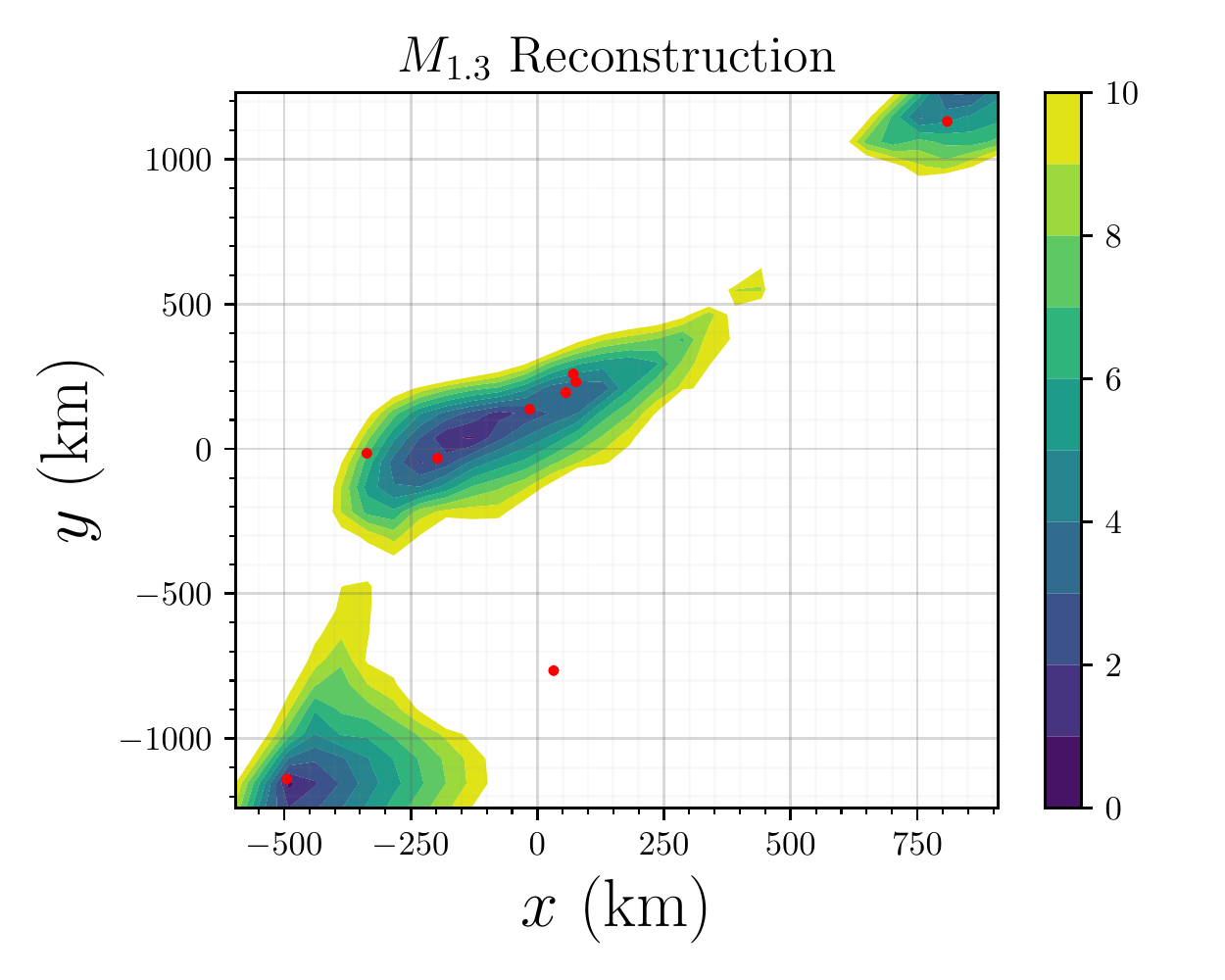}\\
\includegraphics[width=0.3\textwidth]{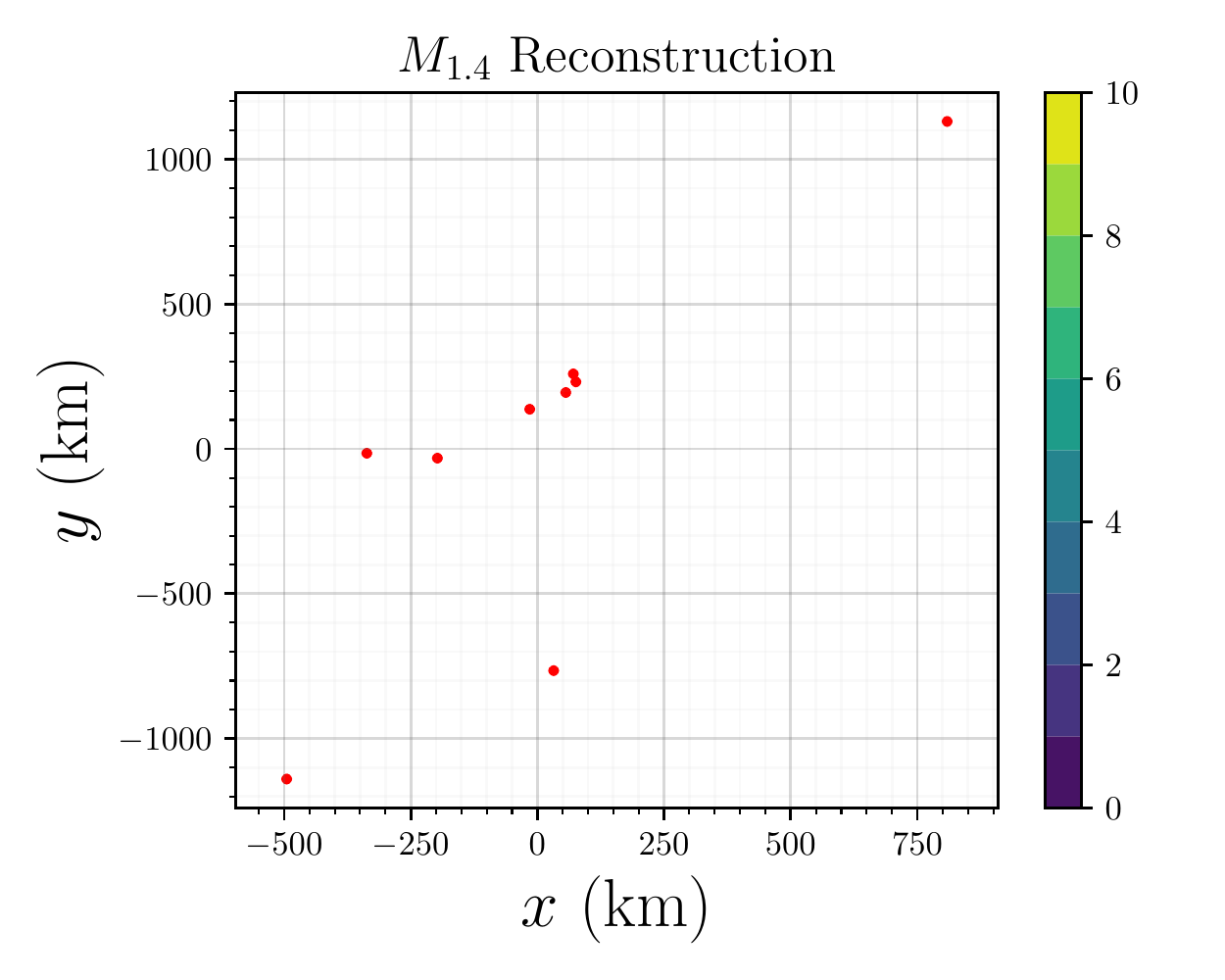}&
\includegraphics[width=0.3\textwidth]{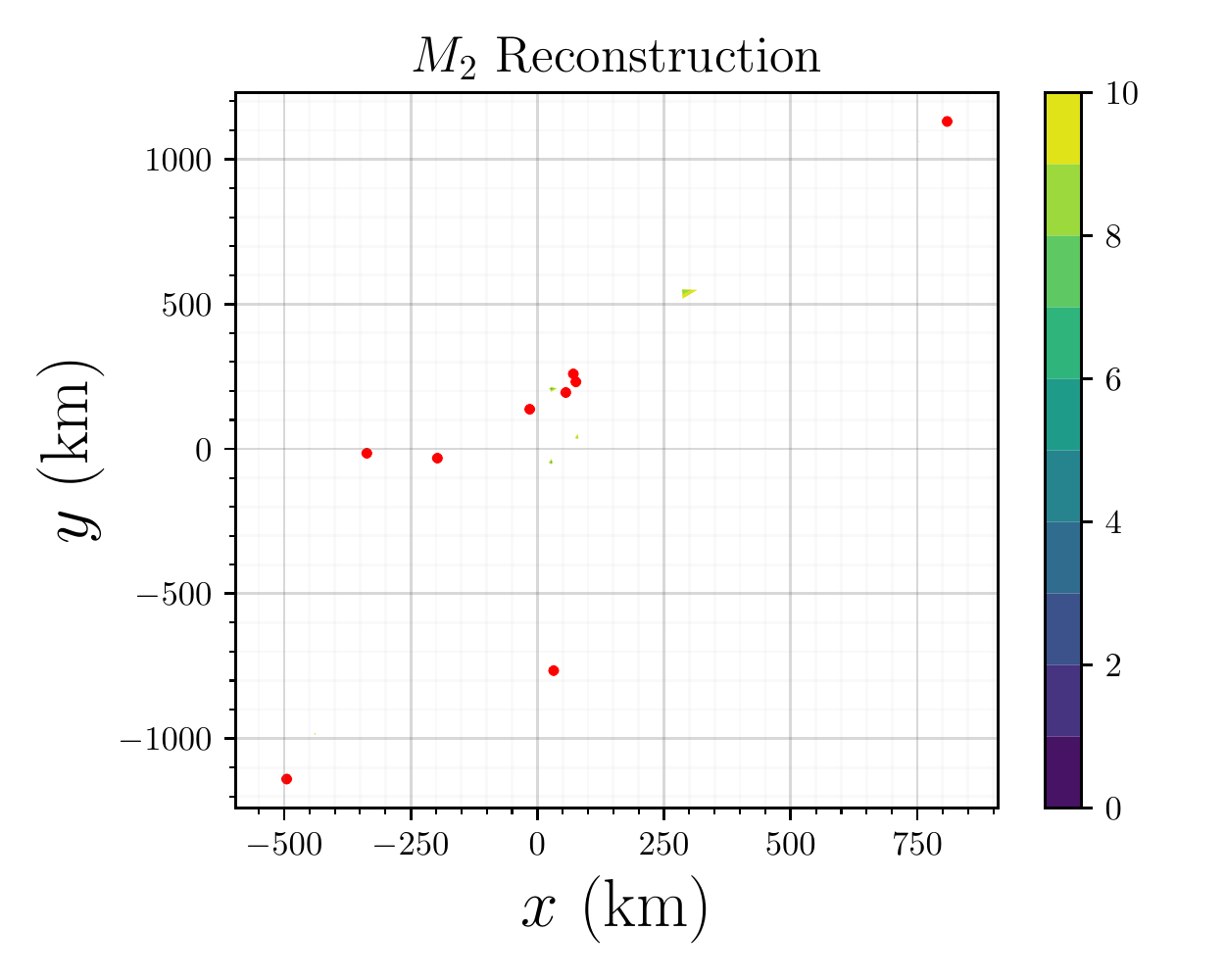}&
\includegraphics[width=0.3\textwidth]{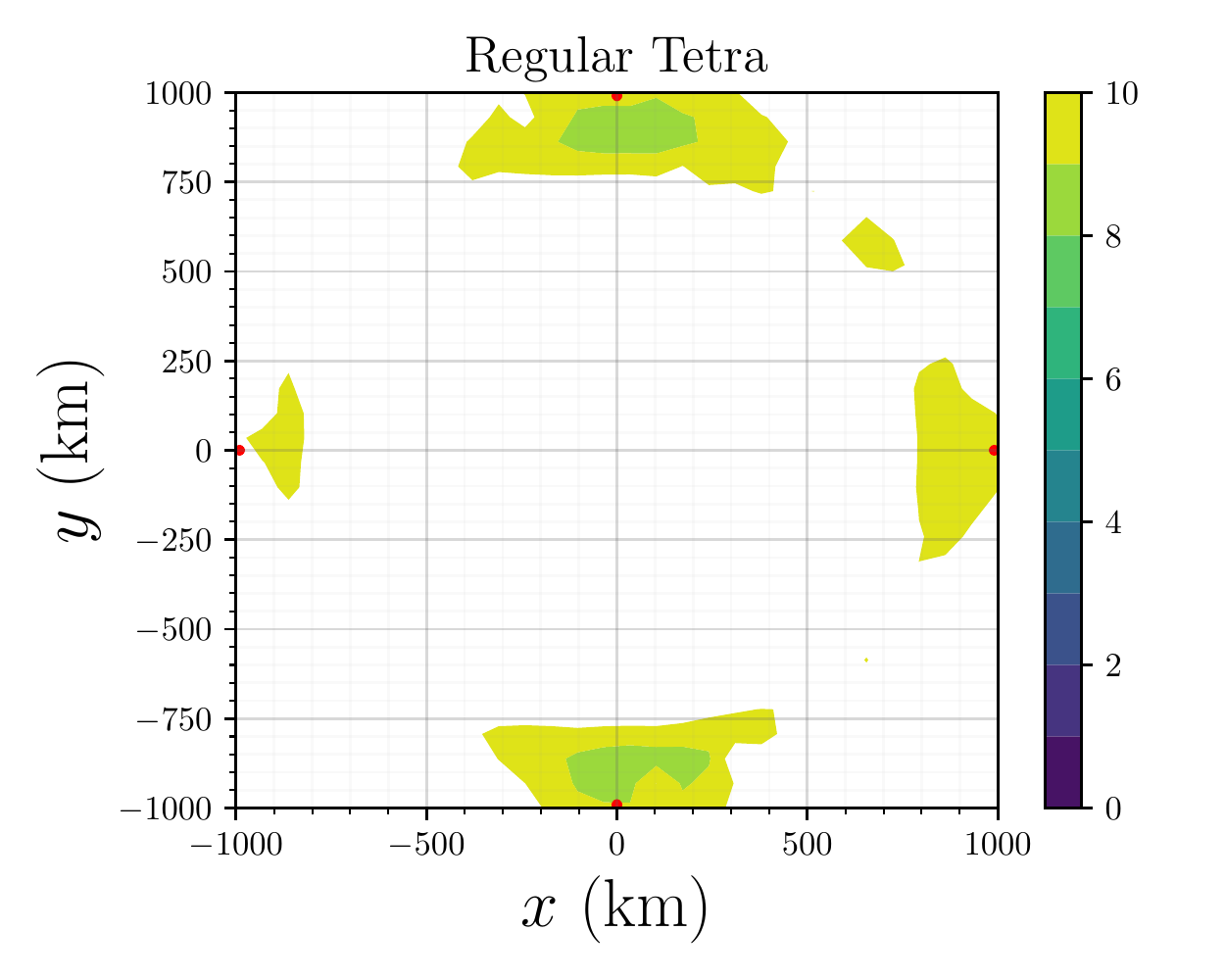}
\end{tabular}
\caption{Computation error (defined in equation \ref{eqn:B_error}) at all points on the $z=0$ plane of the simple current sheet model, using the swarm configuration at hour 205 of the HelioSwarm DRM. The layout is identical to Fig.~\ref{fig:simp_94}.}
\label{fig:simp_205} 
\end{figure*}

The errors computed for the reconstruction of the simple current sheet are displayed in Figures \ref{fig:simp_94}, \ref{fig:simp_144}, and \ref{fig:simp_205} for the hour 94, 144, and 205 configurations respectively. These figures illustrate the ensemble-averaged errors along a 2D plane orthogonal to the current intersecting a given nine-spacecraft configuration's barycenter. The first four panels correspond to the four first-order reconstruction methods, $M_{1,1}$, $M_{1,2}$, $M_{1,3}$ and $M_{1,4}$, the fifth panel corresponds to the second-order method $M_2$, and the final panel corresponds to the reconstruction obtained from the standard first-order method applied to a single regular tetrahedron, with $E=P=0$. This single tetrahedron has the same characteristic size $L$ as the overall nine-spacecraft configuration, calculated as twice the major axis of the volumetric tensor, Eqn~\ref{eqn:R_tensor}, evaluated using all nine points. With four spacecraft, we cannot reconstruct the magnetic field with the second-order method, nor can we select subsets of tetrahedra with advantageous geometric characteristics, so only the first-order reconstruction method from a single tetrahedron is used. 

We see that near the barycenter of each of the nine-spacecraft configurations (located at the origin of Figures \ref{fig:simp_94}, \ref{fig:simp_144}, and \ref{fig:simp_205}) the magnetic field can be reconstructed to within $1 \%$ accuracy. By comparing method $M_{1.1}$ with methods $M_{1.2}$ and $M_{1.3}$ in these figures, we also conclude that leveraging knowledge of the tetrahedral shapes and positions expands the region of high-accuracy reconstruction. Unfortunately, overly restrictive conditions limit the number of tetrahedra available to average over, limiting the size of the reconstructed region. In fact, the bottom left panel of Figure \ref{fig:simp_205} is empty because none of the 126 tetrahedra in the hour 205 configuration satisfy the geometric requirement that $\chi_j \leq 0.6$ demanded by $M_{1.4}$. Additionally, the second-order reconstruction method $M_2$ is accurate for only a small volume when compared with the first-order methods $M_{1.1}$, $M_{1.2}$, $M_{1.3}$, and $M_{1.4}$.

By comparing the bottom right panel to the other five in Figures \ref{fig:simp_94}-\ref{fig:turb_205}, we see that the behavior of methods $M_{1.1}$, $M_{1.2}$, $M_{1.3}$, $M_{1.4}$, and $M_{2}$ is distinct to that of the reconstruction using a single regular tetrahedron. The single regular tetrahedron only accurately reconstructs the magnetic field of the current sheet near each of the four spacecraft. Due to the angular symmetry in the current sheet and the fact that none of the four spacecraft are positioned on the $z=0$ plane, the area of most accurate reconstruction appears to be a ring on the bottom right panel of Figures \ref{fig:simp_94}, \ref{fig:simp_144}, and \ref{fig:simp_205}.

\subsection{Turbulence Simulation}
\begin{figure}[ht]
    \includegraphics[width=\columnwidth]{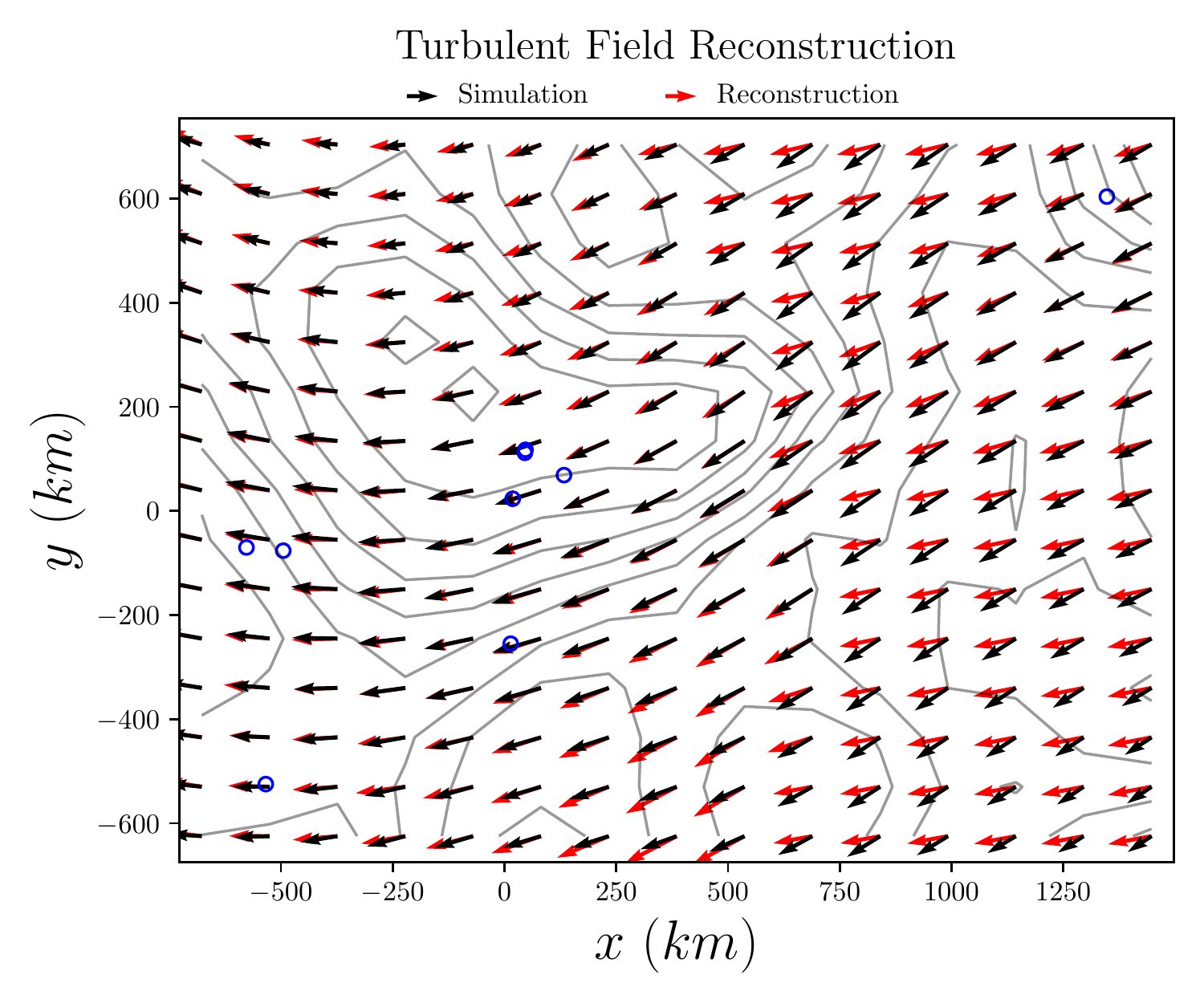}
    \caption{An example of the spacecraft configuration at hour 94, pictured as blue circles, reconstructing the magnetic field associated with a turbulence simulation using first-order method $M_{1.3}$. The simulation's magnetic field is shown as black arrows, and the reconstructed magnetic field is shown as red arrows. Contour lines of the $\hat{z}$ component of current density $\J$ are shown in gray.}
    \label{fig:B_reconst_turb}
\end{figure}

We present an example magnetic field reconstruction of the turbulence simulation (\S \ref{sssec:turb_sim}) in Figure \ref{fig:B_reconst_turb}. Here, we use the first-order method $M_{1.3}$ to reconstruct the magnetic field in the $z=0$ plane in the turbulence simulation. Mirroring the behavior described in Figure \ref{fig:B_reconst_CS}, there is little difference between the reconstructed and original fields near the center of the spacecraft configuration.

\label{ssec:results.turb}

We perform 50 Monte Carlo iterations of reconstruction using each method, observing that 50 was more than enough to point-wise converge in error. The barycenter is chosen as a uniform random variable so that all spacecraft remained in the $31415 \times 31415 \times 157079$ km simulation cube. We then construct a $30\times 30\times 30$ grid of points $\xi$. Each dimension of this grid is selected so that the overall size of the grid extends $100$ km past the furthest spacecraft in all directions.

\begin{figure*}
\centering
\textbf{Errors in Turbulence Simulation: Hour 94} 
\begin{tabular}{ccc}
\includegraphics[width=0.3\textwidth]{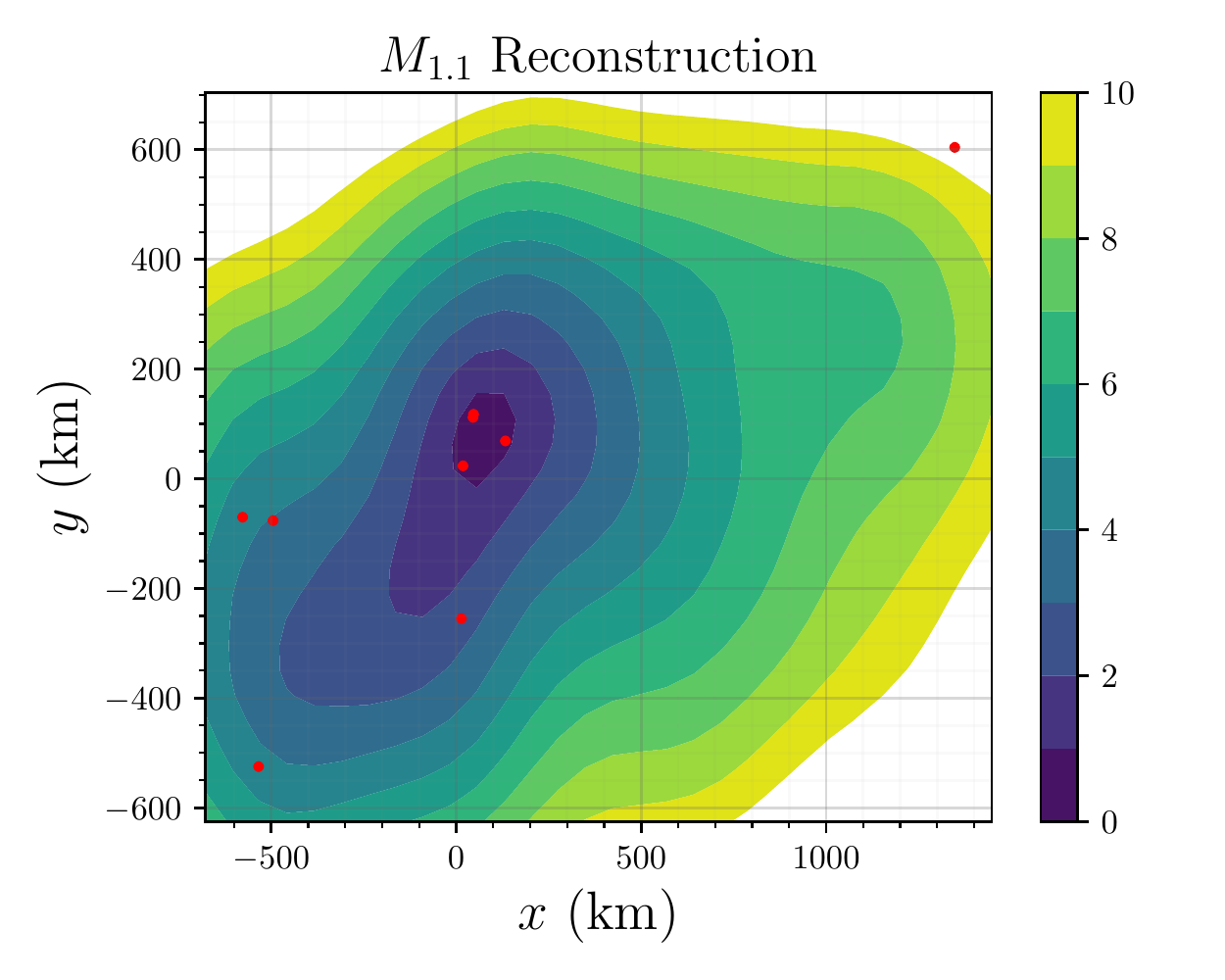}&
\includegraphics[width=0.3\textwidth]{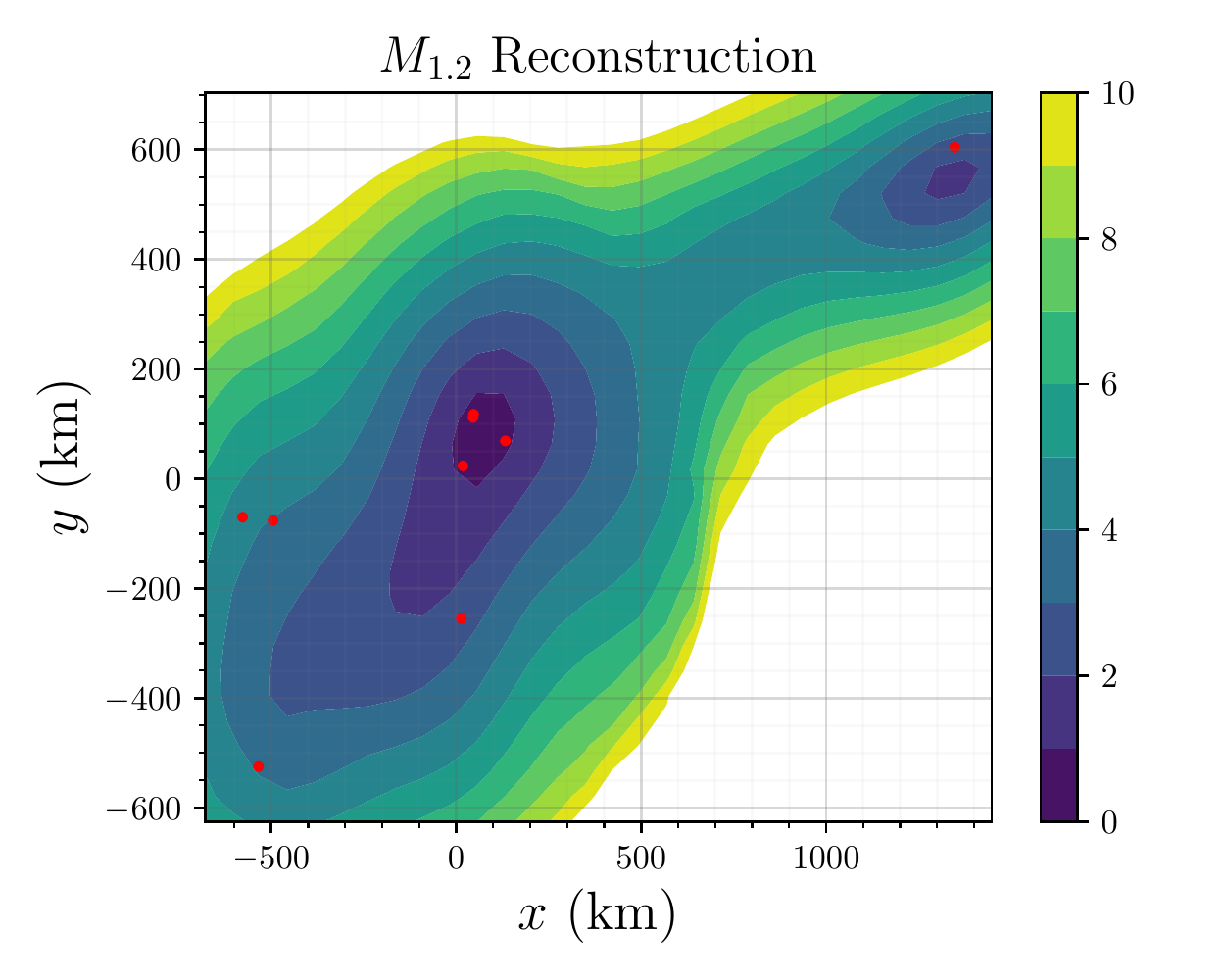}&
\includegraphics[width=0.3\textwidth]{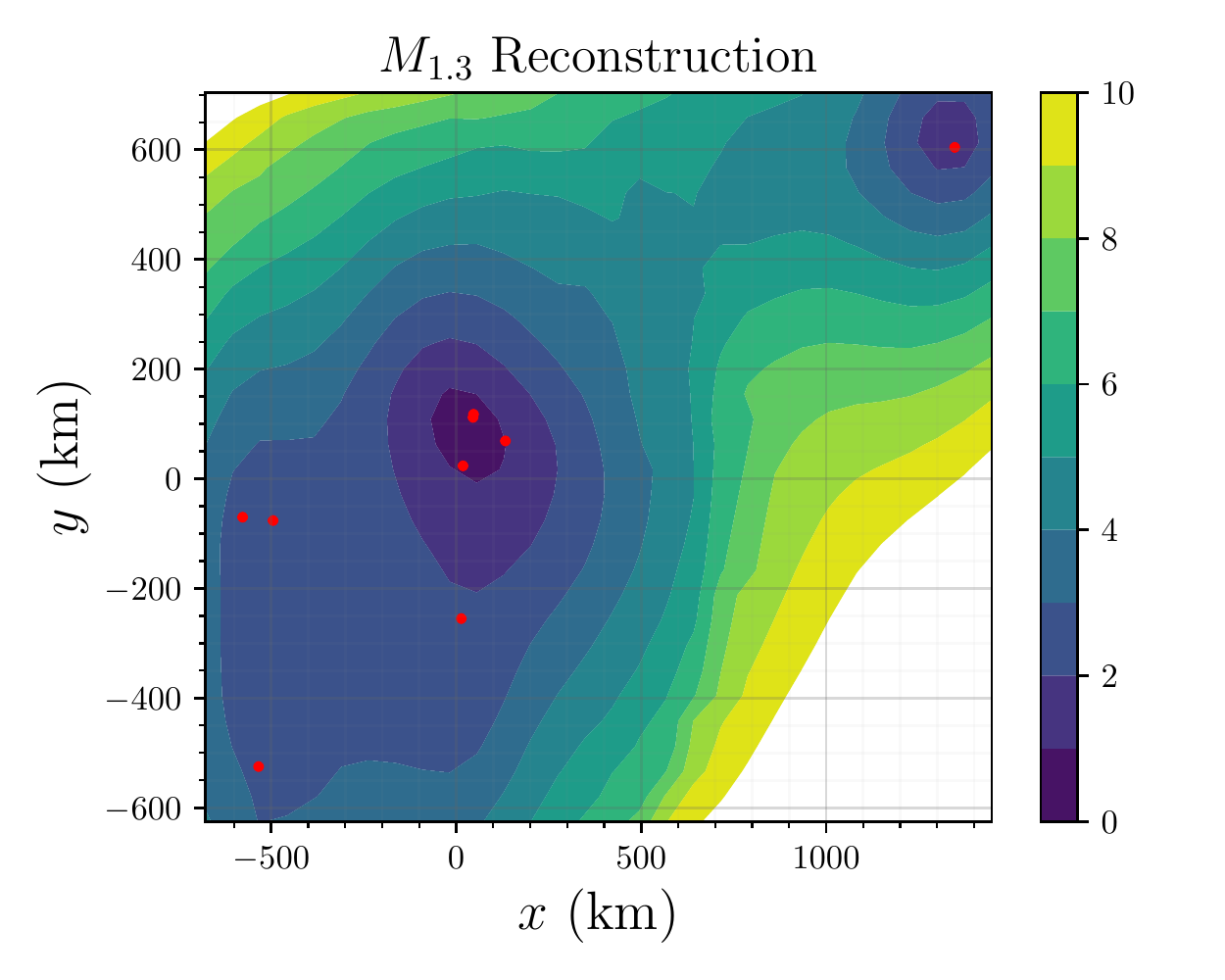}\\
\includegraphics[width=0.3\textwidth]{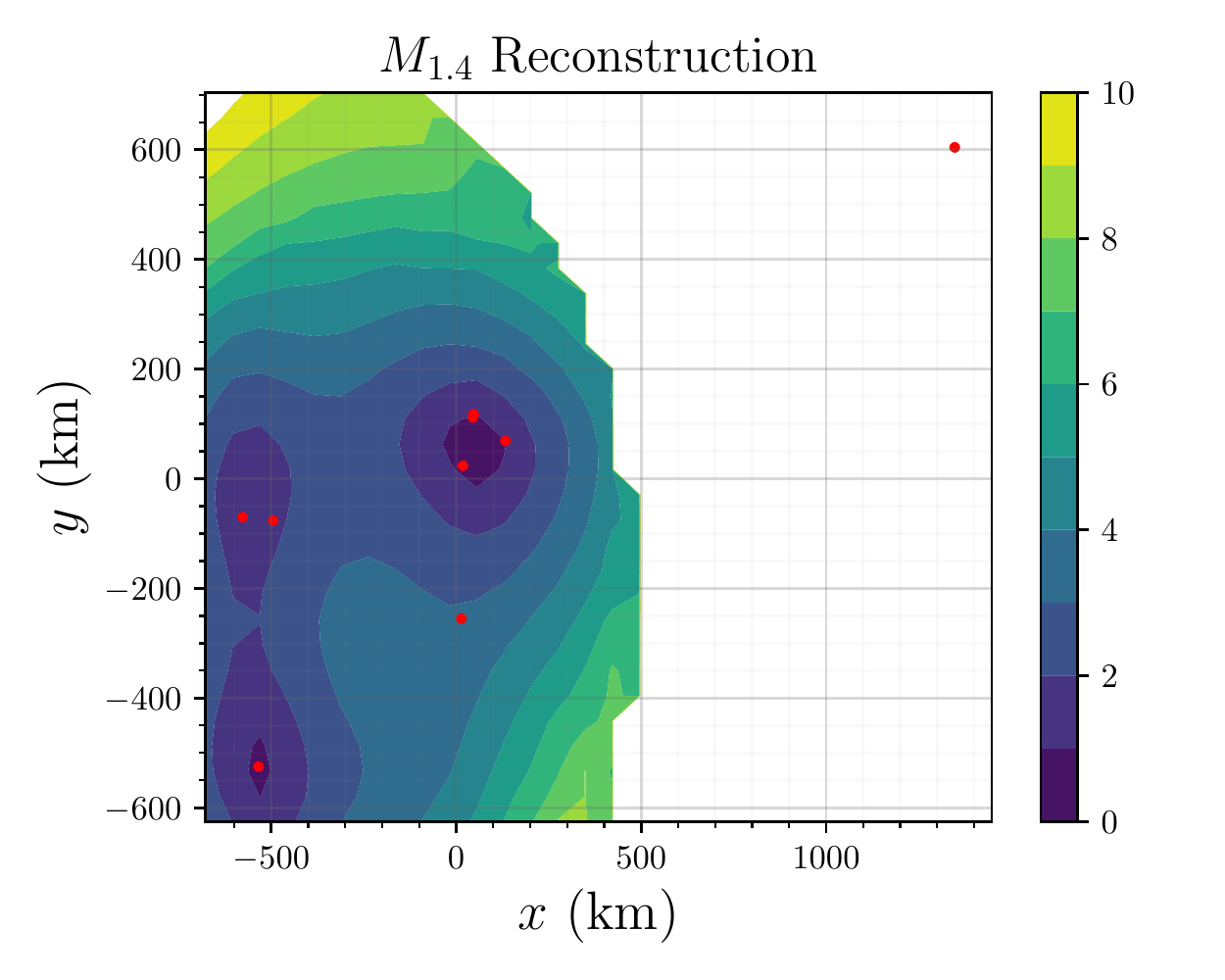}&
\includegraphics[width=0.3\textwidth]{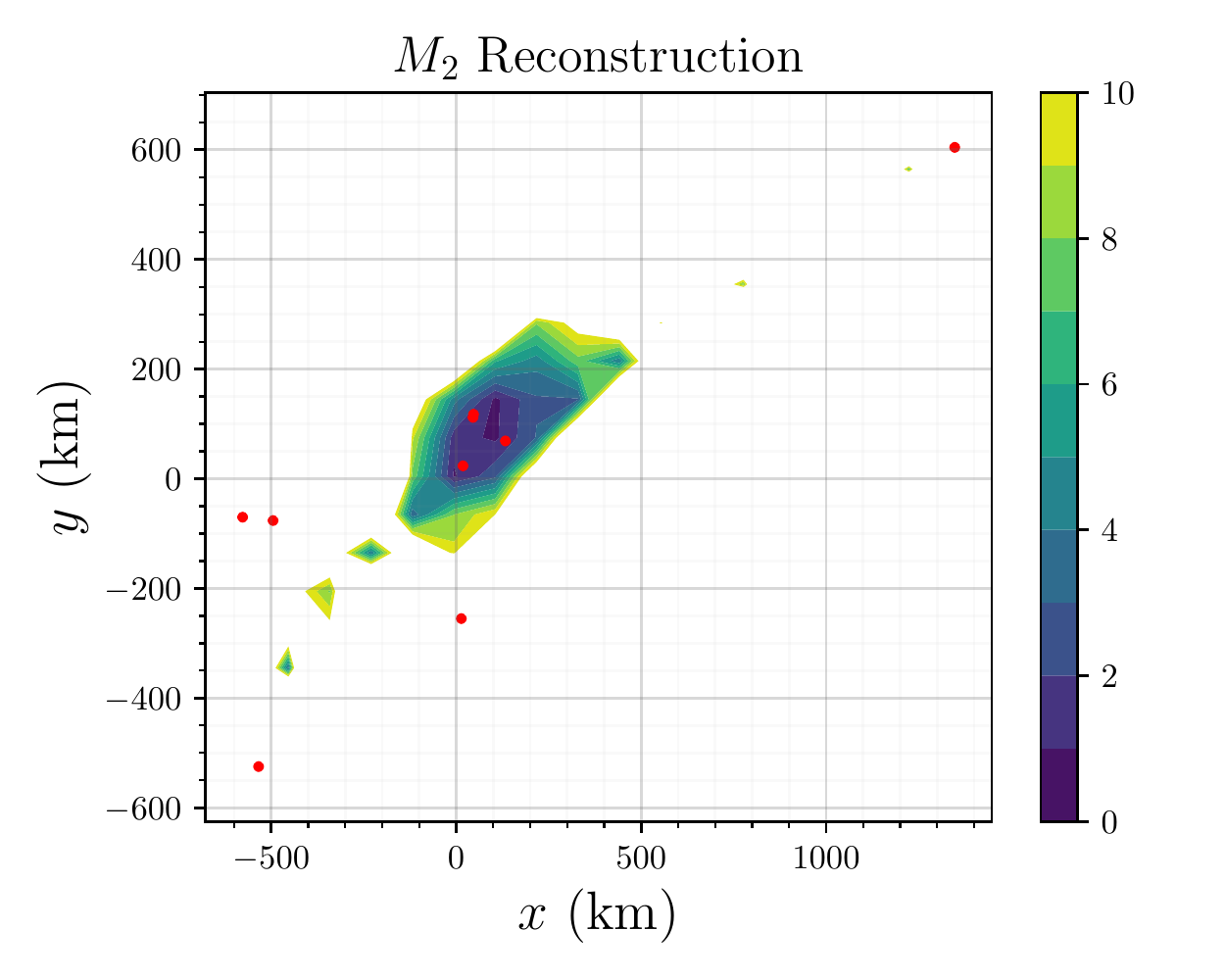}&
\includegraphics[width=0.3\textwidth]{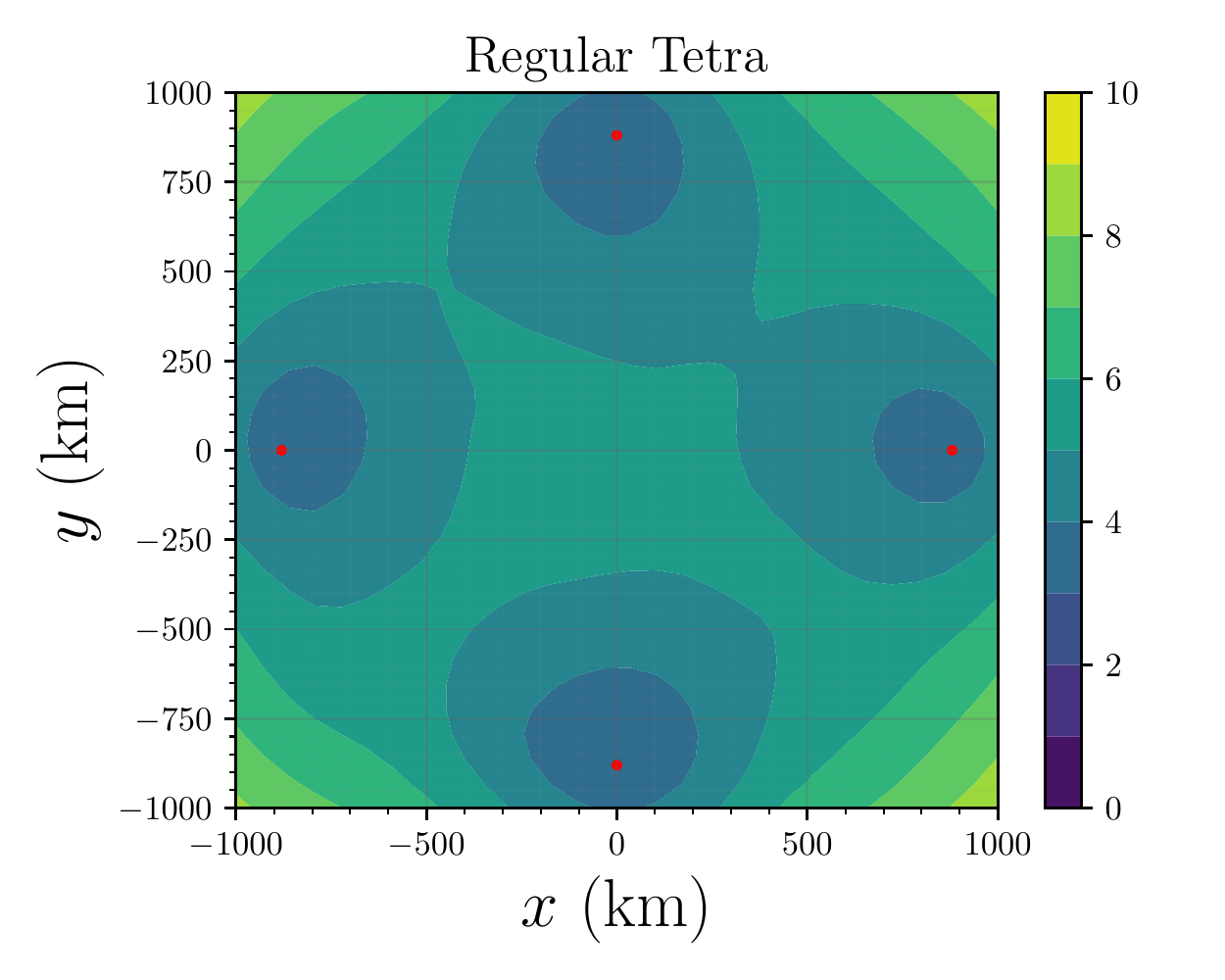}
\end{tabular}
\caption{Computation error (defined in equation \ref{eqn:B_error}) at all points on the $z=0$ plane of the turbulent magnetic field (from the \gkeyll\ Simulation), using the swarm configuration at hour 94 of the HelioSwarm DRM. The layout is identical to Fig.~\ref{fig:simp_94}.}
\label{fig:turb_94} 
\end{figure*}

\begin{figure*}
\centering
\textbf{Errors in Turbulence Simulation: Hour 144} 
\begin{tabular}{ccc}
\includegraphics[width=0.3\textwidth]{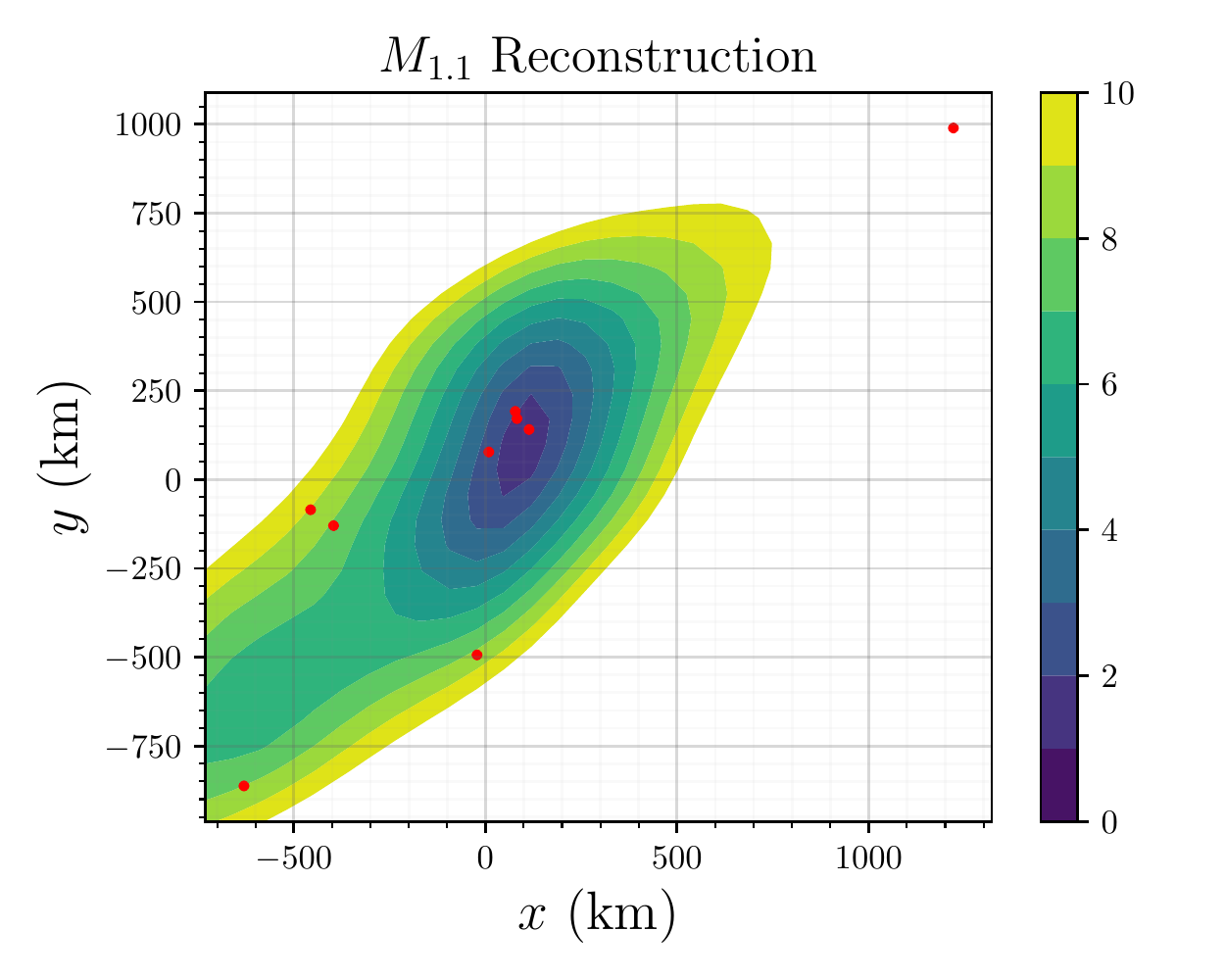}&
\includegraphics[width=0.3\textwidth]{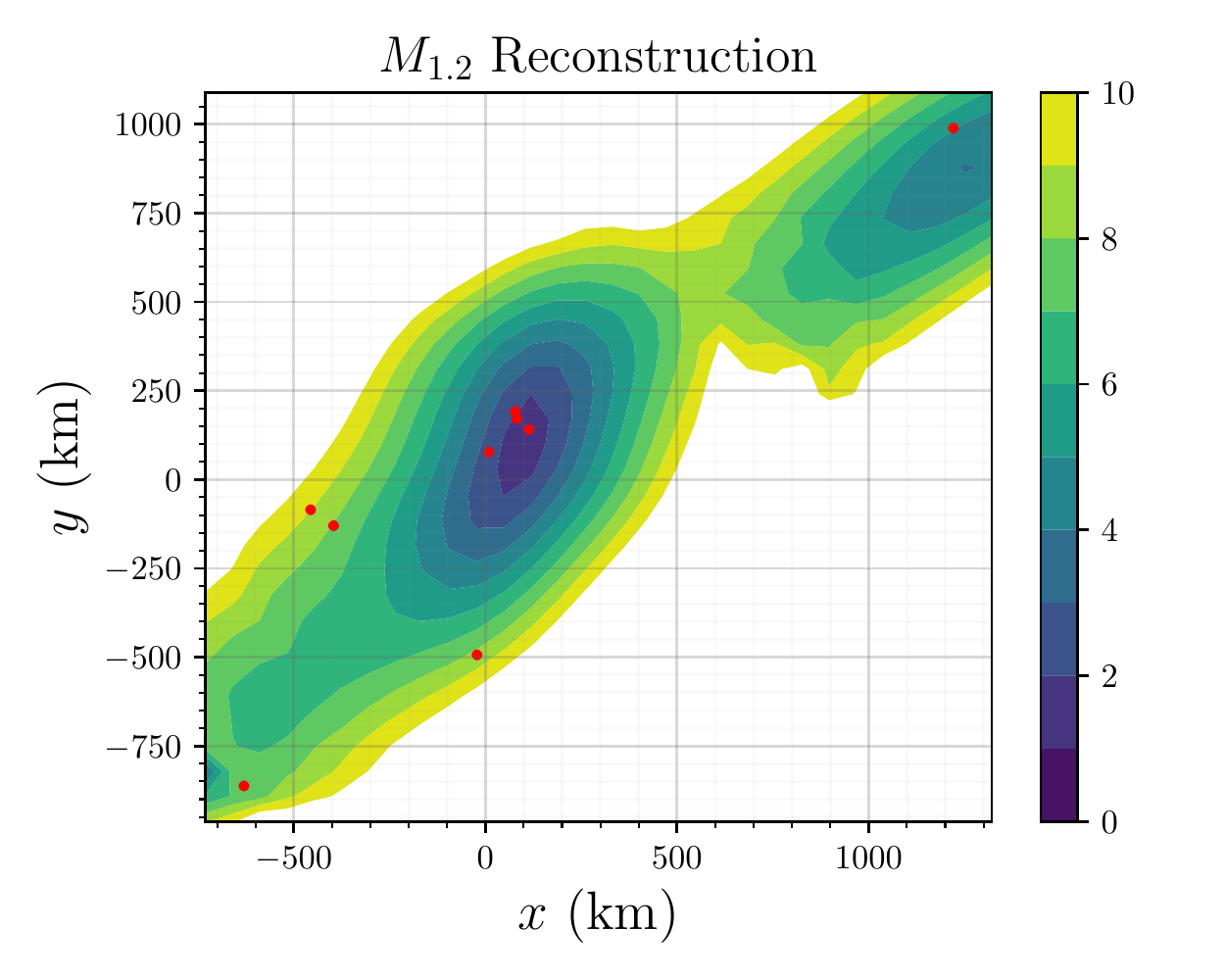}&
\includegraphics[width=0.3\textwidth]{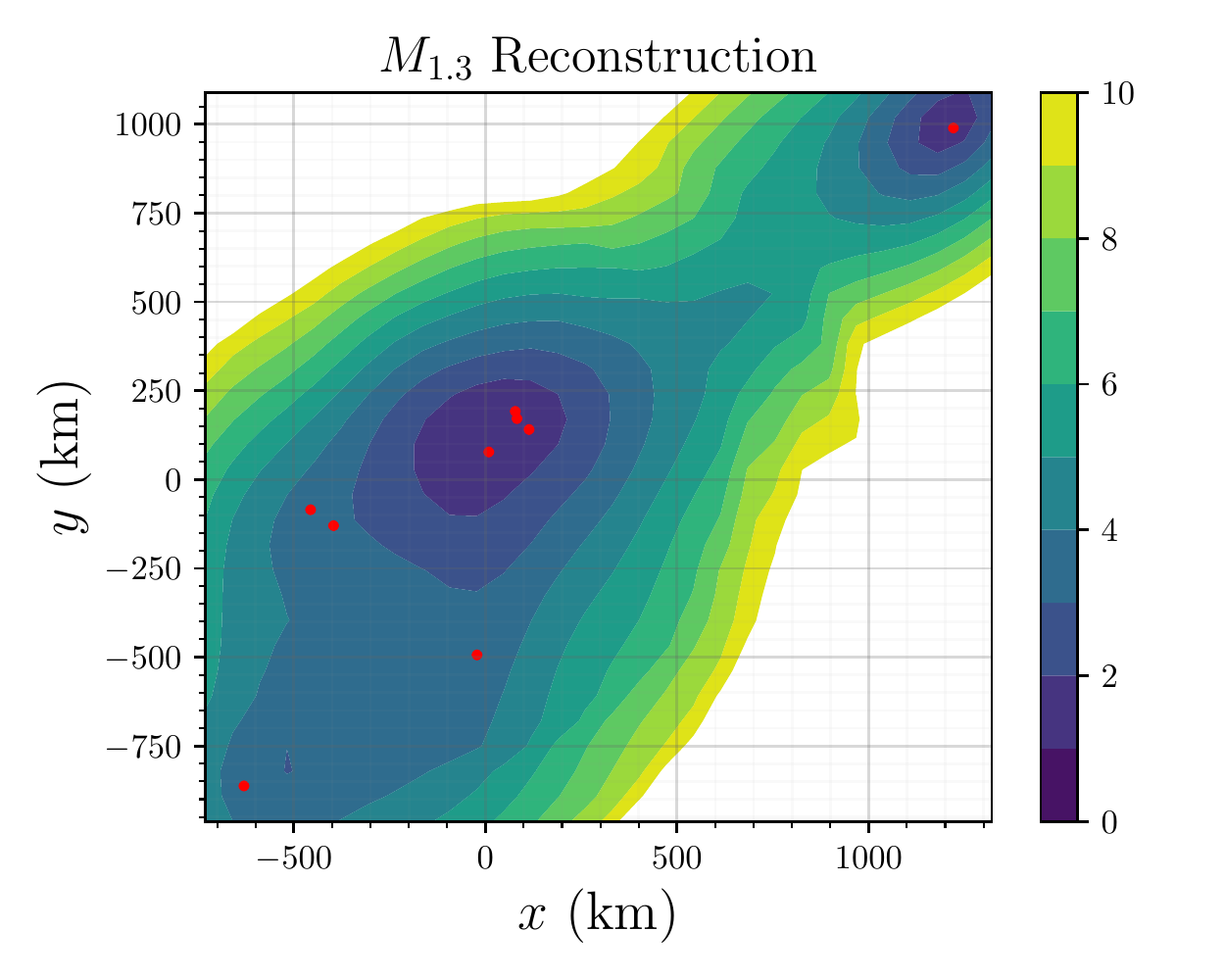}\\
\includegraphics[width=0.3\textwidth]{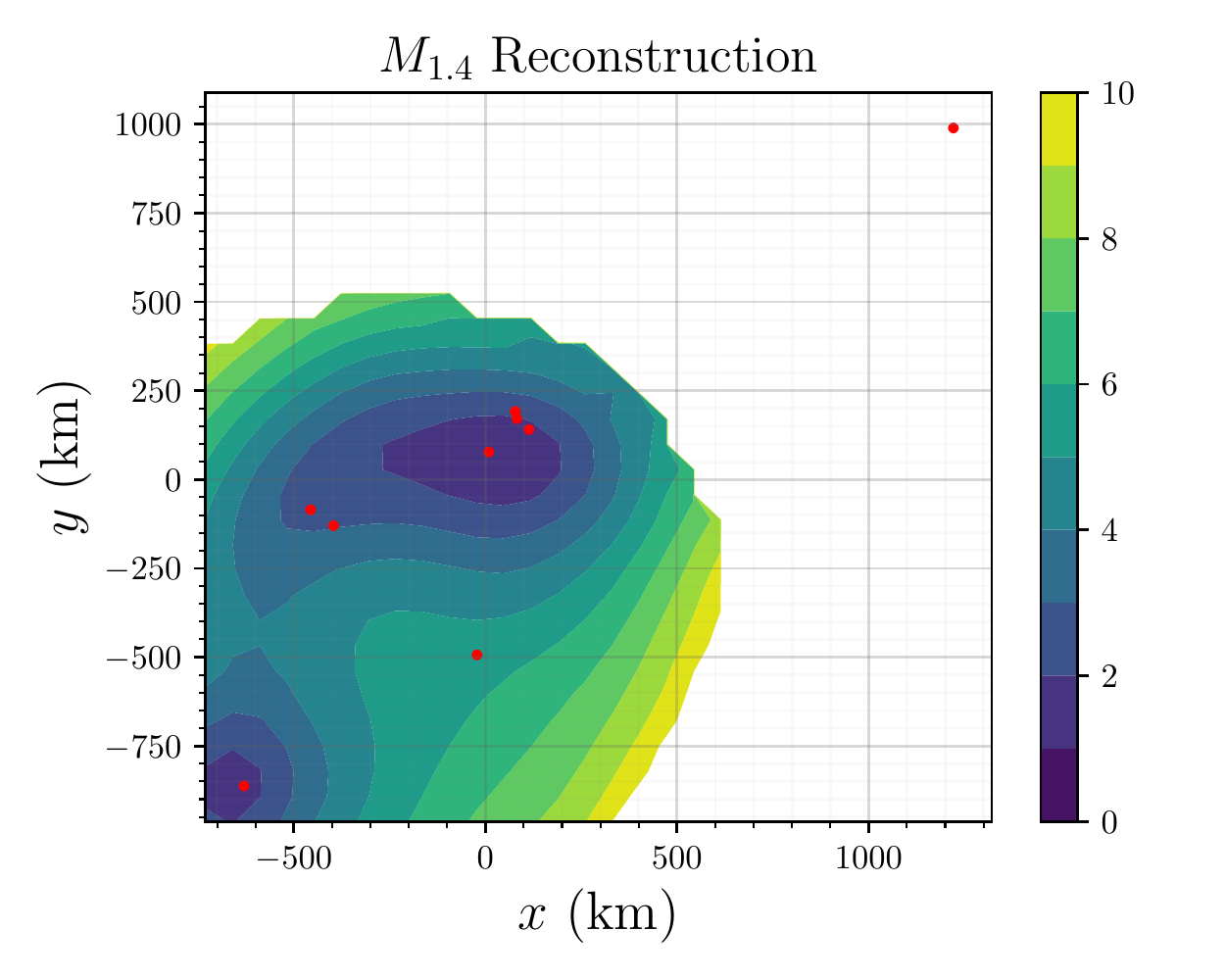}&
\includegraphics[width=0.3\textwidth]{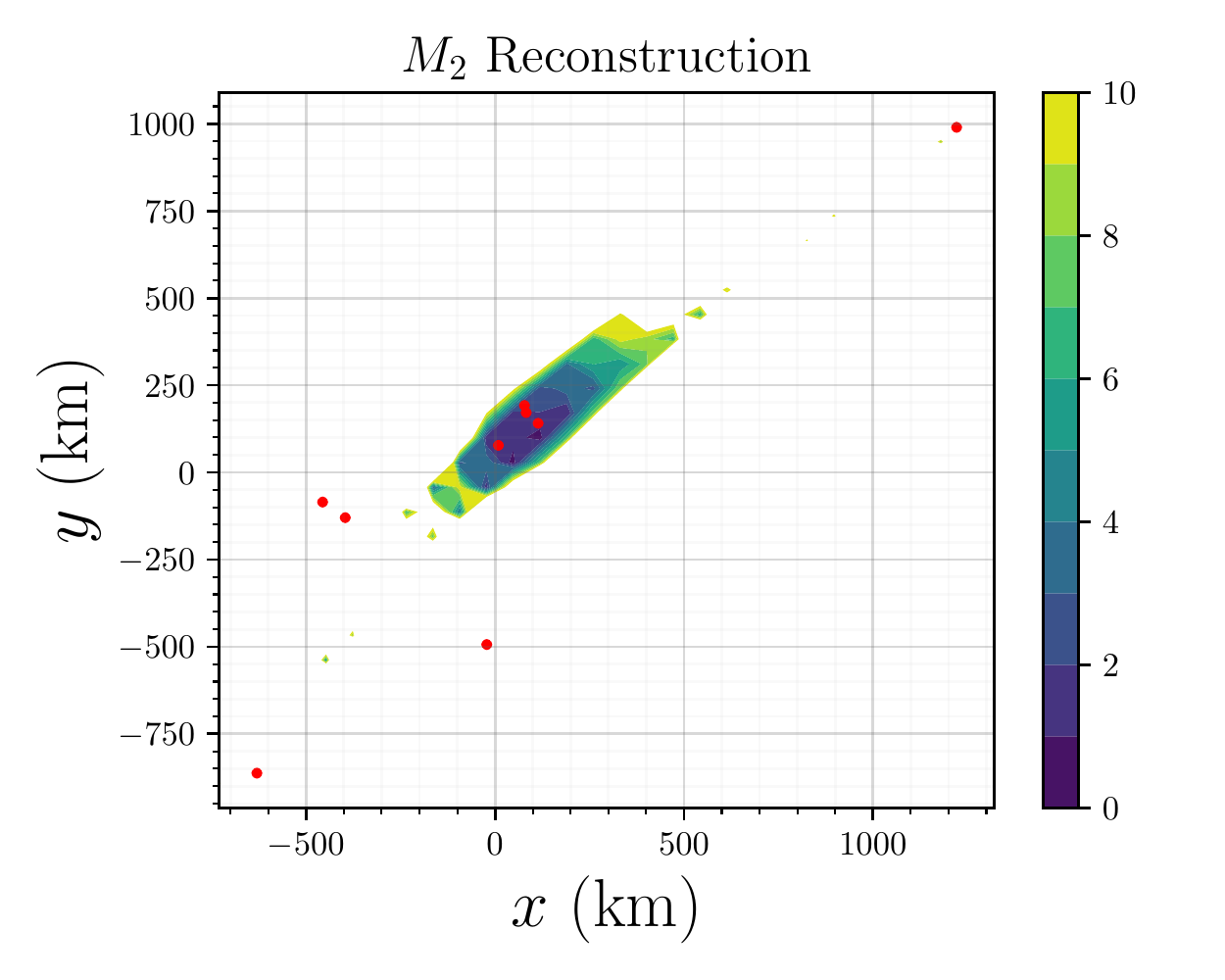}&
\includegraphics[width=0.3\textwidth]{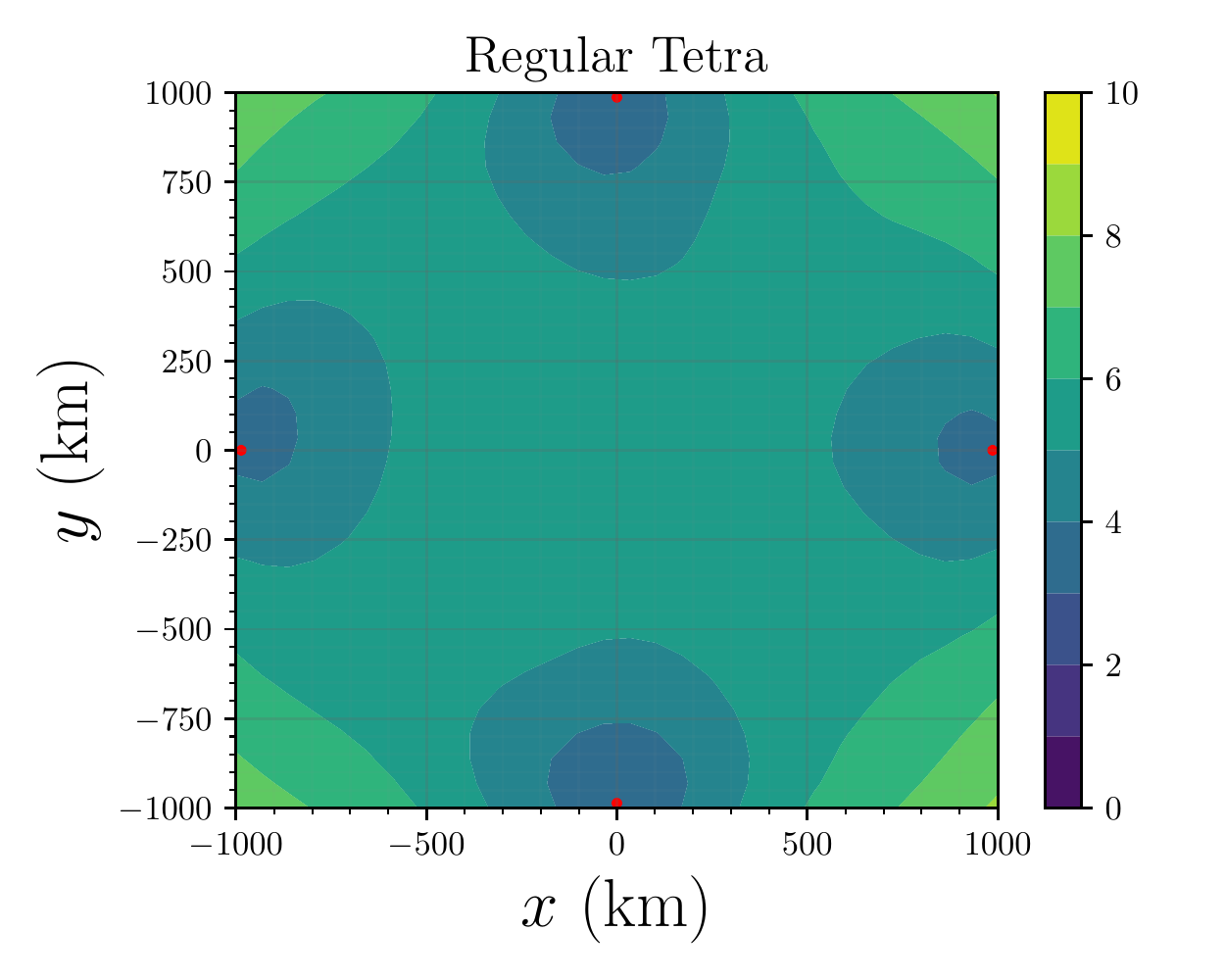}
\end{tabular}
\caption{Computation error (defined in equation \ref{eqn:B_error}) at all points on the $z=0$ plane of the turbulent magnetic field (from the \gkeyll\ Simulation), using the swarm configuration at hour 144 of the HelioSwarm DRM. The layout is identical to Fig.~\ref{fig:simp_94}.}
\label{fig:turb_144} 
\end{figure*}

\begin{figure*}
\centering
\textbf{Errors in Turbulence Simulation: Hour 205} 
\begin{tabular}{ccc}
\includegraphics[width=0.3\textwidth]{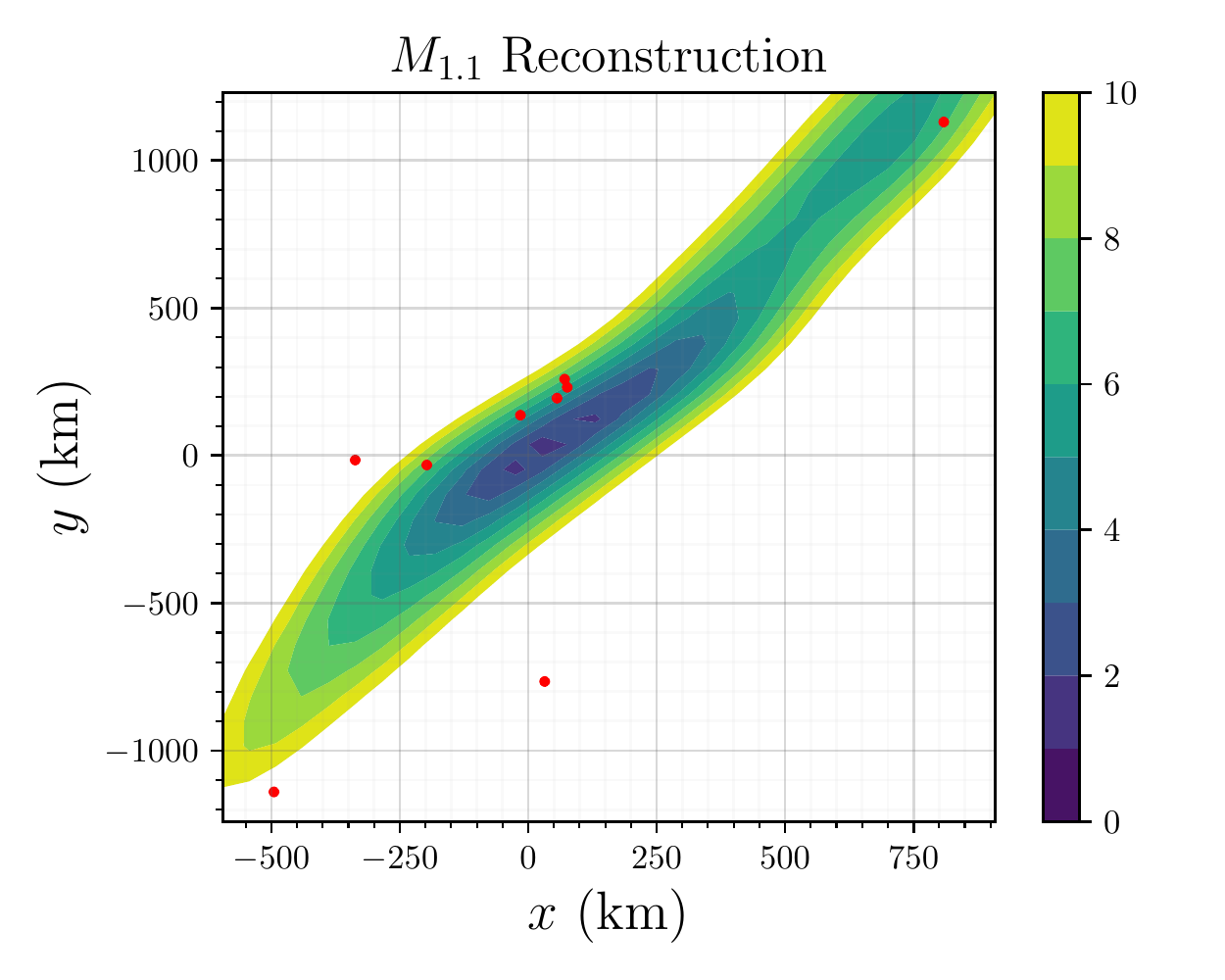}&
\includegraphics[width=0.3\textwidth]{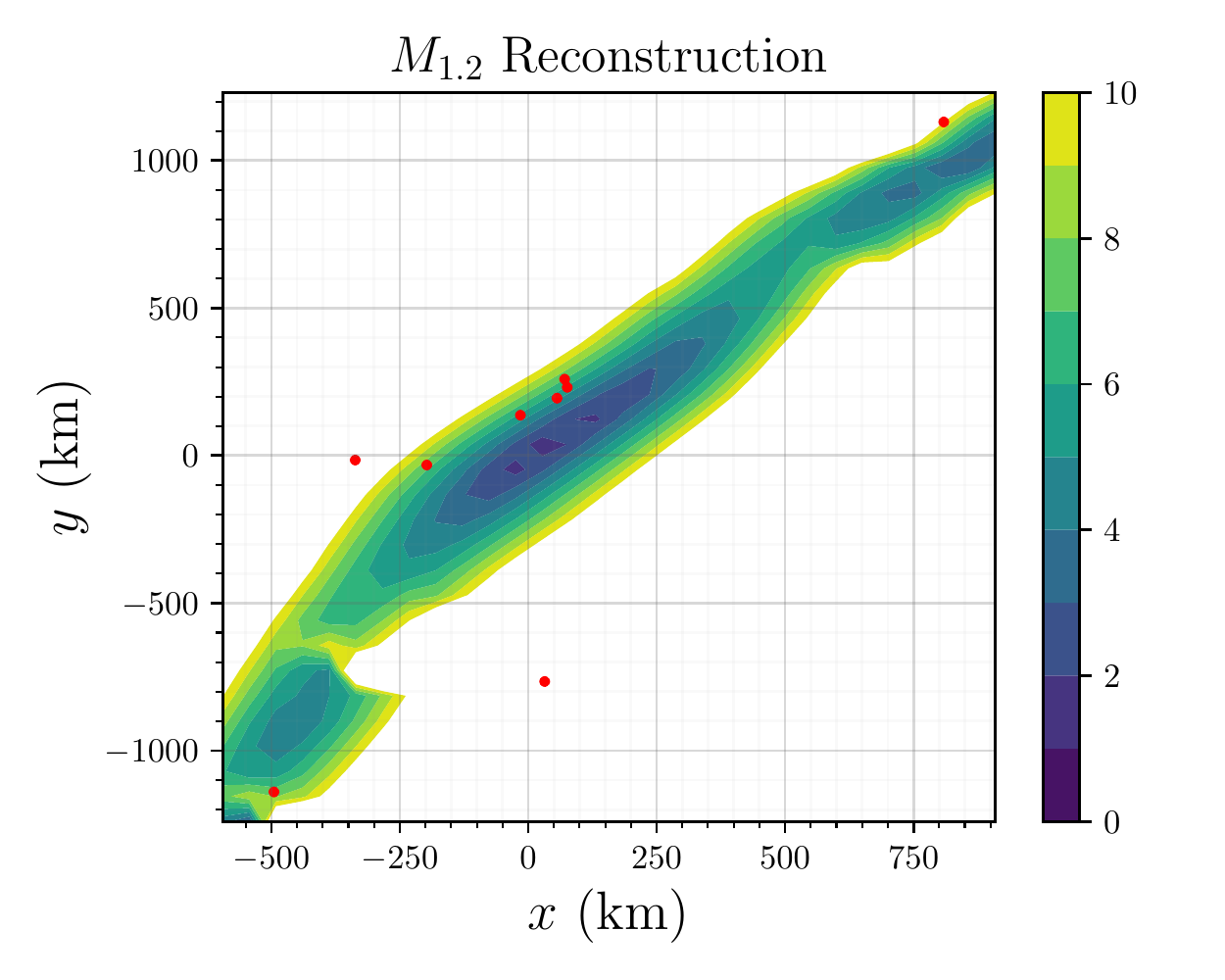}&
\includegraphics[width=0.3\textwidth]{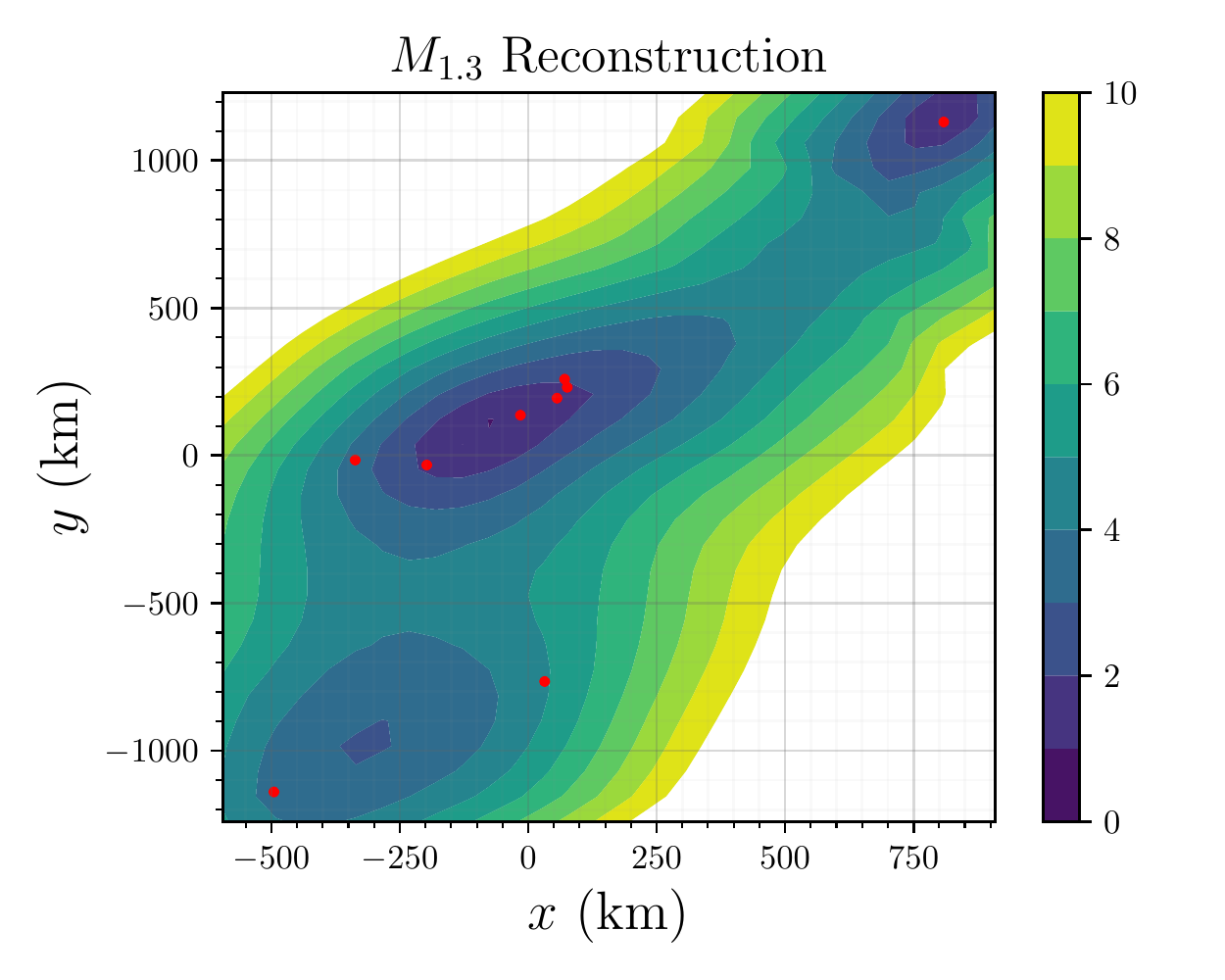}\\
\includegraphics[width=0.3\textwidth]{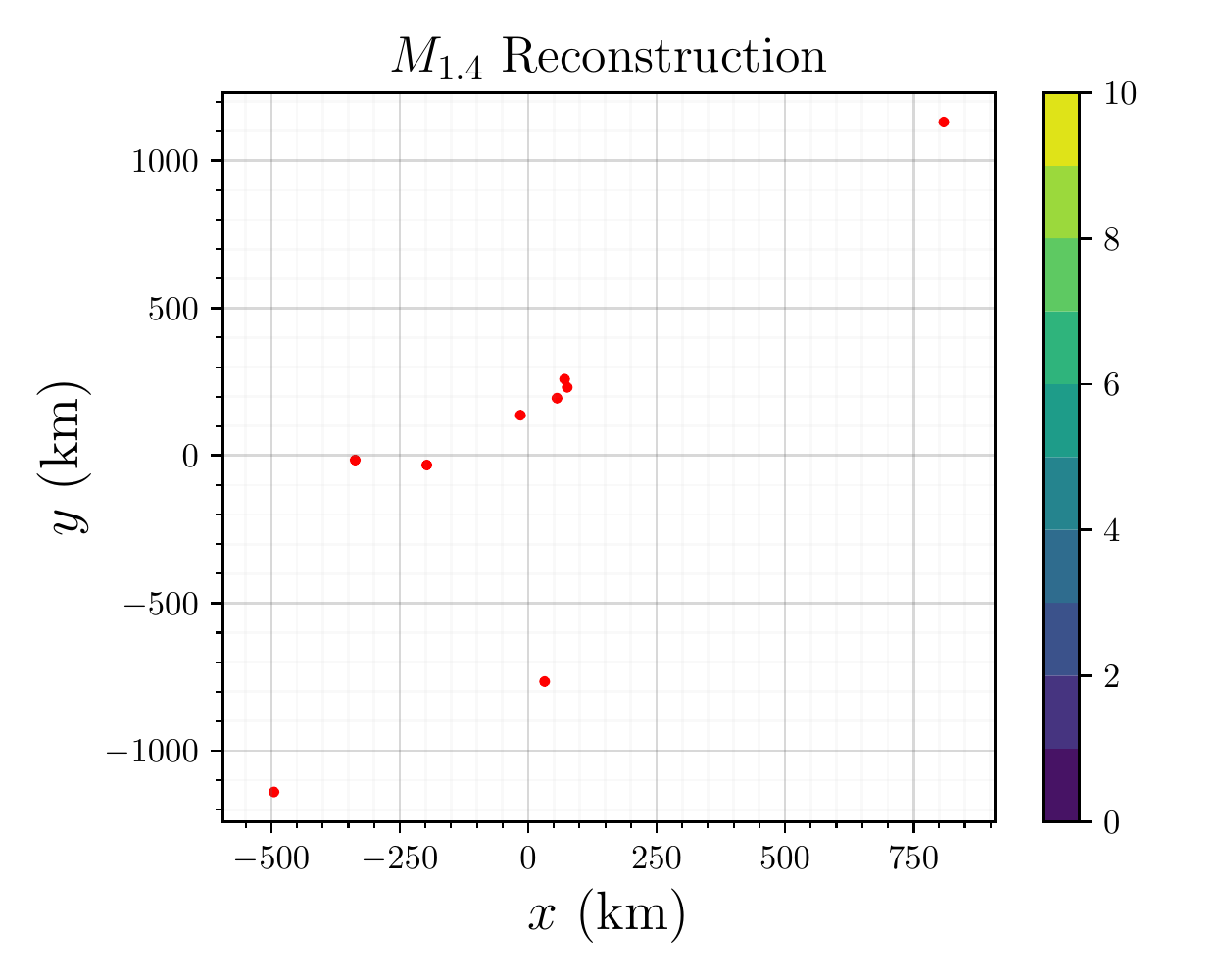}&
\includegraphics[width=0.3\textwidth]{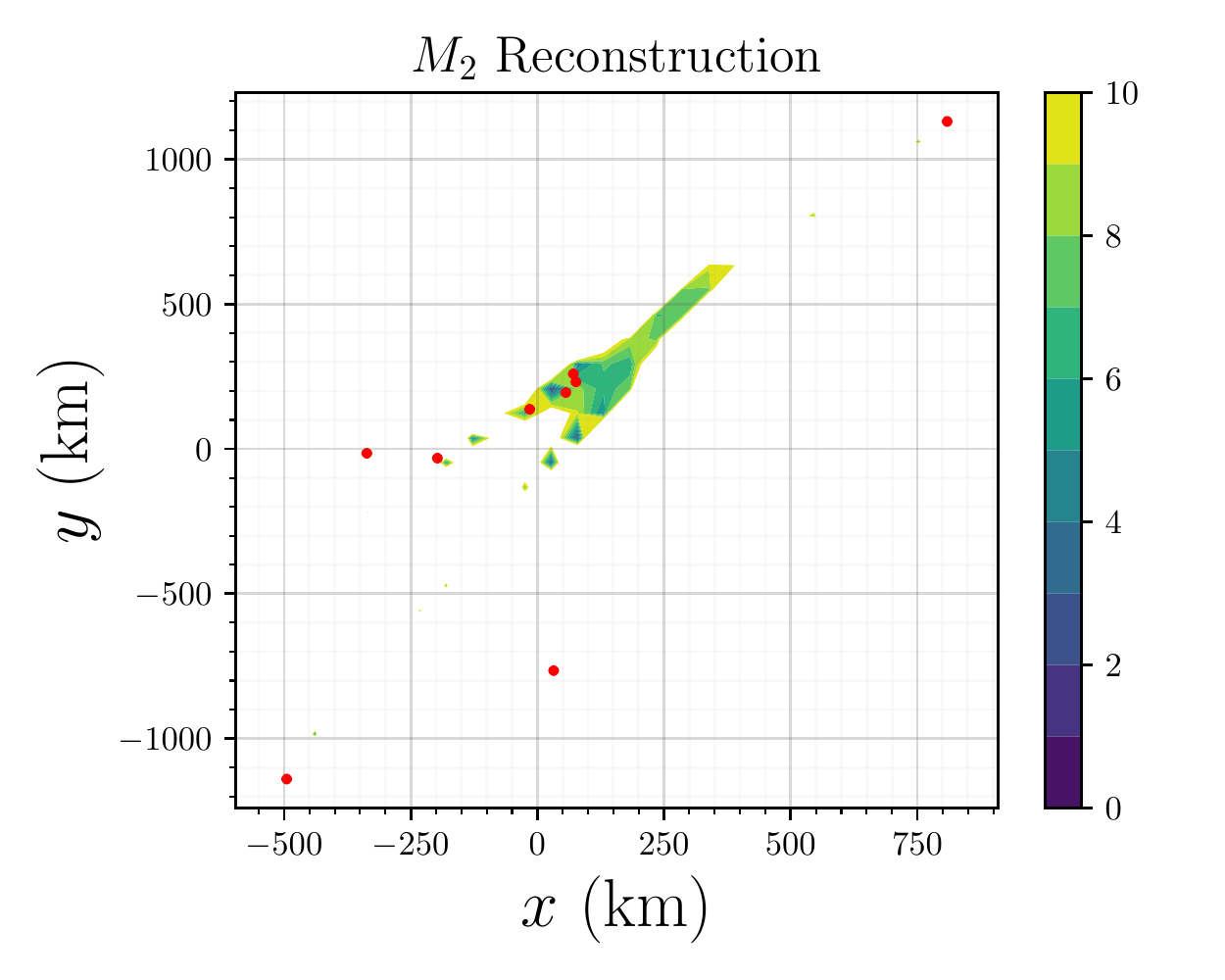}&
\includegraphics[width=0.3\textwidth]{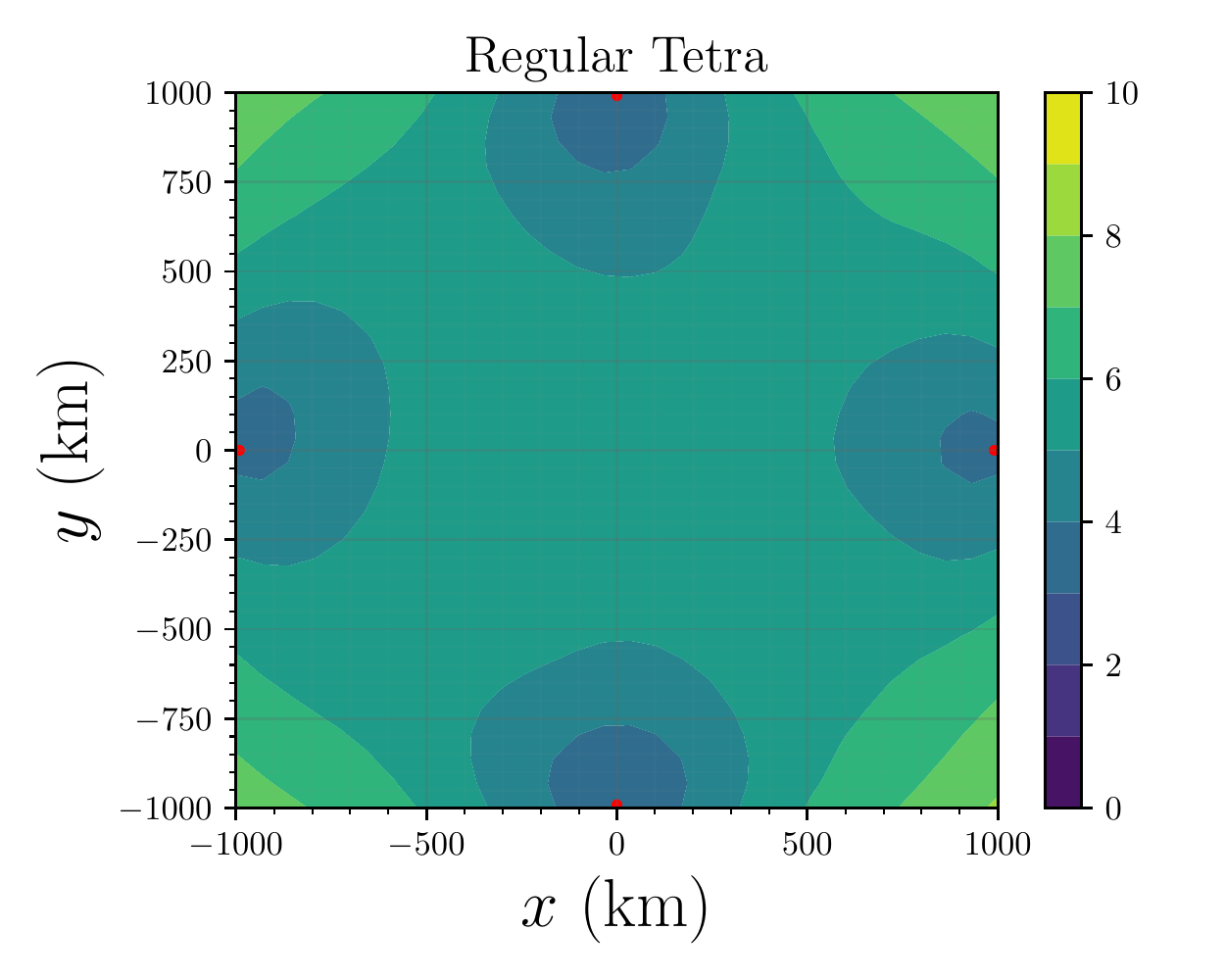}
\end{tabular}
\caption{Computation error (defined in equation \ref{eqn:B_error}) at all points on the $z=0$ plane of the turbulent magnetic field (from the \gkeyll\ Simulation), using the swarm configuration at hour 205 of the HelioSwarm DRM.The layout is identical to Fig.~\ref{fig:simp_94}. }
\label{fig:turb_205} 
\end{figure*}

The errors computed from the turbulence simulation reconstruction are displayed in Figures \ref{fig:turb_94}, \ref{fig:turb_144}, and \ref{fig:turb_205} for the configurations at hours 94, 144, and 205 respectively. The panels shown are organized in the same order as the simple current sheet reconstruction. In Table \ref{tab:error.turb} we show the volume (in units of $ 10^6$ $\text{km}^3$) of the magnetic field that can be reconstructed with errors less than 1\%, 5\%, and 10\%. This is done for all three of the investigated spacecraft configurations, and using all five of the nine-spacecraft reconstruction methods, $M_{1.1}$, $M_{1.2}$, $M_{1.3}$, $M_{1.4}$, and $M_{2}$. In the bottom half of this table, we compare the volume reconstructed using a single regular tetrahedron to that of our five reconstruction methods.

\begin{table*}[ht]
    \centering
    \begin{tabular}{c|ccc|ccc|ccc}
    & Hour 94 & & & Hour 144 & & & Hour 205 & &  \\
    \hline
    Volume & $\epsilon(1)$ &$\epsilon(5)$ &$\epsilon(10)$ & $\epsilon(1)$ &$\epsilon(5)$ &$\epsilon(10)$ & $\epsilon(1)$ &$\epsilon(5)$ &$\epsilon(10)$ \\
        \hline
    $M_{1.1}$ & 17.56 &	1057 &	3095 &	2.95 &	324.4 &	2021 &	1.325 &	145.7 &	1231 \\
    $M_{1.2}$ & 17.93 &	1197 &	2425 &	3.686 &	387.8 &	2330 &	1.325 &	218.6 &	1047 \\
    $M_{1.3}$ & 21.30 &	1816 &	3208 &	9.584 &	2151 &	4722 &	12.59 &	1556 &	4334 \\
    $M_{1.4}$ & 30.26 &	1189 &	1679 &	17.69 &	1169 &	2281 &	0.0 &	0.0 &	0.0 \\
    $M_{2}$ & 3.363 &	47.45 &	118.1 &	5.898 &	48.66 &	137.9 &	3.975 &	33.79 &	103.3 \\
    \hline
        Vs. Regular Tetrahedron (\%) & $\epsilon(1)$ &$\epsilon(5)$ &$\epsilon(10)$ & $\epsilon(1)$ &$\epsilon(5)$ &$\epsilon(10)$ & $\epsilon(1)$ &$\epsilon(5)$ &$\epsilon(10)$ \\
        \hline
    $M_{1.1}$ & 137.20 & 54.31 & 38.70 & 24.58 & 23.75 & 25.26 & 13.80 & 10.77 &	15.39 \\
    $M_{1.2}$ & 140.12 & 61.51 & 30.32 & 30.72 & 28.40 & 29.13 & 13.80 & 16.16 & 13.08 \\
    $M_{1.3}$ & 166.39 & 93.33 & 40.12 & 79.87 & 157.54 & 59.04 & 131.12 & 115.03 & 54.18 \\
    $M_{1.4}$ & 236.44 & 61.06 & 21.00 & 147.45 & 85.62 & 28.51 & 0.0 & 0.0 & 0.0 \\
    $M_{2}$ & 26.27 & 2.44 & 1.48 & 49.15 & 3.56 & 1.72 & 41.41 &	2.50 & 1.29 \\
    \end{tabular}
    \caption{Volumes (in units of $10^6$ km$^3$) with reconstructed magnetic field error less than 1\%, 5\%, or 10\% for the three configurations using the four first-order methods and the second-order method discussed in \S \ref{ssec:reconstruction}. These volumes are compared to the equivalent regions reconstructed from a single regular tetrahedron with the same characteristic size as the overall nine-spacecraft configuration.}
    \label{tab:error.turb}
\end{table*}

Near the barycenter of each of the nine-spacecraft configurations, located at the origin of Figures \ref{fig:turb_94}, \ref{fig:turb_144}, and \ref{fig:turb_205}, the magnetic field can be reconstructed to within $1 \%$ accuracy. The second-order method, $M_2$, can only reconstruct the magnetic field to within $10 \%$ accuracy in a small region near the barycenter of the configuration, while the first-order methods can reconstruct the magnetic field within $10 \%$ over a much greater area. This is the case because the second-order Taylor series expansion diverges quadratically with distance away from the barycenter of the spacecraft configuration, while the first-order Taylor series only diverges linearly with distance. Since our goal is to maximize the volume of accurate reconstruction, the first-order methods are superior. However, the second-order method may be more accurate at reconstructing the values of the magnetic field very close to the barycenter of a spacecraft configuration.

The largest disparity compared to the current sheet simulations occurs for the single regular tetrahedron case, on the bottom right panels of each figure. The magnetic field is again only reconstructed accurately near each of the spacecraft, but because the turbulence simulation lacks angular symmetry, these regions manifest as spheres centered around each spacecraft. These four spheres appear to be the same size on the bottom right panel of each figure because the spacecraft are equidistant from the $z=0$ plane.

Shown in Table \ref{tab:error.turb}, the single regular tetrahedron reconstructs the largest volume with less than 10\% error, however the first-order methods reconstruct larger volumes with smaller errors. To maximize the volume reconstructed with less than 1\% error, it appears it is best to use the first-order method $M_{1.4}$, detailed in \S \ref{sssec:tetra_inclusion} (if a sufficient number of quasi-regular tetrahedra can be formed from the nine spacecraft configuration).

\subsection{Sensitivity to Number of Spacecraft}
\label{ssec:results.robust}
We analyze how the volume reconstructed with less than 5\% error varies as a function of the number of spacecraft. This analysis was completed using the Monte Carlo sampling of the turbulent simulation as described in \S \ref{sec:results}. 

\begin{figure*}
\centering
\textbf{Turbulence Reconstruction: N Spacecraft Comparison} 
\begin{tabular}{ccc}
\includegraphics[width=0.3\textwidth]{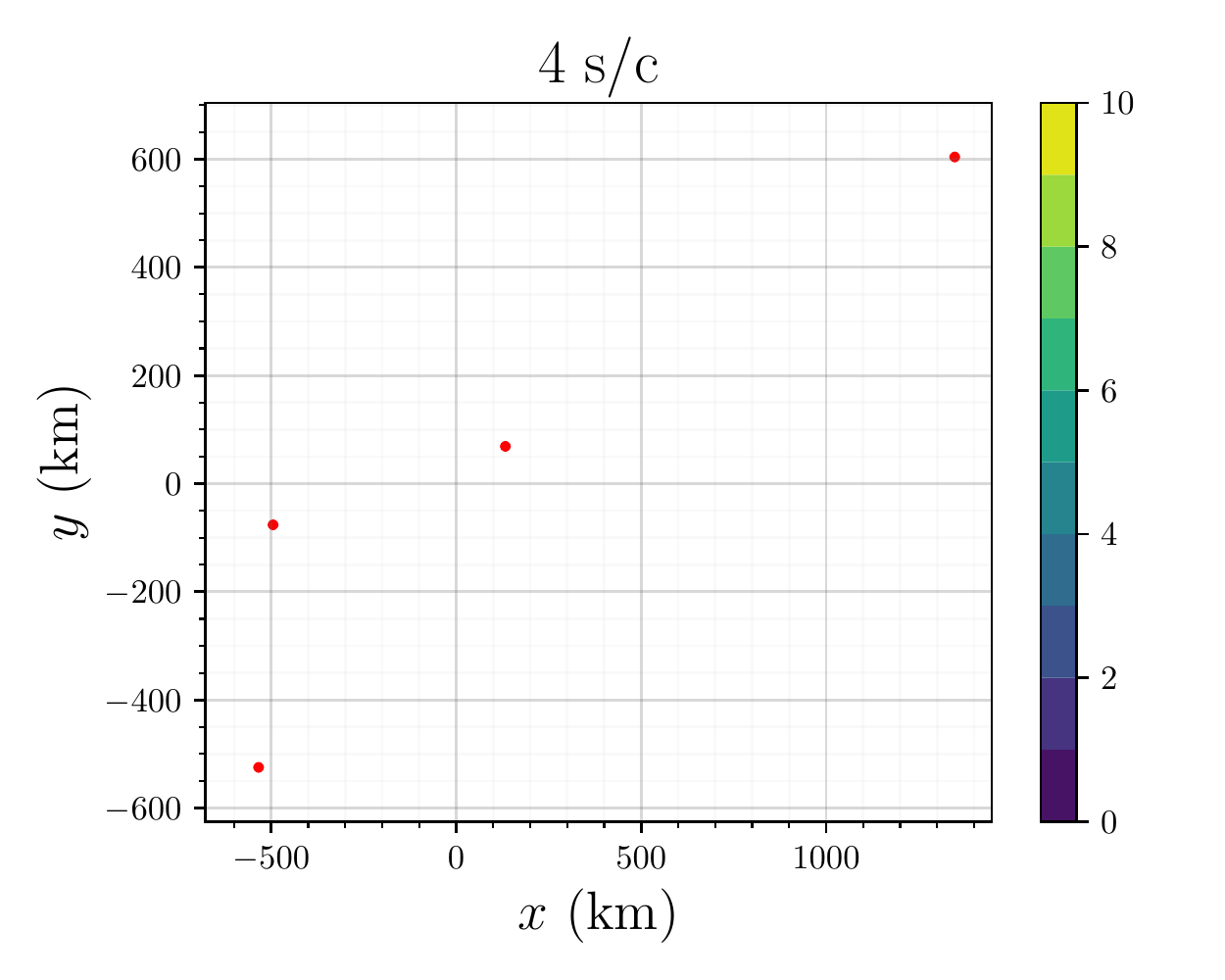}&
\includegraphics[width=0.3\textwidth]{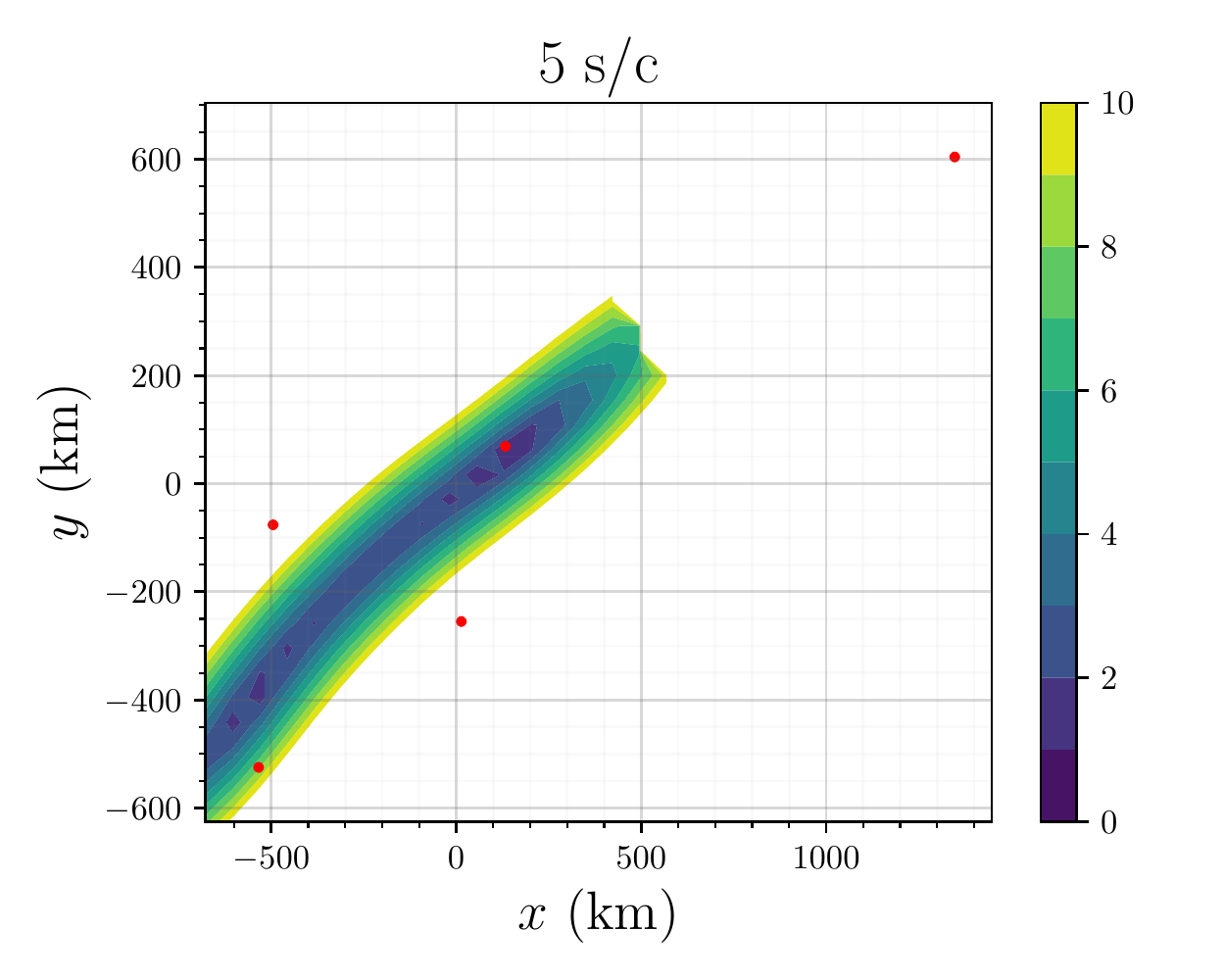}&
\includegraphics[width=0.3\textwidth]{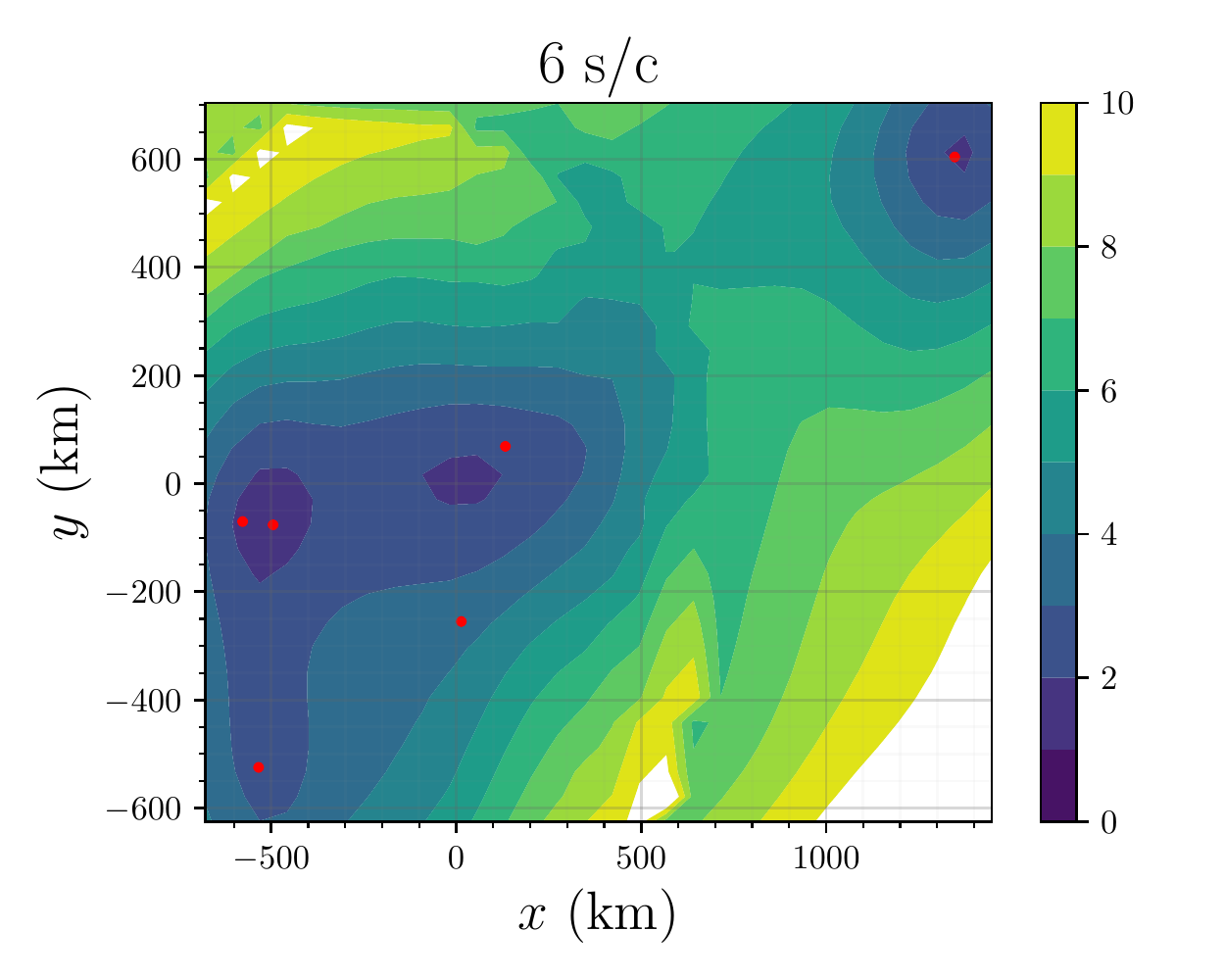}\\
\includegraphics[width=0.3\textwidth]{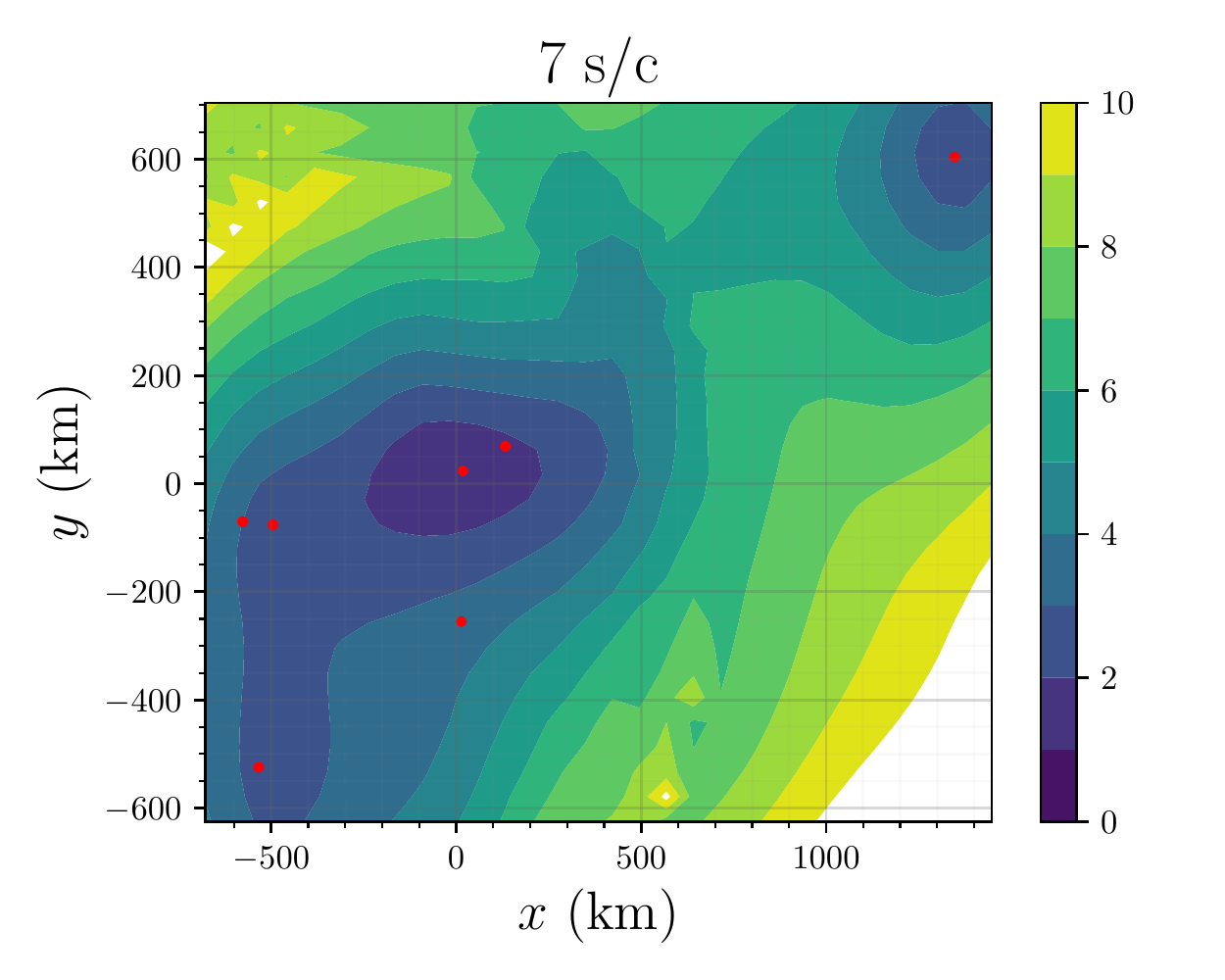}&
\includegraphics[width=0.3\textwidth]{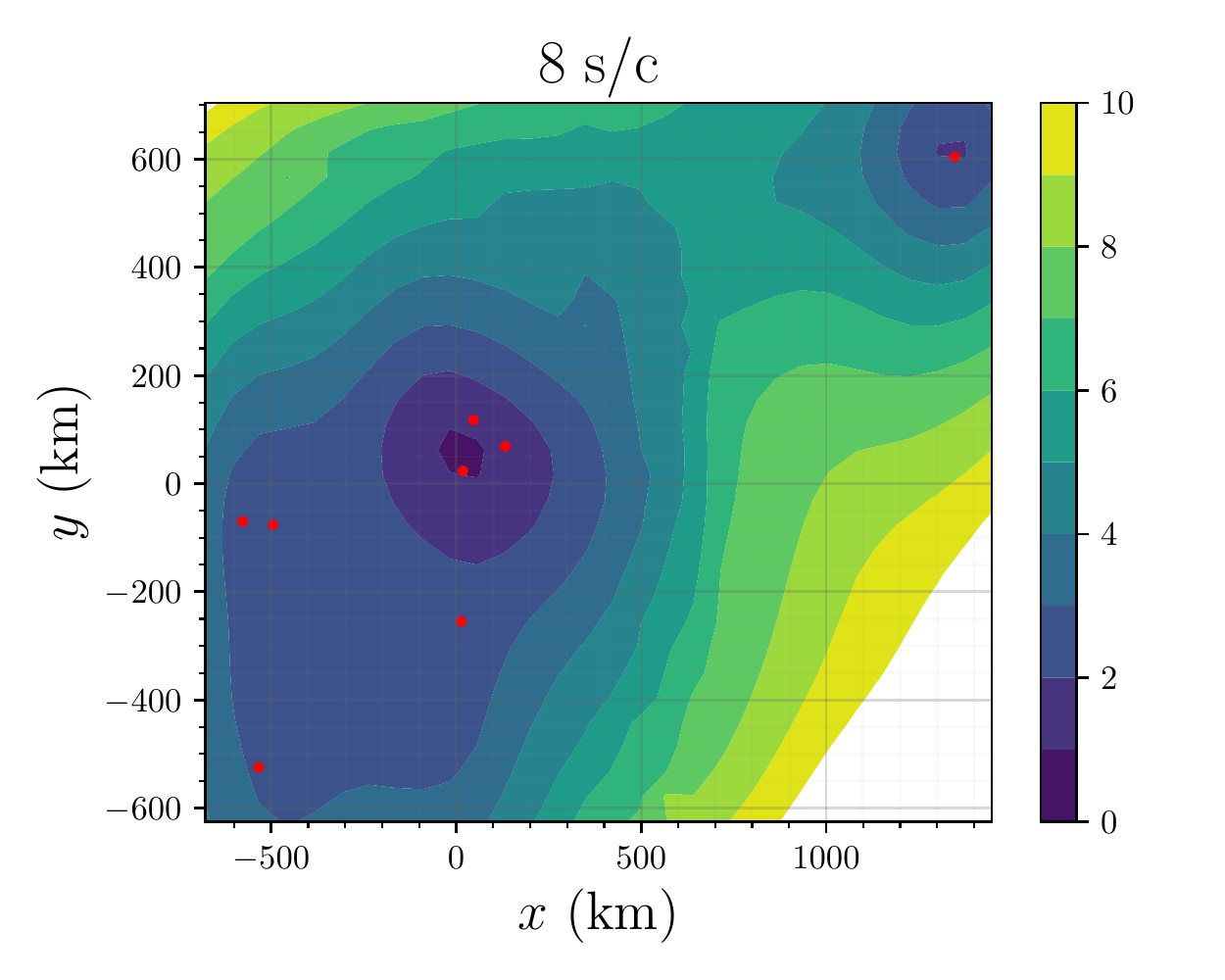}&
\includegraphics[width=0.3\textwidth]{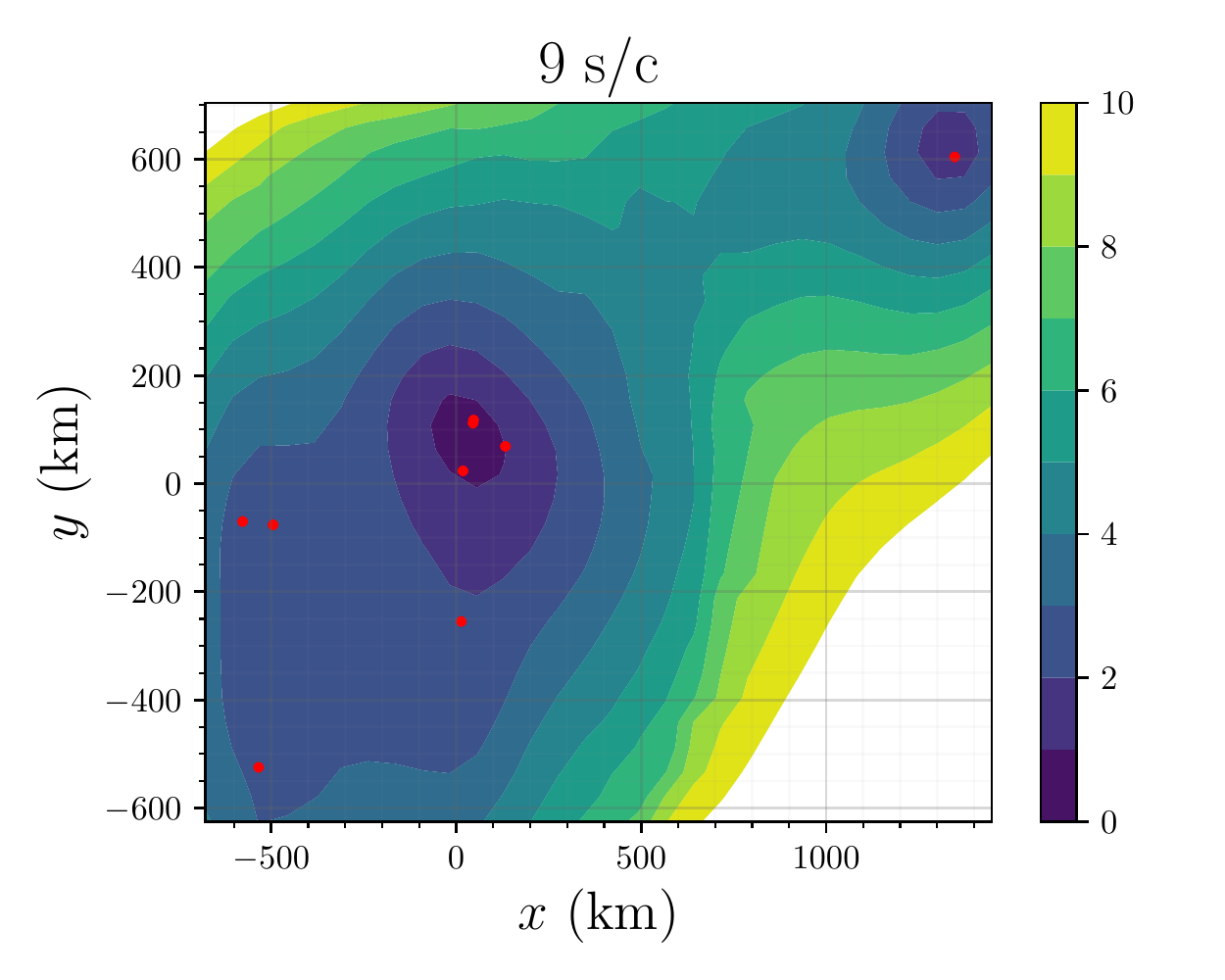}
\end{tabular}
\caption{Computation error (defined in equation \ref{eqn:B_error}) at all points on the $z=0$ plane of the turbulent magnetic field (from the \gkeyll\ Simulation), using the first-order method $M_{1,3}$ with a subset of the spacecraft from the hour 94 configuration of the HelioSwarm DRM.}
\label{fig:N_sc} 
\end{figure*}

For $N \in \{4,5,6,7,8,9\}$ spacecraft, we reconstructed the value of the magnetic field at all $30\times 30\times 30$ points $\xi$ using all $C(N,4)$ tetrahedra. We then use the $M_{1.3}$ first-order reconstruction method of \S \ref{sssec:tetra_inclusion} to reconstruct $\B$ at all points $\xi$. The errors everywhere are computed using equation \ref{eqn:B_error}, and the volume where the error is less than 5\% is computed using equation \ref{eqn:epsilon_3} multiplied by the total reconstructed volume. We visualize the errors of this method for the hour 94 configuration in Figure \ref{fig:N_sc}. In this example, we find that as the number of spacecraft is increased, the area which is reconstructed with a high accuracy also increases. As this result depends on which particular subset of spacecraft are chosen for a given N, we next investigate whether this increase is holds for an arbitrary selection of spacecraft. 

\begin{figure}[ht]
    \includegraphics[width=\columnwidth]{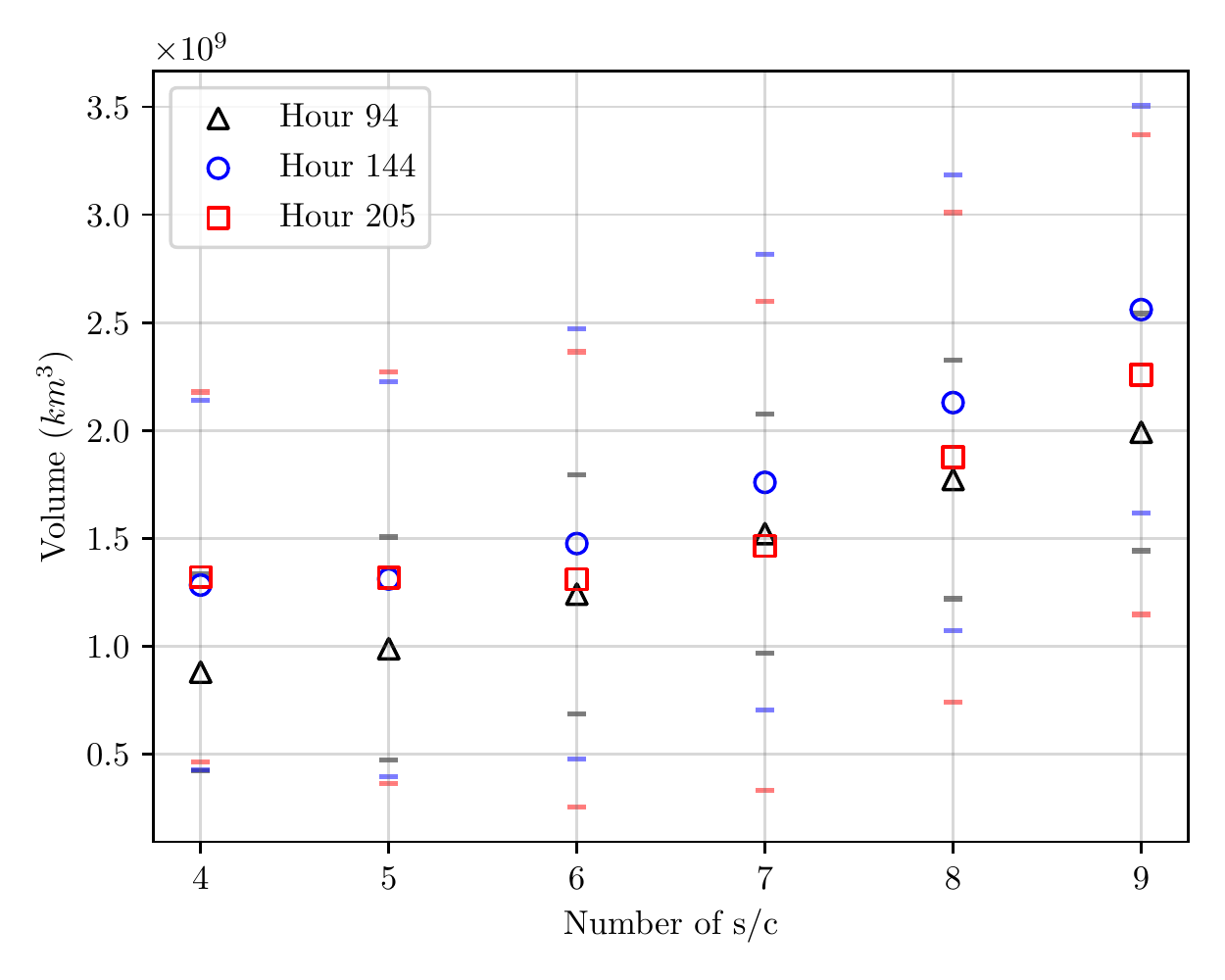}
    \caption{Mean values of volume which were reconstructed with less than 5\% error for the three nine-spacecraft configurations analyzed. The dashes above/below the markers represent one standard deviation away from the mean volume for each configuration.}
    \label{fig:robust}
\end{figure}

We start by choosing 4 out of the 9 spacecraft of the hour 94 configuration. These spacecraft measurements are used to estimate the value of $\B$ everywhere via the first-order reconstruction method $M_{1.3}$. We find the volume over which we can reconstruct $\B$ with an error less than 5\%. This process is repeated for all 126 possible choices of 4 spacecraft. We repeat all of these volume calculations, initializing the spacecraft configuration at 50 different locations within the simulated turbulent $\B$ field. Finally, we take the mean of all $126\times 50$ volume values and plot them in Figure \ref{fig:robust}. In these averages, we omit the instances where no tetrahedra pass the selection criteria of method $M_{1.3}$. We repeat this process for $N=5, 6, 7, 8$, and $9$ spacecraft from the hour 94 configuration, as well as for the hour 144 and 205 configurations.

As shown in Figure~\ref{fig:robust}, we see that increasing the number of spacecraft measurements available increases the volume of the magnetic field reconstructed with less than 5\% error. The variance of this reconstructed volume is smallest for the hour 94 configuration, which contains the most tetrahedra which are quasi-regular ($\chi_j \leq 1$). However, it is not the case that the hour 94 configuration has the highest average volume which is reconstructed with less than 5\% error.

We also track the instances where zero of the available tetrahedra in the set of $N$ spacecraft meet the shape threshold of $\chi_j \leq 1$ for the $M_{1.3}$ method. The percentage of arrangements where this occurs is shown in Table \ref{tab:Percent_poor_tetra} as a function of spacecraft configuration (hour) and number of spacecraft, $N$. We see from this table that for the analyzed configurations, there must be at least seven spacecraft measurements to guarantee that at least one tetrahedron passes the previously stated shape criteria.

\begin{table}
\centering
    \begin{tabular}{l|llllll}
    N & 4 & 5 & 6 & 7 & 8 & 9 \\ \hline
    Hour 94  & 57.1 & 13.5 & 1.2 & 0 & 0 & 0 \\
    Hour 144 & 71.4 & 36.5 & 9.5 & 0 & 0 & 0 \\
    Hour 205 & 77.8 & 51.6 & 20.2 & 0 & 0 & 0
    \end{tabular}
    \caption{From the nine-spacecraft configurations of hours 94, 144, and 205, we select a subset of $N \in \{4,5,6,7,8,9\}$ spacecraft. We determine the probability that this $N$ spacecraft configuration does not contain a tetrahedron which passes the threshold shape requirements of first-order reconstruction method $M_{1.3}$.}
    \label{tab:Percent_poor_tetra}
\end{table}

\section{Discussion}
\label{sec:conclusion}
We have demonstrated that our reconstruction methods are an effective way to leverage magnetometer measurements from a configuration consisting of more than four spacecraft. We have defined a shape metric, $\chi$, for a tetrahedron of spacecraft which can be used as a threshold criterion. Estimates of magnetic field derived from tetrahedron which do meet the threshold value of $\chi$ will be discarded, as they are misshapen and therefore more likely to produce erroneous estimates. Finally, we have shown that increasing the number of spacecraft in a configuration will increase the volume over which the magnetic field can be accurately reconstructed, as well as increase the likelihood that some tetrahedra of spacecraft in the configuration are well shaped.

In Table \ref{tab:error.turb} we demonstrated that our second-order reconstruction method $M_{2}$ does not reconstruct the magnetic field with high accuracy over a large volume. However, we have shown that methods $M_{1.3}$ and $M_{1.4}$, which average over a subset of the many available tetrahedra formed by nine spacecraft, improves the field reconstruction. This work indicates that the subset of tetrahedra which should be averaged over needs to consider each tetrahedron's spacial proximity to the reconstructed point as well as its geometric properties. By comparing results from spacecraft configurations with different tetrahedral geometric configurations, we find that designing spacecraft trajectories which maximize the number of tetrahedra that are quasi-regular (i.e. $\chi \leq 1$) is essential to improving the accuracy of the reconstructed magnetic field.

This work can help optimize future multi-spacecraft missions, such as HelioSwarm. The selection of tetrahedra which are included in the calculation of $\B$ can be tuned to maximize the volume over which the field is reconstructed accurately, or it can be tuned to recreate $\B$ as accurately as possible over a small volume. The first-order methods discussed here can be applied to reconstruct any vector field which is sparsely sampled by in-situ measurements, as no assumptions are made about the physical properties of the field.

The first-order reconstruction method applied to a single tetrahedron reconstructs the magnetic field perfectly at each spacecraft location. However, using any of our proposed composite first-order reconstruction methods, which average over many of these reconstructions, negates this behavior. In future work, we plan to construct a weight function which, when introduced into the tetrahedral averaging, returns this desired limiting behavior. Additional future work could include characterizing methods of predicting the surface inside-of-which we have less than a prescribed error value for an arbitrary configuration of spacecraft. 

The authors would like to thank the HelioSwarm Science and Flight Dynamics teams for discussions and comments during the execution of this project, in particular Laura Plice and Jonathan Niehof.
This material is based upon High Performance Computing (HPC) resources supported by the University of Arizona TRIF, UITS, and Research, Innovation, and Impact (RII) and maintained by the UArizona Research Technologies department. The authors acknowledge the Texas Advanced Computing Center (TACC) at The University of Texas at Austin for providing HPC resources that have contributed to the research results reported within this paper. D.V.~is supported by the Science and Technology Facilities Council (STFC) Ernest Rutherford Fellowship ST/P003826/1 and STFC Consolidated Grant ST/S000240/1. J.M.T~is supported by NSF STR award AGS-1842638.

\bibliographystyle{aasjournal}
\bibliography{main.bib}

\begin{thebibliography}{}
\expandafter\ifx\csname natexlab\endcsname\relax\def\natexlab#1{#1}\fi
\providecommand{\url}[1]{\href{#1}{#1}}
\providecommand{\dodoi}[1]{doi:~\href{http://doi.org/#1}{\nolinkurl{#1}}}
\providecommand{\doeprint}[1]{\href{http://ascl.net/#1}{\nolinkurl{http://ascl.net/#1}}}
\providecommand{\doarXiv}[1]{\href{https://arxiv.org/abs/#1}{\nolinkurl{https://arxiv.org/abs/#1}}}

\bibitem[{Angelopoulos(2008)}]{Angelopoulos:2008}
Angelopoulos, V. 2008, Space Science Reviews, 141, 5,
  \dodoi{10.1007/s11214-008-9336-1}

\bibitem[{Burch {et~al.}(2016)Burch, Moore, Torbert, \& Giles}]{Burch:2016}
Burch, J.~L., Moore, T.~E., Torbert, R.~B., \& Giles, B.~L. 2016, Space Science
  Reviews, 199, \dodoi{10.1007/s11214-015-0164-9}

\bibitem[{Constantinescu {et~al.}(2006)Constantinescu, Glassmeier, Motschmann,
  Treumann, Fornaçon, \& Fränz}]{Constantinescu:2006}
Constantinescu, O.~D., Glassmeier, K.~H., Motschmann, U., {et~al.} 2006,
  Journal of Geophysical Research: Space Physics, 111,
  \dodoi{10.1029/2005JA011550}

\bibitem[{Dai {et~al.}(2020)Dai, Wang, Cai, Gonzalez, Hesse, Escoubet, Phan,
  Vasyliunas, Lu, Li, Kong, Dunlop, Nakamura, He, Fu, Zhou, Huang, Wang,
  Khotyaintsev, Graham, Retino, Zelenyi, Grigorenko, Runov, Angelopoulos,
  Kepko, Hwang, \& Zhang}]{Dai:2020}
Dai, L., Wang, C., Cai, Z., {et~al.} 2020, Frontiers in Physics, 8, 89,
  \dodoi{10.3389/fphy.2020.00089}

\bibitem[{Dunlop {et~al.}(1988)Dunlop, Southwood, Glassmeier, \&
  Neubauer}]{Dunlop:1988}
Dunlop, M.~W., Southwood, D.~J., Glassmeier, K.~H., \& Neubauer, F.~M. 1988,
  Advances in Space Research, 8, \dodoi{10.1016/0273-1177(88)90141-X}

\bibitem[{{Elsasser}(1950)}]{Elsasser:1950}
{Elsasser}, W.~M. 1950, Physical Review, 79, 183,
  \dodoi{10.1103/PhysRev.79.183}

\bibitem[{{Escoubet} {et~al.}(2001){Escoubet}, {Fehringer}, \&
  {Goldstein}}]{Escoubet:2001}
{Escoubet}, C.~P., {Fehringer}, M., \& {Goldstein}, M. 2001, Annales
  Geophysicae, 19, 1197, \dodoi{10.5194/angeo-19-1197-2001}

\bibitem[{{Forsyth} {et~al.}(2011){Forsyth}, {Lester}, {Fazakerley}, {Owen}, \&
  {Walsh}}]{Forsyth:2011}
{Forsyth}, C., {Lester}, M., {Fazakerley}, A.~N., {Owen}, C.~J., \& {Walsh},
  A.~P. 2011, Planet. Space Sci., 59, 598, \dodoi{10.1016/j.pss.2009.12.007}

\bibitem[{Fu {et~al.}(2020)Fu, Wang, Zong, Chen, He, Vaivads, \&
  Olshevsky}]{Fu:2020}
Fu, H.~S., Wang, Z., Zong, Q., {et~al.} 2020, Methods for Finding Magnetic
  Nulls and Reconstructing Field Topology (American Geophysical Union (AGU)),
  153--172, \dodoi{https://doi.org/10.1002/9781119509592.ch9}

\bibitem[{Fu {et~al.}(2015)Fu, Vaivads, Khotyaintsev, Olshevsky, André, Cao,
  Huang, Retinò, \& Lapenta}]{Fu:2015}
Fu, H.~S., Vaivads, A., Khotyaintsev, Y.~V., {et~al.} 2015, Journal of
  Geophysical Research A: Space Physics, 120, \dodoi{10.1002/2015JA021082}

\bibitem[{Goldreich \& Sridhar(1995)}]{Goldreich:1995}
Goldreich, P., \& Sridhar, S. 1995, Astrophys.~J., 438, 763

\bibitem[{{Hakim} {et~al.}(2006){Hakim}, {Loverich}, \& {Shumlak}}]{Hakim:2006}
{Hakim}, A., {Loverich}, J., \& {Shumlak}, U. 2006, 219, 418,
  \dodoi{10.1016/j.jcp.2006.03.036}

\bibitem[{{Klein} {et~al.}(2019){Klein}, {Alexandrova}, {Bookbinder},
  {Caprioli}, {Case}, {Chandran}, {Chen}, {Horbury}, {Jian}, {Kasper}, {Le
  Contel}, {Maruca}, {Matthaeus}, {Retino}, {Roberts}, {Schekochihin}, {Skoug},
  {Smith}, {Steinberg}, {Spence}, {Vasquez}, {TenBarge}, {Verscharen}, \&
  {Whittlesey}}]{Klein:2019:WP}
{Klein}, K.~G., {Alexandrova}, O., {Bookbinder}, J., {et~al.} 2019, arXiv
  e-prints, arXiv:1903.05740.
\newblock \doarXiv{1903.05740}

\bibitem[{{Li} {et~al.}(2016){Li}, {Howes}, {Klein}, \& {TenBarge}}]{Li:2015}
{Li}, T.~C., {Howes}, G.~G., {Klein}, K.~G., \& {TenBarge}, J.~M. 2016,
  Astrophys.~J.~Lett., 832, L24, \dodoi{10.3847/2041-8205/832/2/L24}

\bibitem[{Narita {et~al.}(2013)Narita, Nakamura, \& Baumjohann}]{Narita:2013}
Narita, Y., Nakamura, R., \& Baumjohann, W. 2013, Annales Geophysicae, 31,
  \dodoi{10.5194/angeo-31-1605-2013}

\bibitem[{{Orszag} \& {Tang}(1979)}]{Orszag:1979}
{Orszag}, S.~A., \& {Tang}, C.-M. 1979, J.~Fluid Mech., 90, 129,
  \dodoi{10.1017/S002211207900210X}

\bibitem[{{Paschmann} \& {Daly}(1998)}]{Paschmann:1998}
{Paschmann}, G., \& {Daly}, P.~W. 1998, ISSI Scientific Reports Series, 1

\bibitem[{{Paschmann} \& {Daly}(2008)}]{Paschmann:2008}
---. 2008, {Multi-Spacecraft Analysis Methods Revisited}

\bibitem[{{Pin{\c{c}}on} \& {Motschmann}(1998)}]{Motschmann:1998}
{Pin{\c{c}}on}, J.-L., \& {Motschmann}, U. 1998, ISSI Scientific Reports
  Series, 1, 65

\bibitem[{Plice {et~al.}(2020)Plice, Perez, \& West}]{Plice:2020}
Plice, L., Perez, A.~D., \& West, S. 2020, Advances in the Astronautical
  Sciences, 171

\bibitem[{Robert {et~al.}(1998)Robert, Dunlop, Roux, \& Chanteur}]{Robert:1998}
Robert, P., Dunlop, M.~W., Roux, A., \& Chanteur, G. 1998, ISSI Scientific
  Reports Series, 1

\bibitem[{{Schwartz} {et~al.}(2009){Schwartz}, {Bale}, {Fujimoto}, {Hellinger},
  {Kessel}, {Le}, {Liu}, {Louarn}, {Mann}, {Nakamura}, {Owen}, {Pin{\c{c}}on},
  {Sorriso-Valvo}, {Vaivads}, \& {Wimmer-Schweingruber}}]{Schwartz:2009}
{Schwartz}, S., {Bale}, S.~D., {Fujimoto}, M., {et~al.} 2009, arXiv e-prints,
  arXiv:0912.0856.
\newblock \doarXiv{0912.0856}

\bibitem[{{Srinivasan} \& {Shumlak}(2011)}]{Srinivasan:2011}
{Srinivasan}, B., \& {Shumlak}, U. 2011, Phys.~Plasmas, 18, 092113,
  \dodoi{10.1063/1.3640811}

\bibitem[{{Torbert} {et~al.}(2020){Torbert}, {Dors}, {Argall}, {Genestreti},
  {Burch}, {Farrugia}, {Forbes}, {Giles}, \& {Strangeway}}]{Torbert:2020}
{Torbert}, R.~B., {Dors}, I., {Argall}, M.~R., {et~al.} 2020, \grl, 47, e85542,
  \dodoi{10.1029/2019GL085542}

\bibitem[{{Verscharen} {et~al.}(2019){Verscharen}, {Klein}, \&
  {Maruca}}]{Verscharen:2019}
{Verscharen}, D., {Klein}, K.~G., \& {Maruca}, B.~A. 2019, Living Rev.~Solar
  Phys., 16, 5, \dodoi{10.1007/s41116-019-0021-0}

\bibitem[{{Wang} {et~al.}(2015){Wang}, {Hakim}, {Bhattacharjee}, \&
  {Germaschewski}}]{Wang:2015}
{Wang}, L., {Hakim}, A.~H., {Bhattacharjee}, A., \& {Germaschewski}, K. 2015,
  Phys.~Plasmas, 22, 012108, \dodoi{10.1063/1.4906063}

\bibitem[{{Wang} {et~al.}(2020){Wang}, {Hakim}, {Ng}, {Dong}, \&
  {Germaschewski}}]{Wang:2020}
{Wang}, L., {Hakim}, A.~H., {Ng}, J., {Dong}, C., \& {Germaschewski}, K. 2020,
  J.~Comp.~Phys., 415, 109510, \dodoi{10.1016/j.jcp.2020.109510}

\end{thebibliography}
\appendix

\section{First-Order System}
\label{sec:appendix.method_O1}
We can reformat the Taylor expansion of equation \ref{eqn:curl_O1} into a linear system of form $A\V{x}=\V{b}$. We see this by first explicitly writing out one of the equations associated with \ref{eqn:curl_O1}, i.e. for spacecraft $i$ and component $m$. Therefore, equation \ref{eqn:curl_O1} becomes
\begin{equation}
\hat{B}_m^{i} = B_m + \sum_{k \in \{x,y,z\}} \partial_k B_{m} r_{k}^{i} . \label{eqn:taylor_m}
\end{equation}
By writing out all terms in the sum, we see that this is 
\begin{equation}
\hat{B}_m^{i} = B_m + \partial_x B_{m} r_{x}^{i} +  \partial_y B_{m} r_{y}^{i} +  \partial_z B_{m} r_{z}^{i} . \label{eqn:taylor_m_exp}
\end{equation}
We now write this in vector notation as 
\begin{equation}
    \hat{B}_m^{i} = \begin{bmatrix}
     1 & r_x^{i} & r_y^{i} & r_z^{i}
    \end{bmatrix}
    \begin{bmatrix}
     B_m \\ \partial_x B_{m} \\ \partial_y B_{m} \\ \partial_z B_{m}
    \end{bmatrix}. \label{eqn:taylor_m_vec}
\end{equation}
If we repeat this process for all spacecraft $i \in \{1,2,3,4\}$ (only using component $m$), we find that we can combine the four vector equations into one matrix equation
\begin{equation}
   \begin{bmatrix}
   \hat{B}_m^{1} \\
   \hat{B}_m^{2} \\
   \hat{B}_m^{3} \\
   \hat{B}_m^{4} 
   \end{bmatrix} 
    = \begin{bmatrix}
     1 & r_x^{1} & r_y^{1} & r_z^{1} \\
     1 & r_x^{2} & r_y^{2} & r_z^{2} \\
     1 & r_x^{3} & r_y^{3} & r_z^{3} \\
     1 & r_x^{4} & r_y^{4} & r_z^{4}
    \end{bmatrix}
    \begin{bmatrix}
     B_m \\ \partial_x B_{m} \\ \partial_y B_{m} \\ \partial_z B_{m}
    \end{bmatrix}. \label{eqn:taylor_m_matrix}
\end{equation}
Now we need to generalize this to include all values of $m$. We first note that the matrix from equation \ref{eqn:taylor_m_matrix} is independent of $m$. This independence means that if we replace $m$ in equation \ref{eqn:taylor_m_matrix} with $x$, $y$, or $z$, this matrix will stay the same. By doing this procedure, we see the three systems are:
\begin{equation}
   \begin{bmatrix}
   \hat{B}_x^{1} \\
   \hat{B}_x^{2} \\
   \hat{B}_x^{3} \\
   \hat{B}_x^{4} 
   \end{bmatrix} 
    = \begin{bmatrix}
     1 & r_x^{1} & r_y^{1} & r_z^{1} \\
     1 & r_x^{2} & r_y^{2} & r_z^{2} \\
     1 & r_x^{3} & r_y^{3} & r_z^{3} \\
     1 & r_x^{4} & r_y^{4} & r_z^{4}
    \end{bmatrix}
    \begin{bmatrix}
     B_x \\ \partial_x B_{x} \\ \partial_y B_{x} \\ \partial_z B_{x}
    \end{bmatrix} \label{eqn:taylor_x_matrix}
\end{equation}
\begin{equation}
   \begin{bmatrix}
   \hat{B}_y^{1} \\
   \hat{B}_y^{2} \\
   \hat{B}_y^{3} \\
   \hat{B}_y^{4} 
   \end{bmatrix} 
    = \begin{bmatrix}
     1 & r_x^{1} & r_y^{1} & r_z^{1} \\
     1 & r_x^{2} & r_y^{2} & r_z^{2} \\
     1 & r_x^{3} & r_y^{3} & r_z^{3} \\
     1 & r_x^{4} & r_y^{4} & r_z^{4}
    \end{bmatrix}
    \begin{bmatrix}
     B_y \\ \partial_x B_{y} \\ \partial_y B_{y} \\ \partial_z B_{y}
    \end{bmatrix} \label{eqn:taylor_y_matrix}
\end{equation}
\begin{equation}
   \begin{bmatrix}
   \hat{B}_z^{1} \\
   \hat{B}_z^{2} \\
   \hat{B}_z^{3} \\
   \hat{B}_z^{4} 
   \end{bmatrix} 
    = \begin{bmatrix}
     1 & r_x^{1} & r_y^{1} & r_z^{1} \\
     1 & r_x^{2} & r_y^{2} & r_z^{2} \\
     1 & r_x^{3} & r_y^{3} & r_z^{3} \\
     1 & r_x^{4} & r_y^{4} & r_z^{4}
    \end{bmatrix}
    \begin{bmatrix}
     B_z \\ \partial_x B_{z} \\ \partial_y B_{z} \\ \partial_z B_{z}
    \end{bmatrix}. \label{eqn:taylor_z_matrix}
\end{equation}
We now combine the three linear equations of \ref{eqn:taylor_x_matrix}, \ref{eqn:taylor_y_matrix}, and \ref{eqn:taylor_z_matrix} into a single linear system of $A \V{x} = \V{b}$ form. This single system's independent variable $\V{x}$ (and dependent variable $\V{b}$) will be a concatenation of the independent (and dependent) variables of systems \ref{eqn:taylor_x_matrix}, \ref{eqn:taylor_y_matrix}, and \ref{eqn:taylor_z_matrix}. The matrix of this combined linear system is a block matrix where the matrix from equation \ref{eqn:taylor_m_matrix} is a block on the diagonal of larger ($12 \times 12$) matrix. These three matrix blocks on the diagonal ensure that the matrix from equation \ref{eqn:taylor_m_matrix} is being applied to the $x$, $y$, and $z$ components independently. This system written out is:
\begin{equation}
 \begin{bmatrix} \hat{B}_x^{1} \\ \hat{B}_x^{2} \\ \hat{B}_x^{3} \\ \hat{B}_x^{4} \\ \hat{B}_y^{1} \\ \hat{B}_y^{2} \\ \hat{B}_y^{3} \\ \hat{B}_y^{4} \\ \hat{B}_z^{1} \\ \hat{B}_z^{2} \\ \hat{B}_z^{3} \\ \hat{B}_z^{4}  \end{bmatrix}  
=
\begin{bmatrix}
1 & r_x^{1} & r_y^{1} & r_z^{1} & 0 & 0 & 0 & 0 & 0 & 0 & 0 & 0 \\
1 & r_x^{2} & r_y^{2} & r_z^{2} & 0 & 0 & 0 & 0 & 0 & 0 & 0 & 0 \\
1 & r_x^{3} & r_y^{3} & r_z^{3} & 0 & 0 & 0 & 0 & 0 & 0 & 0 & 0 \\
1 & r_x^{4} & r_y^{4} & r_z^{4} & 0 & 0 & 0 & 0 & 0 & 0 & 0 & 0 \\
0 & 0 & 0 & 0 & 1 & r_x^{1} & r_y^{1} & r_z^{1} & 0 & 0 & 0 & 0 \\
0 & 0 & 0 & 0 & 1 & r_x^{2} & r_y^{2} & r_z^{2} & 0 & 0 & 0 & 0 \\
0 & 0 & 0 & 0 & 1 & r_x^{3} & r_y^{3} & r_z^{3} & 0 & 0 & 0 & 0 \\
0 & 0 & 0 & 0 & 1 & r_x^{4} & r_y^{4} & r_z^{4} & 0 & 0 & 0 & 0 \\
0 & 0 & 0 & 0 & 0 & 0 & 0 & 0 & 1 & r_x^{1} & r_y^{1} & r_z^{1} \\
0 & 0 & 0 & 0 & 0 & 0 & 0 & 0 & 1 & r_x^{2} & r_y^{2} & r_z^{2} \\
0 & 0 & 0 & 0 & 0 & 0 & 0 & 0 & 1 & r_x^{3} & r_y^{3} & r_z^{3} \\
0 & 0 & 0 & 0 & 0 & 0 & 0 & 0 & 1 & r_x^{4} & r_y^{4} & r_z^{4}
\end{bmatrix}
\begin{bmatrix} B_x \\ \partial_x B_{x} \\ \partial_y B_{x} \\ \partial_z B_{x} \\ B_y \\ \partial_x B_{y} \\ \partial_y B_{y} \\ \partial_z B_{y} \\ B_z \\ \partial_x B_{z} \\ \partial_y B_{z} \\ \partial_z B_{z}  \end{bmatrix} .
\label{eqn:order_1_system}
\end{equation}
We now solve this linear system and find the values of $B_x, \partial_x B_{x}, \partial_y B_{x},...$ to apply our first-order reconstruction method.

\section{Second-Order System}
\label{sec:appendix.method_O2}
Making a few adjustments, the second order system can be derived in exactly the same manner as the first. We see that for spacecraft $i$ and component $m$, equation \ref{eqn:curl_O2} is 
\begin{equation}
\hat{B}_m^{i} = B_m + \sum_{k \in \{x,y,z\}} \partial_k B_{m} r_{k}^{i} + \frac{1}{2}\sum_{j,k \in \{x,y,z\}} \partial_j \partial_k B_{m} r_{k}^{i} r_{j}^{i} . \label{eqn:taylor2_m}
\end{equation}
By writing out all of the terms in the sums, we see
\begin{align}
    \hat{B}_m^{i} = B_m + &\partial_x B_{m} r_{x}^{i} + \partial_y B_{m} r_{y}^{i} + \partial_z B_{m} r_{z}^{i} \label{eqn:taylor2_exp} \\
    + \frac{1}{2} \Bigg ( &\partial_x \partial_x B_{m} r_{x}^{i} r_{x}^{i} + \partial_x \partial_y B_{m} r_{x}^{i} r_{y}^{i} + \partial_x \partial_z B_{m} r_{x}^{i} r_{z}^{i} \nonumber\\
    + &\partial_y \partial_x B_{m} r_{y}^{i} r_{x}^{i} + \partial_y \partial_y B_{m} r_{y}^{i} r_{y}^{i} + \partial_y \partial_z B_{m} r_{y}^{i} r_{z}^{i} \nonumber\\
    + &\partial_z \partial_x B_{m} r_{z}^{i} r_{x}^{i} + \partial_z \partial_y B_{m} r_{z}^{i} r_{y}^{i} + \partial_z \partial_z B_{m} r_{z}^{i} r_{z}^{i} \Bigg ). \nonumber
\end{align}
We now use the fact that second derivatives are symmetric to see that
there are three redundant terms in this equation. This is because
\begin{align*}
    \partial_x \partial_y B_{m} r_{x}^{i} r_{y}^{i} &= \partial_y \partial_x B_{m} r_{y}^{i} r_{x}^{i} \\
    \partial_x \partial_z B_{m} r_{x}^{i} r_{z}^{i} &= \partial_z \partial_x B_{m} r_{z}^{i} r_{x}^{i} \\
    \partial_y \partial_z B_{m} r_{y}^{i} r_{z}^{i} &= \partial_z \partial_y B_{m} r_{z}^{i} r_{y}^{i} .
\end{align*}
Therefore equation \ref{eqn:taylor2_exp} can be simplified to 
\begin{align}
    \hat{B}_m^{i} = B_m + &\partial_x B_{m} r_{x}^{i} + \partial_y B_{m} r_{y}^{i} + \partial_z B_{m} r_{z}^{i} \label{eqn:taylor2_simple} \\
    + & \partial_x \partial_y B_{m} r_{x}^{i} r_{y}^{i} + \partial_x \partial_z B_{m} r_{x}^{i} r_{z}^{i} + \partial_y \partial_z B_{m} r_{y}^{i} r_{z}^{i} \nonumber \\
     + \frac{1}{2} \Bigg (&\partial_x \partial_x B_{m} r_{x}^{i} r_{x}^{i} + \partial_y \partial_y B_{m} r_{y}^{i} r_{y}^{i} + \partial_z \partial_z B_{m} r_{z}^{i} r_{z}^{i} \Bigg ). \nonumber
\end{align}

We write equation \ref{eqn:taylor2_simple} in vector notation
\begin{equation}
     \hat{B}_m^{i} = \frac{1}{2} \begin{bmatrix}
     2 & 2r_x^{i} & 2r_y^{i} & 2r_z^{i} & r_x^{i}r_x^{i} & 2r_x^{i}r_y^{i} & 2r_x^{i}r_z^{i} & r_y^{i}r_y^{i} & 2r_y^{i}r_z^{i} & r_z^{i}r_z^{i}
     \end{bmatrix} 
     \begin{bmatrix}
     B_m \\ \partial_x B_{m} \\ \partial_y B_{m} \\ \partial_z B_{m} \\ \partial_x\partial_x B_{m} \\ \partial_x\partial_y B_{m} \\ \partial_x\partial_z B_{m} \\ \partial_y\partial_y B_{m} \\ \partial_y\partial_z B_{m} \\ \partial_z\partial_z B_{m}
     \end{bmatrix}.
\label{eqn:taylor2_vec}
\end{equation}

We repeat this calculation for all nine spacecraft $i$ and combine the vector equations in the following matrix equation:
\begin{equation}
    \begin{bmatrix} \hat{B}_m^{1} \\ \hat{B}_m^{2} \\ \hat{B}_m^{3} \\ \hat{B}_m^{4} \\ \hat{B}_m^{5} \\ \hat{B}_m^{6} \\ \hat{B}_m^{7} \\ \hat{B}_m^{8} \\  \hat{B}_m^{9} \end{bmatrix}
    =
    \frac{1}{2}
\begin{bmatrix}
2 & 2r_x^{1} & 2r_y^{1} & 2r_z^{1} & r_x^{1}r_x^{1} & 2r_x^{1}r_y^{1} & 2r_x^{1}r_z^{1} & r_y^{1}r_y^{1} & 2r_y^{1}r_z^{1} & r_z^{1}r_z^{1} \\
2 & 2r_x^{2} & 2r_y^{2} & 2r_z^{2} & r_x^{2}r_x^{2} & 2r_x^{2}r_y^{2} & 2r_x^{2}r_z^{2} & r_y^{2}r_y^{2} & 2r_y^{2}r_z^{2} & r_z^{2}r_z^{2} \\
2 & 2r_x^{3} & 2r_y^{3} & 2r_z^{3} & r_x^{3}r_x^{3} & 2r_x^{3}r_y^{3} & 2r_x^{3}r_z^{3} & r_y^{3}r_y^{3} & 2r_y^{3}r_z^{3} & r_z^{3}r_z^{3}  \\
2 & 2r_x^{4} & 2r_y^{4} & 2r_z^{4} & r_x^{4}r_x^{4} & 2r_x^{4}r_y^{4} & 2r_x^{4}r_z^{4} & r_y^{4}r_y^{4} & 2r_y^{4}r_z^{4} & r_z^{4}r_z^{4} \\
2 & 2r_x^{5} & 2r_y^{5} & 2r_z^{5} & r_x^{5}r_x^{5} & 2r_x^{5}r_y^{5} & 2r_x^{5}r_z^{5} & r_y^{5}r_y^{5} & 2r_y^{5}r_z^{5} & r_z^{5}r_z^{5} \\
2 & 2r_x^{6} & 2r_y^{6} & 2r_z^{6} & r_x^{6}r_x^{6} & 2r_x^{6}r_y^{6} & 2r_x^{6}r_z^{6} & r_y^{6}r_y^{6} & 2r_y^{6}r_z^{6} & r_z^{6}r_z^{6} \\
2 & 2r_x^{7} & 2r_y^{7} & 2r_z^{7} & r_x^{7}r_x^{7} & 2r_x^{7}r_y^{7} & 2r_x^{7}r_z^{7} & r_y^{7}r_y^{7} & 2r_y^{7}r_z^{7} & r_z^{7}r_z^{7}  \\
2 & 2r_x^{8} & 2r_y^{8} & 2r_z^{8} & r_x^{8}r_x^{8} & 2r_x^{8}r_y^{8} & 2r_x^{8}r_z^{8} & r_y^{8}r_y^{8} & 2r_y^{8}r_z^{8} & r_z^{8}r_z^{8} \\
2 & 2r_x^{9} & 2r_y^{9} & 2r_z^{9} & r_x^{9}r_x^{9} & 2r_x^{9}r_y^{9} & 2r_x^{9}r_z^{9} & r_y^{9}r_y^{9} & 2r_y^{9}r_z^{9} & r_z^{9}r_z^{9}
\end{bmatrix}
\begin{bmatrix}
B_m \\ \partial_x B_{m} \\ \partial_y B_{m} \\ \partial_z B_{m} \\ \partial_x\partial_x B_{m} \\ \partial_x\partial_y B_{m} \\ \partial_x\partial_z B_{m} \\ \partial_y\partial_y B_{m} \\ \partial_y\partial_z B_{m} \\ \partial_z\partial_z B_{m}
\end{bmatrix}
\label{eqn:taylor2_matrix}
\end{equation}
We repeat the matrix in equation \ref{eqn:taylor2_matrix} three times as blocks on the diagonal of a larger matrix, one for each component $m \in \{x,y,z \}$. However, because the matrix from equation \ref{eqn:taylor2_matrix} is $9 \times 10$, this results in a matrix which is $27 \times 30$. A linear system with a matrix of dimension $27 \times 30$ is under-determined, and therefore, cannot be solved. We remedy this by using the known physical properties of the magnetic field $\B$. We know that magnetic fields have no monopoles, therefore the divergence of $\B$ and its gradient is zero, i.e.
\begin{align*}
    \nabla ( \nabla \cdot \B) &= \mathbf{0} \\
    \nabla \cdot \B &= 0 .
\end{align*}
Applying this fact, we must satisfy the following four equations:
\begin{align}
    \partial_x\partial_x B_{x} + \partial_x\partial_y B_{y} + \partial_x\partial_z B_{z} &= 0 \label{eqn:taylor2_constraints}\\
    \partial_y\partial_x B_{x} + \partial_y\partial_y B_{y} + \partial_y\partial_z B_{z} &= 0 \nonumber\\
    \partial_z\partial_x B_{x} + \partial_z\partial_y B_{y} + \partial_z\partial_z B_{z} &= 0 \nonumber\\
    \partial_x B_{x} + \partial_y B_{y} + \partial_z B_{z} &= 0 . \nonumber
\end{align}
Because these constraints are in terms of the same independent variables as the overall linear system (ie $\B$ and its partial derivatives), we can include these four constraints directly into our linear system. This is done by concatenating four rows onto the bottom of our $27 \times 30$ matrix, and adding four zeros onto the bottom of our dependent variable vector. These four rows are:
\begin{equation}
    \begin{bmatrix}
0 & 0 & 0 & 0 & 1 & 0 & 0 & 0 & 0 & 0 & 0 & 0 & 0 & 0 & 1 & 0 & 0 & 0 & 0 & 0 & 0 & 0 & 0 & 0 & 0 & 1 & 0 & 0 & 0 \\
0 & 0 & 0 & 0 & 0 & 1 & 0 & 0 & 0 & 0 & 0 & 0 & 0 & 0 & 0 & 0 & 1 & 0 & 0 & 0 & 0 & 0 & 0 & 0 & 0 & 0 & 0 & 1 & 0 \\
0 & 0 & 0 & 0 & 0 & 0 & 1 & 0 & 0 & 0 & 0 & 0 & 0 & 0 & 0 & 0 & 0 & 1 & 0 & 0 & 0 & 0 & 0 & 0 & 0 & 0 & 0 & 0 & 1 \\
0 & 1 & 0 & 0 & 0 & 0 & 0 & 0 & 0 & 0 & 0 & 1 & 0 & 0 & 0 & 0 & 0 & 0 & 0 & 0 & 0 & 0 & 1 & 0 & 0 & 0 & 0 & 0 & 0
\end{bmatrix}_{4 \times 30}.
\label{eqn:taylor2_constraint_01s}
\end{equation}
These four rows are constructed so that 1's are in positions corresponding to the partial derivatives in constraints \ref{eqn:taylor2_constraints}, and there are 0's everywhere else. 

In conclusion, by solving the linear system 
{\small
\begin{equation}
 \begin{bmatrix} \hat{B}_x^{1} \\ \hat{B}_x^{2} \\ \hat{B}_x^{3} \\ \hat{B}_x^{4} \\ \hat{B}_x^{5} \\ \hat{B}_x^{6} \\ \hat{B}_x^{7} \\ \hat{B}_x^{8} \\  \hat{B}_x^{9} \\ \hat{B}_y^{1} \\ \hat{B}_y^{2} \\ \hat{B}_y^{3} \\ \hat{B}_y^{4} \\ \hat{B}_y^{5} \\ \hat{B}_y^{6} \\ \hat{B}_y^{7} \\ \hat{B}_y^{8} \\ \hat{B}_y^{9} \\ \hat{B}_z^{1} \\ \hat{B}_z^{2} \\ \hat{B}_z^{3} \\ \hat{B}_z^{4} \\ \hat{B}_z^{5} \\ \hat{B}_z^{6} \\ \hat{B}_z^{7} \\ \hat{B}_z^{8} \\ \hat{B}_z^{9} \\ 0 \\ 0 \\ 0 \\ 0 \end{bmatrix}_{31 \times 1}
= \begin{bmatrix}
A & 0 & 0 \\
0 & A & 0 \\
0 & 0 & A \\
* & * & * \end{bmatrix}_{31 \times 30}
\begin{bmatrix} B_x \\ \partial_x B_{x} \\ \partial_y B_{x} \\ \partial_z B_{x} \\ \partial_x\partial_x B_{x} \\ \partial_x\partial_y B_{x} \\ \partial_x\partial_z B_{x} \\ \partial_y\partial_y B_{x} \\ \partial_y\partial_z B_{x} \\ \partial_z\partial_z B_{x} \\
B_y \\ \partial_x B_{y} \\ \partial_y B_{y} \\ \partial_z B_{y} \\  \partial_x\partial_x B_{y} \\ \partial_x\partial_y B_{y} \\ \partial_x\partial_z B_{y} \\ \partial_y\partial_y B_{y} \\ \partial_y\partial_z B_{y} \\ \partial_z\partial_z B_{y} \\ 
B_z \\ \partial_x B_{z} \\ \partial_y B_{z} \\ \partial_z B_{z} \\ \partial_x\partial_x B_{z} \\ \partial_x\partial_y B_{z} \\ \partial_x\partial_z B_{z} \\ \partial_y\partial_y B_{z} \\ \partial_y\partial_z B_{z} \\ \partial_z\partial_z B_{z} \\ \end{bmatrix}_{30 \times 1} .
\label{eqn:taylor2_system}
\end{equation}
}
for $B_x, \partial_x B_{x}, \partial_y B_{x},...$, we can apply the second-order reconstruction method. In system \ref{eqn:taylor2_system}:
\begin{equation*}
A = \frac{1}{2}
\begin{bmatrix}
2 & 2r_x^{1} & 2r_y^{1} & 2r_z^{1} & r_x^{1}r_x^{1} & 2r_x^{1}r_y^{1} & 2r_x^{1}r_z^{1} & r_y^{1}r_y^{1} & 2r_y^{1}r_z^{1} & r_z^{1}r_z^{1} \\
2 & 2r_x^{2} & 2r_y^{2} & 2r_z^{2} & r_x^{2}r_x^{2} & 2r_x^{2}r_y^{2} & 2r_x^{2}r_z^{2} & r_y^{2}r_y^{2} & 2r_y^{2}r_z^{2} & r_z^{2}r_z^{2} \\
2 & 2r_x^{3} & 2r_y^{3} & 2r_z^{3} & r_x^{3}r_x^{3} & 2r_x^{3}r_y^{3} & 2r_x^{3}r_z^{3} & r_y^{3}r_y^{3} & 2r_y^{3}r_z^{3} & r_z^{3}r_z^{3}  \\
2 & 2r_x^{4} & 2r_y^{4} & 2r_z^{4} & r_x^{4}r_x^{4} & 2r_x^{4}r_y^{4} & 2r_x^{4}r_z^{4} & r_y^{4}r_y^{4} & 2r_y^{4}r_z^{4} & r_z^{4}r_z^{4} \\
2 & 2r_x^{5} & 2r_y^{5} & 2r_z^{5} & r_x^{5}r_x^{5} & 2r_x^{5}r_y^{5} & 2r_x^{5}r_z^{5} & r_y^{5}r_y^{5} & 2r_y^{5}r_z^{5} & r_z^{5}r_z^{5} \\
2 & 2r_x^{6} & 2r_y^{6} & 2r_z^{6} & r_x^{6}r_x^{6} & 2r_x^{6}r_y^{6} & 2r_x^{6}r_z^{6} & r_y^{6}r_y^{6} & 2r_y^{6}r_z^{6} & r_z^{6}r_z^{6} \\
2 & 2r_x^{7} & 2r_y^{7} & 2r_z^{7} & r_x^{7}r_x^{7} & 2r_x^{7}r_y^{7} & 2r_x^{7}r_z^{7} & r_y^{7}r_y^{7} & 2r_y^{7}r_z^{7} & r_z^{7}r_z^{7}  \\
2 & 2r_x^{8} & 2r_y^{8} & 2r_z^{8} & r_x^{8}r_x^{8} & 2r_x^{8}r_y^{8} & 2r_x^{8}r_z^{8} & r_y^{8}r_y^{8} & 2r_y^{8}r_z^{8} & r_z^{8}r_z^{8} \\
2 & 2r_x^{9} & 2r_y^{9} & 2r_z^{9} & r_x^{9}r_x^{9} & 2r_x^{9}r_y^{9} & 2r_x^{9}r_z^{9} & r_y^{9}r_y^{9} & 2r_y^{9}r_z^{9} & r_z^{9}r_z^{9}
\end{bmatrix}_{9 \times 10} ,
\end{equation*}
and the $*$'s indicate the four rows representing the four constraints, written out in equation \ref{eqn:taylor2_constraint_01s}.

\newpage

\end{document}